\documentclass[12pt,a4paper]{article}
\pdfoutput=1

\usepackage[T1]{fontenc}
\usepackage[nomath]{lmodern}
\usepackage{microtype}
\usepackage{amsmath}
\usepackage{amsfonts}
\usepackage{amssymb}
\usepackage{mathrsfs}
\usepackage{physics}
\usepackage{graphicx}
\usepackage{caption}
\usepackage{subcaption}
\usepackage{jheppub}
\usepackage{enumitem}
\usepackage{framed,float}
\usepackage{booktabs}
\usepackage{array}
\usepackage{soul} 
\usepackage{tikz}
\usetikzlibrary{calc,arrows,decorations.markings}
\usepackage{multirow}

\newcommand{\PE}{\operatorname{PE}}

\newcolumntype{L}[1]{>{\raggedright\let\newline\\\arraybackslash\hspace{0pt}}m{#1}}
\newcolumntype{C}[1]{>{\centering\let\newline\\\arraybackslash\hspace{0pt}}m{#1}}
\newcolumntype{R}[1]{>{\raggedleft\let\newline\\\arraybackslash\hspace{0pt}}m{#1}}

\title{5d/6d Wilson loops from blowups}
\author[a,b]{Hee-Cheol Kim,}
\author[a]{Minsung Kim,}
\author[c]{Sung-Soo Kim,}

\affiliation[a]{Department of Physics, POSTECH, Pohang 790-784, Korea}
\affiliation[b]{Asia Pacific Center for Theoretical Physics, Postech, Pohang 37673, Korea}
\affiliation[c]{School of Physics, University of Electronic Science and Technology of China,\\
No. 2006 Xiyuan Ave, West Hi-Tech Zone, Chengdu, Sichuan 611731, China}

\abstract
{We generalize Nakajima-Yoshioka's blowup formula to calculate the partition functions counting the spectrum of bound states to half-BPS Wilson loop operators in 5d (and 6d) supersymmetric field theories. The partition function in the presence of a Wilson loop operator on the $\Omega$-background is factorized when put on the blowup $\hat{\mathbb{C}}^2$ into two Wilson loop partition functions under the localization. This structure provides a set of blowup equations for Wilson loop operators. We explain how to formulate the blowup equations and solve them to compute the partition functions of Wilson loop operators. We test this idea by explicitly calculating the Wilson loop partition functions in various 5d/6d field theories and comparing them against known results and expected dualities.
}

\begin{document}

\maketitle


\section{Introduction}\label{sec:intro}

The classification programs of 5d and 6d supersymmetric field theories, in the recent past, have provided a new avenue for improving our understanding of the quantum phenomena of higher dimensional field theories. Such programs have been executed by means of various methods such as M-/F-theory compactifications on local (elliptic) Calabi-Yau (CY) three-folds to 6d \cite{Heckman:2013pva, Heckman:2015bfa} and to 5d \cite{Douglas:1996xp, Intriligator:1997pq, DelZotto:2017pti, Xie:2017pfl, Jefferson:2017ahm, Bhardwaj:2019jtr, Bhardwaj:2019fzv, Apruzzi:2019vpe, Apruzzi:2019opn, Apruzzi:2019enx, Apruzzi:2019kgb} and the standard gauge theoretic technique in 6d \cite{Bhardwaj:2015xxa,Bhardwaj:2019hhd} and in 5d \cite{Seiberg:1996bd, Intriligator:1997pq, Jefferson:2017ahm, Bhardwaj:2020gyu}. 

Along with these programs, there has been quite a bit of progress recently in exploring the rich physics of 5d and 6d field theories via various observables protected by supersymmetry. The most notable examples are supersymmetric partition functions such as the spectrum of BPS particles on the $\Omega$-background, which are Nekrasov's instanton partition functions in 5d gauge theories \cite{Nekrasov:2002qd,Nekrasov:2003rj}, the elliptic genera of self-dual strings in 6d SCFTs, and the 5d/6d superconformal indices \cite{Bhattacharya:2008zy,Kim:2012gu,Kim:2012qf,Kim:2012tr,Kim:2013nva,Bergman:2013koa}. The ADHM constructions of the instanton moduli space have been used to calculate the Nekrasov's instanton partition functions in 5d \cite{Kim:2011mv,Rodriguez-Gomez:2013dpa,Bergman:2013ala,Hwang:2014uwa} and the elliptic genera of the self-dual strings in 6d \cite{Haghighat:2013gba,Kim:2014dza,Haghighat:2014vxa,Gadde:2015tra,Kim:2015fxa,Kim:2016foj,Kim:2018gjo,Kim:2018gak}. These partition functions can also be computed by the topological vertex method for the 5d/6d field theories realized by Type IIB 5-brane webs \cite{Aganagic:2003db,Iqbal:2007ii}. Recently, by extending the blowup formalism in \cite{Nakajima:2003pg,Nakajima:2005fg}, a universal blowup approach for computing the BPS spectra of arbitrary 5d/6d field theories was developed in \cite{Kim:2020hhh}. See also \cite{Gottsche:2006bm,Keller:2012da,Huang:2017mis,Gu:2018gmy,Gu:2019dan,Kim:2019uqw,Gu:2019pqj,Gu:2020fem,Duan:2021ges} for recent developments of the blowup formalism.

Supersymmetric Wilson loop operators are yet another protected observables that play an indispensable role in the inspection of strong coupling physics in supersymmetric gauge theories. The expectation values of BPS Wilson loop operators in 5d gauge theories on the $\Omega$-deformed $\mathbb{C}^2$ were calculated in \cite{Gaiotto:2014ina,Nekrasov:2015wsu,Kim:2016qqs,Assel:2018rcw,Gaiotto:2015una} using the ADHM quantum mechanics in the presence of Wilson loops. The analogous observables in 6d SCFTs are the BPS Wilson surface operators realized by probe strings carrying tensor charges as well as gauge charges \cite{Ganor:1996nf,Chen:2007ir}. The partition functions of the Wilson surfaces have been evaluated by generalizing the ADHM approaches in \cite{Bullimore:2014upa,Bullimore:2014awa,Agarwal:2018tso,Chen:2020jla}. Such loop (and surface) operators
\footnote{{\it Wilson surface operators} in 6d SCFTs are codimension-4 defect operators carrying tensor charges. In this paper we focus on the Wilson surface operators wrapping the 6d circle along which the 6d SCFT is compactified. These operators in the 5d KK theories become charged line operators under the gauge symmetry groups which inherit the 6d tensor symmetries. From the perspective of 5d KK theory, these line operators are on a par with ordinary Wilson loop operators for 6d gauge symmetries. 
Indeed, they are mapped to Wilson loop operators in dual 5d gauge theories, when such a duality exists, for 5d gauge symmetries. For this reason, we will often refer to them as Wilson loop operators in this paper.} 
have been used for non-trivial checks of dualities among a large class of 5d/6d gauge theories \cite{Gaiotto:2015una,Agarwal:2018tso,Assel:2018rcw}.

The main objective of this paper is to develop blowup equations for the partition functions enriched by Wilson loop operators (or Wilson surface operators in 6d theories) as new tools to calculate the spectrum of 1d BPS bound states with the loop operators in 5d and 6d supersymmetric field theories. More specifically, we consider the Wilson loop operators in 5d field theories on the blowup $\hat{\mathbb{C}}^2$ and the computation of the partition functions using localization. The factorization structure of the blowup partition functions without loop operators under the localization implies that the Wilson loop partition functions on the blowup $\hat{\mathbb{C}}^2$ will also be factorized into a pair of Wilson loop partition functions localized near two fixed points on $\mathbb{P}^1$ at the origin of $\hat{\mathbb{C}}^2$. A smooth blow-down transition back to the original $\mathbb{C}^2$ gives rise to a novel blowup equation for Wilson loop operators that relates the partition function of a Wilson loop operator on the $\Omega$-deformed $\mathbb{C}^2$ to a pair of Wilson loop partition functions on the same $\mathbb{C}^2$. A systematic algorithm for formulating these blowup equations will be explained in section \ref{sec:blowup-loop}.

We use the blowup equations for Wilson loop operators to calculate vacuum expectation values (VEVs) of Wilson loops in various representations that capture the spectrum of 1d BPS states bound to the loop operators. The ordinary blowup equations without loops can be solved by an iterative procedure \cite{Nakajima:2003pg,Nakajima:2005fg,Gottsche:2006bm}. The seeds for this iteration are the effective prepotential on the $\Omega$-background and a set of consistent magnetic fluxes \cite{Kim:2020hhh}. Likewise, we find that the blowup equations for Wilson loop operators can be solved iteratively by a similar procedure. In this case, the representation ${\bf r}$ of the Wilson loop operator will be used for an additional seed for the iteration process. We suggest that when the result from this process takes the right form of a 1d particle index, the solution of the blowup equation correctly produces the spectrum of 1d BPS bound states of the Wilson loop and therefore the VEV of the Wilson loop operator. We propose that the VEVs of Wilson loops (or Wilson surfaces) in generic 5d/6d field theories can be computed by employing this blowup approach (at least for the minimal representations). In addition, we illustrate that the blowup approach can be used to calculate partition functions of codimension-4 defects introduced by coupling 2d degrees of freedom to 6d theories on a circle with/without a twist. This will provide new connections between the codimension-4 defects and the Wilson surface operators in the 6d theories.

We test this proposal with several concrete examples including Wilson loops in the 5d CFT of a local $\mathbb{P}^2$, which will give the first example of evaluation of the VEVs of loop operators in 5d non-Lagrangian theories; Wilson loops in 5d gauge theories with exceptional gauge groups $G_2$ and $F_4$; and Wilson loops (or Wilson surfaces) in various 6d SCFTs. As we will see, this computation allows us to establish non-trivial maps between Wilson loop operators of different types in the pairs of dual gauge theories.

The paper is organized as follows. In section \ref{sec:Loop operators}, we discuss Wilson loop operators from the perspective of gauge theory as well as Calabi-Yau geometry. In section \ref{sec:Blowup}, we review the blowup equation which provides a systematic way of producing BPS spectrum of 5d/6d field theories and generalize it to bound states to Wilson loops. Namely, we discuss how to formulate the blowup equations in the presence of Wilson loop operators. We then test our proposal with various interesting 5d/6d field theories in section \ref{sec:examples}. In section \ref{sec:conclusion}, we summarize the result and discuss subtle issues and interesting directions to pursue.


\section{Loop operators}\label{sec:Loop operators}
In this section, we shall introduce 1/2 BPS loop operators on the Coulomb branch of 5d field theories. We will define the loop operators using both gauge theory descriptions and geometric descriptions.

\subsection{Loops in gauge theories}
In gauge theory, a natural loop operator is a Wilson loop which is a gauge-invariant observable defined by the trace of a path-ordered exponential of a gauge field $A_\mu$ around a 1d loop $C$. In a 5d $\mathcal{N}=1$ theory, we can define a 1/2 BPS Wilson loop operator as \cite{Young:2011aa,Assel:2012nf}
\begin{align}
	W_{\bf r}[C] = {\rm Tr}_{\bf r} \, \mathcal{P}\, {\rm exp}\int_{C}\left(iA_\mu \dot{x}^\mu + |\dot{x}|\phi\right)ds \, ,
\end{align}
where $\phi$ is the real scalar field in the vector multiplet, $x^\mu(s)$ is the worldline of the loop operator parametrized by $s$ and $\dot{x}^\mu\equiv dx^\mu/ds$. We shall consider the 5d gauge theory on $S^1\times \mathbb{R}^4$ with the $\Omega$-deformation. To preserve the supersymmetry, the Wilson loop operator is placed at the origin of $\mathbb{R}^4$ and stretched along the time circle $S^1$. This operator is now labeled by the representation ${\bf r}$ of the gauge group $G$\footnote{We will assume in this paper that all gauge groups are simply connected.}.

In this paper, we denote the representation ${\bf r}$ by its {\it lowest weight} contrary to a usual convention that uses the highest weight for this. The reason is that this convention for Wilson loops can be directly generalized to loop operators for the cases without a gauge description. This convention is also more convenient to discuss loop operators on the Coulomb branch. We will see this more clearly in the subsequent discussions.

On the Coulomb branch of the moduli space, the scalar field $\phi$ takes non-zero vacuum expectation values in the Cartan subalgebra of the gauge group $G$. Then the gauge group will be broken to its Abelian subgroup $G\rightarrow U(1)^r$ where $r$ is the rank of the gauge group $G$. The Wilson loop operators defined in the original non-Abelian gauge theory will now be labeled by their charges under the Abelian subgroup on the Coulomb branch. For example, a Wilson loop in the fundamental representation of $SU(2)$ gauge group can be perturbatively considered as a sum of two Wilson loops with charges $-1$ and $+1$ under the $U(1)$ gauge group on the Coulomb branch.

We are interested in the BPS spectrum in the presence of such Wilson loops on the Coulomb branch of a 5d field theory. The BPS spectrum without insertion of Wilson loop operators can be calculated by a Witten index defined as \cite{Nekrasov:2002qd}
\begin{align}\label{eq:index}
Z(\phi,m;\epsilon_1,\epsilon_2) = {\rm Tr}\left[(-1)^F e^{-\epsilon_1(J_1+J_R)} e^{-\epsilon_2(J_2+J_R)}e^{-\phi\cdot \Pi} e^{-m\cdot H}\right] \, .
\end{align}
This can also be regarded as the partition function of the 5d theory on $\Omega$-deformed $\mathbb{R}^4\times S^1$ (up to an overall regularization factor which we will explain shortly). Here $J_1,J_2$ are the Cartan generators of the $SO(4)$ Lorentz symmetry along $\mathbb{R}^4$, $J_R$ is the Cartan of the $SU(2)_R$ R-symmetry, and $\Pi$ and $H$ are the gauge and the flavor charges respectively. $\epsilon_1,\epsilon_2$ are the $SO(4)$ chemical potentials which are identified with the $\Omega$-deformation parameters, and $\phi$ and $m$ are the chemical potentials for the gauge and the flavor symmetries, respectively. $F$ is the fermion number operator.

This partition function is defined on the (extended) Coulomb branch where the scalar fields $\phi$ in the vector multiplets and the flavor mass parameters are turned on. The chemical potential $\phi$ on the Coulomb branch is in fact complexified by combining the scalar expectation value, which parametrizes the Coulomb branch of the moduli space, and the gauge holonomy around the time circle. However, we will take the chemical potentials $\phi$ as pure real values in the discussions below. Likewise, $m$ are also the complexified background flavor holonomies, but we regard them as real values.

We can recast the partition function as a composition of the regularization factor and the index factor:
\begin{align}\label{eq:partitionftn}
	Z(\phi,m;\epsilon_1,\epsilon_2) &= e^{\mathcal{E}(\phi,m;\epsilon_1,\epsilon_2)}Z_{GV}(\phi,m;\epsilon_1,\epsilon_2) \, ,  \\
	Z_{GV} (\phi,m;\epsilon_1,\epsilon_2) &= {\rm PE}\left[\,\sum_{j_l,j_r,{\bf d}}(-1)^{2(j_l+j_r)} N^{\bf d}_{j_l,j_r} \frac{\chi_{j_l}(p_1/p_2)\,\chi_{j_r}(p_1p_2)}{(p_1^{1/2}-p_1^{-1/2})(p_2^{1/2}-p_2^{-1/2})}e^{-{\bf d}\cdot {\bf m}}\right] \, , \nonumber
\end{align}
where PE$[f(\mu)]$ stands for the Plethystic exponential of a letter index $f(\mu)$ with a chemical potential $\mu$, ${\bf d}$ denotes the charge of a BPS state, ${\bf m}$ stands collectively for the chemical potentials $\phi,m$, and $N^{\bf d}_{j_l,j_r}$ is the degeneracy of a single-particle BPS state with spin $(j_l,j_r)$ and charge ${\bf d}$, and $\chi_j$ is the $SU(2)$ character of spin $j$. Also, $j_l= \tfrac{J_1-J_2}{2}$, $j_r= \tfrac{J_1+J_2}{2}$, and $p_{1,2}=e^{-\epsilon_{1,2}}$. 

In this expression, $Z_{GV}$ is the refined Gopakumar-Vafa (GV) invariants in \cite{Gopakumar:1998ii,Gopakumar:1998jq} that is the index part of the partition function capturing the spectrum of charged BPS particles. $\mathcal{E}$ in the prefactor is the {\it effective prepotential} in the Coulomb phase with background fields turned on \cite{Kim:2020hhh}. It is a collection of cubic and mixed Chern-Simons terms (and their SUSY completions) evaluated on the $\Omega$-background. Following the definition in \cite{Kim:2020hhh}, one finds
\begin{align}\label{eq:effprepot}
	\mathcal{E}(\phi,m;\epsilon_1,\epsilon_2)= \frac{1}{\epsilon_1\epsilon_2}\left[\mathcal{F}(\phi,m)+\frac{1}{48}C_i^G \phi^i (\epsilon_1^2+\epsilon_2^2)+\frac{1}{2}C^R_i \phi^i \epsilon_+^2\right] \, .
\end{align}
Here, $\mathcal{F}$ is the cubic prepotential in the Coulomb branch given by \cite{Witten:1996qb, Intriligator:1997pq}
\begin{equation}\label{eq:preF}
	\mathcal{F} \!= \sum_a\!\Big(\frac{m_a}{2}K^a_{ij}\phi^a_i\phi^a_j + \frac{\kappa_a}{6}d^a_{ijk}\phi^a_i\phi^a_j\phi^a_k\Big) +\frac{1}{12}\bigg(\sum_{e\in{\bf R}}|e\cdot \phi|^3-\sum_f\!\sum_{w\in{\bf w}_f}|w\cdot\phi+m_f|^3\!\bigg) \, ,
\end{equation}
where $a$ runs over all non-Abelian gauge groups $G_a\subset G$, $m_a= 1/g_a^2$ is the inverse gauge coupling squared associated with gauge group $G_a$, and $\kappa_a$ is the classical Chern-Simons level, which are non-zero only for $G_a=SU(N)$ with $N\ge3$. $K^a_{ij}={\rm Tr} (T^a_iT^a_j)$ is the Killing form of $G_a$ and $d^a_{ijk}=\frac{1}{2}{\rm Tr}T^a_i\{T^a_j,T^a_k\}$ with the generator $T^a_i$ in the fundamental representation of $G_a$. ${\bf R}$ and ${\bf w}_f$ are the roots and the weights of $G$, respectively, for the $f$-th hypermultiplet with masses $m_f$. The second and third terms in the square bracket in \eqref{eq:effprepot} are the contributions from mixed Chern-Simons terms with the gauge/gravitational Chern-Simons coefficient \cite{Bonetti:2013ela,Grimm:2015zea,Kim:2020hhh}
\begin{align}
	C_i^{G} = -\partial_i\bigg(\sum_{e\in {\bf R}}|e\cdot \phi| - \sum_f\sum_{w\in{\bf w}_f}|w\cdot\phi+m_f|\bigg) \, ,
\end{align}
and with the gauge/$SU(2)_R$ Chern-Simons coefficient \cite{BenettiGenolini:2019zth,Kim:2020hhh}
\begin{align}
	C^R_i = \frac{1}{2} \partial_i \sum_{e\in {\bf R}}|e\cdot \phi| \, ,
\end{align}
respectively. We refer to \cite{Kim:2020hhh} for more detailed discussions about the effective prepotential $\mathcal{E}$ and the Chern-Simons contributions, and also about their counterparts in local Calabi-Yau geometries and in twisted compactifications of 6d SCFTs.

Now we move on to the vacuum expectation value (VEV) of a BPS Wilson loop, which is defined as
\begin{align}\label{eq:VEV of BPS Wilson loop}
	\langle W_{\bf r}\rangle(\phi,m;\epsilon_1,\epsilon_2) = \frac{Z_{W_{\bf r}}(\phi,m;\epsilon_1,\epsilon_2)}{Z(\phi,m;\epsilon_1,\epsilon_2)} \, ,
\end{align}
where $Z_{W_{\bf r}}$ is the partition function on the $\Omega$-background with an insertion of the Wilson loop operator $W_{\bf r}$ wrapping the time circle at the origin of $\mathbb{R}^4$, whereas $Z$ is the bare partition function. The VEV of the Wilson loop operator turns out to be an Witten index counting 1d BPS bound states of the Wilson loop with the bulk BPS particles, including instantonic particles, supported along the worldline of the loop. Regarding this fact, we can recast the VEV of a Wilson loop as an index form of
\begin{align}\label{eq:wilson-state}
	\langle W_{\bf r}\rangle(\phi,m;\epsilon_1,\epsilon_2) = \sum_{j_l,j_r,{\bf d}}(-1)^{2(j_l+j_r)}\tilde{N}^{{\bf d}}_{j_l,j_r}\,\chi_{j_l}(p_1/p_2)\,\chi_{j_r}(p_1p_2) \,e^{-{\bf d}\cdot {\bf m}} \, ,
\end{align}
where $\tilde{N}^{{\bf d}}_{j_l,j_r}$ is the degeneracy of the 1d BPS state with charge ${\bf d}$ and spins $(j_l,j_r)$ bound to the Wilson loop. Note that unlike the form of the GV-invariant, this expression has no $\mathbb{R}^4$ momentum factor appearing in the denominator since these 1d states cannot move along the $\mathbb{R}^4$ direction.

Classically, the VEV of a Wilson loop $W_{\bf r}$ will be given by the character of the representation ${\bf r}$ written in terms of the gauge chemical potential $e^{-\phi}$. For example, the classical VEVs of Wilson loops in the fundamental and the symmetric representations of $SU(2)$ group, which are denoted by their lowest weight ${\bf r}=[-1]$ and ${\bf r}=[-2]$ respectively, are given by
\begin{align}\label{eq:su2-cls}
W^{\mathrm{cls}}_{[-1]} = e^{\phi} + e^{-\phi} \, , \qquad
W^{\mathrm{cls}}_{[-2]} = e^{2\phi}+1+e^{-2\phi} \, .
\end{align}
These VEVs can be considered as collections of 1d BPS bound states of an Abelian Wilson loop with bulk charged W-bosons in the Coulomb phase. Two states captured in $W^{\mathrm{cls}}_{[-1]}$ are, for example, from the 1d state with $U(1)$ charge $-1$ inserted by the Wilson loop operator and a single W-boson with $U(1)$ charge $+2$ bound to the 1d Wilson loop state respectively. Similarly, the first state in $W^{\mathrm{cls}}_{[-2]}$ is the 1d state with $U(1)$ charge $-2$ corresponding to the lowest weight of the $\mathbf{r}=[-2]$ representation, and the other two states are bound states of the 1d lowest weight state with one and two W-bosons respectively. Note here that the $U(1)$ charge of the bound states is weighted with respect to the chemical potential $e^{-\phi}$.

Wilson loop VEVs receive non-perturbative contribution on the instanton background. The instanton contributions to Wilson loop VEVs in various classical gauge groups have been computed using supersymmetric localization based on ADHM (or brane) constructions of instanton moduli space \cite{Bullimore:2014upa, Bullimore:2014awa, Gaiotto:2015una, Kim:2016qqs, Assel:2018rcw} or using auxiliary loop observables called qq-characters \cite{Nekrasov:2015wsu, Kimura:2015rgi, Bourgine:2016vsq, Bourgine:2019phm, Haouzi:2020yxy}. For instance, the 1-instanton correction to the fundamental Wilson loop in the pure $SU(2)$ gauge theory at $\theta=0$ is \cite{Bullimore:2014upa, Gaiotto:2015una, Assel:2018rcw}
\begin{align}\label{eq:su2-1-inst}
	\langle W^{k=1}_{[-1]} \rangle =- \frac{p_1p_2 (e^{-\phi}+e^{\phi})}{(1-p_1p_2e^{-2\phi})(1-p_1p_2e^{2\phi})} \, ,
\end{align}
and that for the symmetric Wilson loop is \cite{Assel:2018rcw}
\begin{align}\label{eq:su2_adj_adhm}
	\langle W^{k=1}_{[-2]} \rangle = \frac{(1-p_1)(1-p_2)(1+p_1p_2)-2p_1p_2( e^{-2\phi}+2+e^{2\phi})}{(1-p_1p_2e^{2\phi})(1-p_1p_2e^{-2\phi})} \, .
\end{align}
However, Wilson loop VEVs for exceptional gauge groups and also those in gauge theories with a large number of matters or with matters in general representations haven't been discussed so far. There is currently no method to compute Wilson loop VEVs for general gauge groups and matter content. The main purpose of this paper is to provide a systematic method for defining and computing partition functions of loop operators that can be applicable to arbitrary 5d QFTs, including gauge theories with exceptional gauge group and non-Lagrangian theories, as well as 5d Kaluza-Klein theories.

\subsection{Loops in Calabi-Yau threefolds}\label{sec:loop_cy}
Wilson loop operators are natural loop operators in gauge theories. However, a huge class of 5d field theories does not admit any mass deformations leading to gauge theory descriptions. Most of QFTs engineered by M-theory compactifications on local Calabi-Yau 3-folds are such theories. In those non-Lagrangian theories, the Wilson loops characterized by gauge invariant loop operators are not well-defined. Hence, we need more general definition of loop operators that can be broadly defined in arbitrary 5d field theories.

For a broader definition of Wilson loop operators, we will discuss how loop operators can arise from M-theory compactifications. In 5d field theories, Wilson loop operators can be considered as 1-dimensional defects arising from heavy BPS charged particles. At low energy below its mass scale, the charged particle will lose mobility along transverse directions and become a 1d defect stretched along its worldline.

In M-theory compactification, a charged BPS state comes from an M2-brane wrapping a holomorphic 2-cycle $C$ inside a Calabi-Yau 3-fold $X$. The mass of the BPS state is proportional to the volume of the 2-cycle, which is measured with respect to the K\"ahler form $J$ in geometry, and electric charge of the state under the low-energy Abelian symmetry group is given by the intersection number of the 2-cycle with holomorphic 4-cycles $D_I$: 
\begin{align}\label{eq:vol-charge}
	{\rm vol}(C) = -J\cdot C \, , \qquad e_I(C) = -C\cdot D_I \, ,
\end{align}
with
\begin{align}
	J = \sum_{I=1}^{h^{1,1}(X)}\phi_I D_I = \sum_{i=1}^r \phi_i S_i + \sum_{j=1}^{r_F}m_j N_j \, ,
\end{align}
where $D_{I=1,\cdots ,r} = S_{i=1,\cdots ,r}$ and $D_{I=r+1,\cdots ,h^{1,1}(X)} = N_{j=1,\cdots ,r_F}$ are the divisors for the compact and the non-compact 4-cycles inside $X$, respectively. $\phi_I$ are K\"ahler parameters for the divisors. $r$ is the dimension of dynamical K\"ahler moduli space of $X$, which is identified with the dimension of Coulomb branch moduli in the field theory, and $r_F=h^{1,1}(X)-r$ is the dimension of non-dynamical K\"ahler deformations corresponding to the rank of flavor symmetry group.

Loop operators then correspond to heavy wrapped M2-brane states in M-theory. Therefore, we interpret a Wilson loop in a field theory as an M2-brane state with infinite volume wrapping a non-compact holomorphic 2-cycle in the Calabi-Yau geometry. In geometry, any non-compact 2-cycle can always be decomposed into a linear sum of multiple compact 2-cycles and a {\it primitive non-compact 2-cycle}. If a non-compact 2-cycle is not decomposable, we call it a primitive non-compact 2-cycle. One distinguished property of the primitive non-compact 2-cycle ${\sf C}$ is that it always intersects non-negatively with compact divisors, i.e. ${\sf C}\cdot S_i \ge0$ for all $i$. Roughly speaking, this means the primitive non-compact 2-cycles are placed strictly outside any compact 4-cycles $S_i$ in $X$.

Then we can geometrically define a Wilson loop operator as an operator creating an M2-brane state wrapping a primitive non-compact 2-cycle ${\sf C}$. Note that this state from the primitive 2-cycle ${\sf C}$ carries negative electric charges for all Abelian gauge groups in the low energy field theory due to the fact that ${\sf C}\cdot S_i \ge0$. In fact, this state corresponds to the lowest weight of the representation of the Wilson loop when defined in a gauge theory. We will label this type of Wilson loop operators by the electric charges $e_i({\sf C})$ defined in (\ref{eq:vol-charge}) for the primitive 2-cycle ${\sf C}$, which is equivalent to labelling Wilson loops by their lowest weights in gauge theories. Namely, $W_{\bf r}$ for ${\bf r}=e_i({\sf C})$ refers to the loop operator coming from an M2-brane wrapping a primitive non-compact curve ${\sf C}$. In this geometric construction, BPS bound states to the loop operator are the M2-brane states wrapping non-primitive 2-cycles which are given by a linear combination of the primitive curve ${\sf C}$ and other compact curves $C_i$ as ${\sf C}+\sum_i n_i C_i$ with $n_i\ge0$. 

\begin{figure}[t]
\includegraphics[width=13cm]{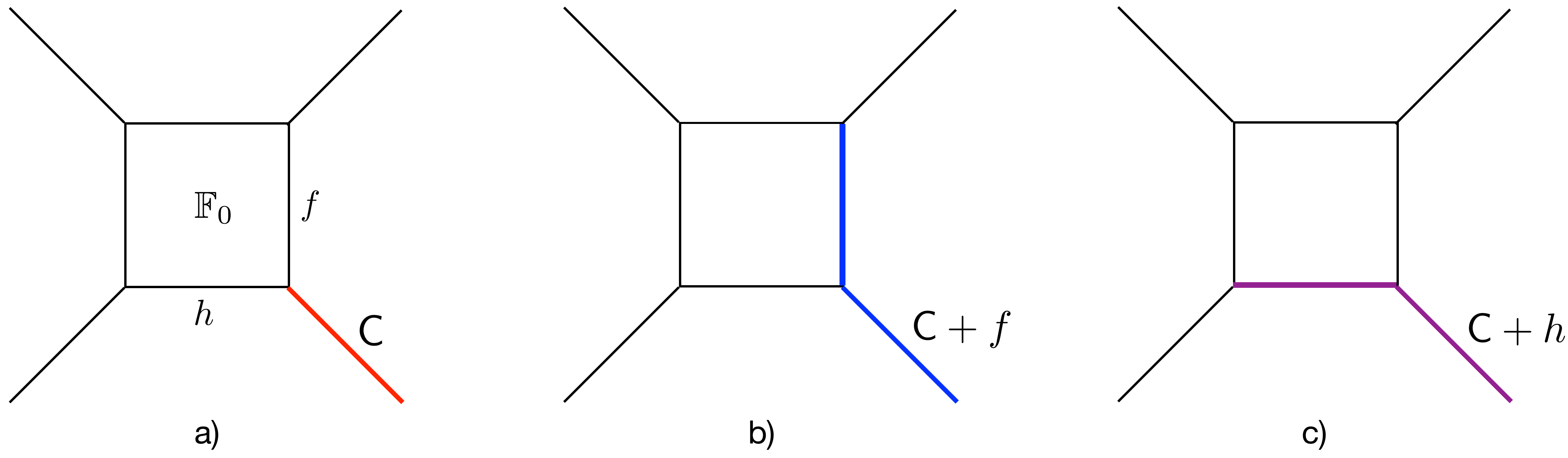}
\centering	
\caption{Geometric realization of Wilson loop states in a Hirzebruch surface $\mathbb{F}_0$ for the $SU(2)$ gauge theory at $\theta=0$ on the Coulomb branch. Red, blue and purple lines in a), b), and c) correspond to BPS bound states to the Wilson loop with $U(1)$ electric charges $-1, +1, +1$ respectively.}
\label{fig:F0}
\end{figure}

For example, the fundamental Wilson loop operator in the $SU(2)$ gauge theory at the discrete theta angle $\theta=0$ can be realized in geometry as follows. The $SU(2)_{\theta=0}$ gauge theory is engineered by compactifying M-theory on a local Calabi-Yau 3-fold containing a compact Hirzebruch surface $\mathbb{F}_0$. The fundamental Wilson loop of the $SU(2)$ gauge group can be realized by inserting a heavy M2-brane state wrapping a primitive non-compact 2-cycle ${\sf C}$ intersecting $\mathbb{F}_0$ at one point, which for example corresponds to the red line in Figure \ref{fig:F0} a). This operator will introduce a 1d defect with electric charge $-1$ to the low-energy Abelian gauge theory. Coupling this 1d state to the theory then induces other BPS bound states. Figure \ref{fig:F0} b) and c) illustrate two non-primitive 2-cycles ${\sf C}+f$ and ${\sf C}+h$, respectively, where $f$ is the fiber curve and $h$ is the base curve of $\mathbb{F}_0$. The M2-brane wrapping the non-compact curve ${\sf C}+f$ provides a bound state of the heavy M2-brane state of ${\sf C}$ and a W-boson with charge $+2$ coming from the fiber curve $f$. So the low-energy theory will have a 1d BPS bound state with electric charge $-1+2 = +1$ supported on the loop operator. Similarly, the M2-brane wrapping ${\sf C}+h$ introduces another 1d bound state to the Wilson loop. This state is now an instantonic loop state carrying both the gauge charge and the instanton charge $(+1,+1)$ as it comes from a combination of 1-instanton state from the base curve $h$ and the primitive state of ${\sf C}$. In this way, we can geometrically understand all other 1d BPS bound states to the Wilson loop state with larger electric charges.

We can now deduce that the Witten index in the presence of the Wilson loop operator can be expanded as
\begin{align}
\langle W_{[-1]}\rangle = e^{\phi} + e^{-\phi} + e^{-m - \phi} + \mathcal{O}(e^{-2\phi}) \, ,
\end{align}
where $m=1/g^2$ is the inverse of the $SU(2)$ gauge coupling being identified with the mass parameter for the $U(1)_T$ topological symmetry. The first term comes from the state of a), and the second and the third terms are from the states of b), c) respectively in Figure~\ref{fig:F0}. The exponent of each term amounts to the normalized volume of the non-compact curves wrapped by M2-branes after subtracting the infinite volume factor. Note that the first two terms are precisely the classical contribution of the $SU(2)$ fundamental Wilson loop operator in the gauge theory in \eqref{eq:su2-cls}, and the third term is the leading 1-instanton correction given in \eqref{eq:su2-1-inst} when expanded in terms of the Coulomb branch parameter $e^{-\phi}$.

Another interesting example is loop operators in the 5d SCFT arising from a local $\mathbb{P}^2$. The 5d SCFT engineered by a Calabi-Yau 3-fold containing a local $\mathbb{P}^2$ has no mass deformation and thus it has no gauge theory description. Hence we cannot define conventional Wilson loop operators in this theory. Instead, we can define loop operators in a geometric fashion in the same way as done for the $\mathbb{F}_0$ theory.

\begin{figure}[t]
\includegraphics[width=10cm]{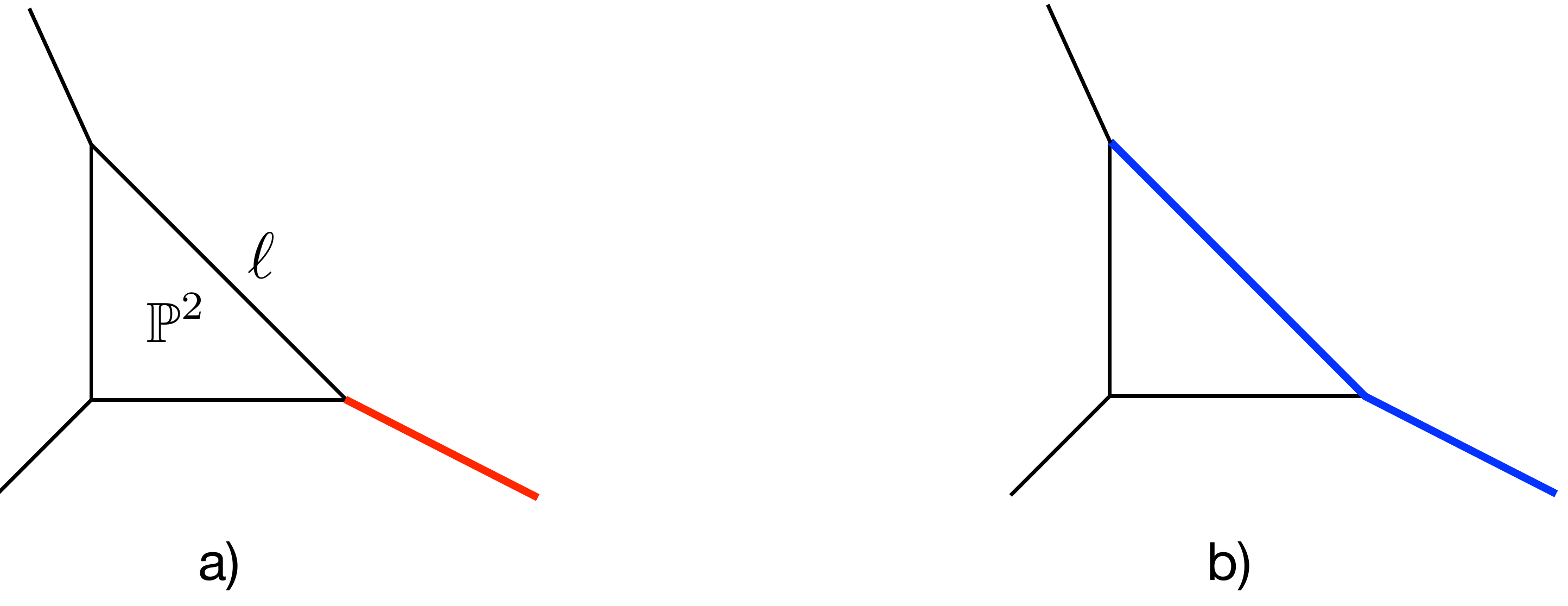}
\centering	
\caption{Geometric realization of Wilson loop states in a local $\mathbb{P}^2$. Red and blue lines in a) and b) correspond to BPS bound states to the Wilson loop with $U(1)$ electric charges $-1$ and $+2$ respectively on the Coulomb branch.}
\label{fig:P2}
\end{figure}

Figure~\ref{fig:P2} a) illustrates a loop operator from a primitive non-compact 2-cycle denoted by a red line intersecting with the $\mathbb{P}^2$ at one point. An M2-brane wrapped on this non-compact 2-cycle will induce a 1d Wilson loop state with $U(1)$ charge $-1$ in the Coulomb phase of the 5d theory. The blue line in Figure~\ref{fig:P2} b), which is a non-primitive 2-cycle, illustrates the first bound state of the 1d loop state with a bulk BPS particle with charge $+3$ coming from a curve $\ell$ in $\mathbb{P}^2$. Thus we expect the Witten index of the Wilson loop operator corresponding to the red line in Figure~\ref{fig:P2} a) has the following form:
\begin{align}
	\langle W_{[-1]}\rangle = e^{\phi} + c(\epsilon_1,\epsilon_2) e^{-2\phi} + \mathcal{O}(e^{-5\phi}) \, .
\end{align}
The coefficient $c(\epsilon_1,\epsilon_2)$, which will be computed below, encodes degeneracy and spins $(j_1,j_2)$ of the 1d bound state with charge $+2$.

Based on the geometric considerations as well as the discussions in gauge theories we propose a universal definition of our 1/2 BPS Wilson loop operators on the Coulomb branches of 5d QFTs as
\begin{framed}
\noindent {\bf Definition:} Wilson loop operator in a 5d supersymmetric field theory is a defect operator introduced by coupling the 5d bulk theory to an infinitely heavy {\it primitive 1d BPS state} with electric charge $e_i\le 0$ for all Abelian gauge groups on the Coulomb branch. This Wilson loop operator will be denoted by ${\bf r}=[e_1,\cdots, e_r]$, the electric charge obeying the Dirac quantization condition.
\end{framed}
This definition covers Wilson loops defined in gauge theories and geometric loop operators that we discussed above. Moreover, it allows us to define Wilson loop operators in more general 5d QFTs which do not admit either gauge theory descriptions or geometric realizations.

In a gauge theory, the 1d primitive state is defined as the Wilson loop state with electric charge given by the lowest weight (in Dynkin basis) of a representation ${\bf r}$ of the loop operator in the low-energy Abelian theory on the Coulomb branch. For consistency with geometric descriptions, in this paper we use Dynkin basis to label the representations of non-Abelian gauge group in the UV gauge theory. For a Wilson loop operator $W_{\bf r}$, the classical spectrum of 1d BPS bound states are simply fixed to be a sum of the states taking weights in the representation ${\bf r}$. We will discuss non-perturbative contributions to the Wilson loop spectrum in gauge theories in the next section.

By abuse of notation, we use $W_{\bf r}$ in gauge theories to refer to an insertion of a primitive 1d state with charge ${\bf r}$ and also other associated perturbative and non-perturbative bound states. We can also consider a collection of Wilson loops in representations ${\bf r}_1, {\bf r}_2, \cdots$ put on top of each other. In this case, we will denote them by $W_{{\bf r}_1\otimes{\bf r}_2\otimes \cdots}$ or $W_{{\bf r}_1}\otimes W_{{\bf r}_2}\otimes \cdots$.

On the other hand, when the 5d theory has no gauge theory description, there is no distinction between the classical and non-perturbative spectrum of loop operators. Moreover, it is possible that a state of a primitive curve ${\sf C}$ has the same charge as that of a non-primitive curve ${\sf C}'+\sum_i n_iC_i$ with spin $(0,0)$ from another primitive curve ${\sf C}'$. In this case, the spectrum itself cannot distinguish these two states. Thus, for the theories without gauge theory description, we define a Wilson loop operator by specifying all the spectrum of non-compact curves positively intersecting all compact divisors and carrying spin $(0,0)$. For example, we can insert to a local $\mathbb{P}^2$ a Wilson loop operator of ``representation'' ${\bf r}=[-4] + n[-1]$ with $n\ge0$. This means a loop operator with spectrum involving a primitive non-compact curve intersecting $\mathbb{P}^2$ at four points and $n$ non-compact curves, regardless of being primitive or not, intersecting $\mathbb{P}^2$ at one point.


\section{Blowup formula and loop operators}\label{sec:Blowup}
In this section, we introduce a systematic procedure to compute VEVs of Wilson loop operators using the blowup equations. We begin with a brief review of Nakajima-Yoshioka's blowup formula for computing the BPS spectra of 5d/6d QFTs without loops. We then explain how to generalize the blowup approach to VEVs of Wilson loops (or Wilson surfaces) in 5d/6d supersymmetric field theories.

\subsection{Review : Blowup equations}
The blowup partition function $\hat{Z}$ is defined on a blowup $\hat{\mathbb{C}}^2$ where the origin of the $\mathbb{C}^2$ is replaced by a 2-sphere $\mathbb{P}^1$. The blowup partition function of 4d and 5d $SU(N)$ SYMs was first computed by (Gottsche and) Nakajima and Yoshioka using localization in \cite{Nakajima:2003pg,Nakajima:2005fg,Gottsche:2006bm}. They have shown that upon the localization the blowup partition function is factorized into two partition functions on local $\hat{\mathbb{C}}^2$'s around the North and the South poles on $\mathbb{P}^1$. This leads to the celebrated blowup equation of the form
\begin{align}\label{eq:bleq}
&\Lambda(m_j;\epsilon_1,\epsilon_2)\hat{Z}(\phi_i, m_j; \epsilon_1,\epsilon_2) = \sum_{\vec{n}}(-1)^{|\vec{n}|}\hat{Z}^{(N)}(\vec{n},\vec{B}) \times \hat{Z}^{(S)}(\vec{n},\vec{B}) \, , 
\end{align}
where $\vec{n}=\{n_i\}$ (with $|\vec{n}|=\sum_i n_i$) denotes the magnetic fluxes on $\mathbb{P}^1$ for gauge group $G$, and $\vec{B}=\{B_j\}$ denotes the background magnetic fluxes for global symmetries. Here two local partition functions at the North and South poles $\hat{Z}^{(N)}$ and $\hat{Z}^{(S)}$ take the same form as the partition function $\hat{Z}$ with the shifted chemical potentials 
\begin{align}
	\hat{Z}^{(N)}(\vec{n},\vec{B}) &\equiv \hat{Z}(\phi_i\!+\!n_i\epsilon_1,m_j\!+\!B_j\epsilon_1;\epsilon_1,\epsilon_2\!-\!\epsilon_1) \, , \nonumber \\
	\hat{Z}^{(S)}(\vec{n},\vec{B}) &\equiv \hat{Z}(\phi_i\!+\!n_i\epsilon_2,m_j\!+\!B_j\epsilon_2;\epsilon_1\!-\!\epsilon_2,\epsilon_2) \, .
\end{align}
The prefactor $\Lambda$ on the LHS of \eqref{eq:bleq} is independent of the dynamical parameters $\phi_i$. When $\Lambda=0$, the blowup equation is called the vanishing blowup equation, and the blowup equation with non-trivial $\Lambda$ is called the unity blowup equation.

As explained in \cite{Kim:2020hhh}, the partition function $\hat{Z}$ is related to the usual partition function $Z$ defined in \eqref{eq:index} and \eqref{eq:partitionftn} by a simple replacement $(-1)^F \rightarrow (-1)^{2J_R}$, which is equivalent to the redefinition of the chemical potential for angular momentum in $Z_{GV}$ as $\epsilon_1\rightarrow \epsilon_1+2\pi i$. Therefore the blowup partition function $\hat{Z}$ can be recast in terms of the refined GV-invariant and the effective prepotential as
\begin{align}\label{eq:blowupptn}
\hat{Z}(\phi,m;\epsilon_1,\epsilon_2) &= e^{\mathcal{E}(\phi,m;\epsilon_1,\epsilon_2)}\hat{Z}_{GV}(\phi,m;\epsilon_1,\epsilon_2) \, , \nonumber \\
\hat{Z}_{GV}(\phi,m;\epsilon_1,\epsilon_2) &\equiv Z_{GV}(\phi,m;\epsilon_1+2\pi i,\epsilon_2) \, .
\end{align}

Magnetic fluxes $(\vec{n},\vec{B})$ appearing in the blowup equation must be properly quantized. The quantization condition that allows the theory to be consistently put on the blowup $\hat{\mathbb{C}}^2$ is as follows \cite{Huang:2017mis}:
\begin{align}
(\vec{n},\vec{B})\cdot e \ \text{is integral/half-integral, when } 2(j_l+j_r) \text{ is odd/even} \, ,
\end{align}
for all BPS particles, where $e$ here denotes electric gauge and flavor charges for a BPS particle with a spin $(j_l,j_r)$. For each choice of the magnetic fluxes, one can find a corresponding blowup equation. So, in principle, we can find more than one blowup equation for a given theory. However, there are special sets of magnetic fluxes, called {\it consistent magnetic fluxes}, which give rise to consistent blowup equations. Blowup equations with consistent magnetic fluxes are solvable and their solutions provide the correct spectrum of BPS states on the Coulomb branch of the theory. See \cite{Kim:2020hhh} for more detailed discussions on the quantification of magnetic fluxes and the conditions for these special magnetic fluxes, and also for various examples.

The basic strategy for solving the blowup equations is as follows. Since the blowup partition function takes the form of \eqref{eq:blowupptn}, we can first rewrite the blowup equations in terms of the refined GV-invariant as
\begin{align}\label{eq:bleq-GV}
	&\Lambda(m_j;\epsilon_{1},\epsilon_2) \hat{Z}_{GV}(\phi_i,m_j;\epsilon_1,\epsilon_2) = \sum_{\vec{n}}(-1)^{|\vec{n}|}e^{-V(\phi_i,m_j,\vec{n},\vec{B};\epsilon_1,\epsilon_2) } \nonumber \\
	&\times \hat{Z}_{GV} (\phi_i\!+\!n_i\epsilon_1,m_j\!+\!B_j\epsilon_1;\epsilon_1,\epsilon_2\!-\!\epsilon_1) \cdot \hat{Z}_{GV}(\phi_i\!+\!n_i\epsilon_2,m_j\!+\!B_j\epsilon_2;\epsilon_1\!-\!\epsilon_2,\epsilon_2) \, ,
\end{align}
where 
\begin{align}\label{eq:GV-V}
	&V(\phi_i,m_j,\vec{n},\vec{B};\epsilon_1,\epsilon_2) \equiv \  \mathcal{E}(\phi_i,m_j;\epsilon_1,\epsilon_2) \\
	& \qquad \quad - \mathcal{E}(\phi_i\!+\!n_i\epsilon_1,m_j\!+\!B_j\epsilon_1;\epsilon_1,\epsilon_2-\epsilon_1) - \mathcal{E}(\phi_i\!+\!n_i\epsilon_2,m_j\!+\!B_j\epsilon_2;\epsilon_1-\epsilon_2,\epsilon_2)\nonumber \, .
\end{align}
Now we use the fact that the GV-invariant should take the form of an index given in \eqref{eq:partitionftn} and hence it should be given as an expansion of the power series in terms of masses (or K\"ahler parameters) of BPS states $e^{-{\bf d}\cdot{\bf m}}$. Since the masses of BPS states are all non-negative on the Coulomb branch, the power series expansion is well-defined.\footnote{More precisely, ${\bf d}\cdot {\bf m}$ is the central charge of a BPS state of charge ${\bf d}$ in the Coulomb branch. The central charges for all the BPS states are non-negative on a sub-chamber of the Coulomb branch. Thanks to the BPS relation, the mass of a particle will be related to its central charge.} So we expand both sides of the blowup equation in \eqref{eq:bleq-GV}, and solve the equation at each order iteratively to determine the BPS degeneracies $N_{j_l,j_r}^{\bf d}$. For a given set of consistent magnetic fluxes, we expect the solution of the blowup equations to correctly yield the BPS spectrum of the theory.

\subsection{Blowup equations for loop operators}\label{sec:blowup-loop}

We shall now discuss blowup equations for the Wilson loop operators. On a blowup $\hat{\mathbb{C}}^2$, the partition function without loop operators is factorized into two local partition functions around two poles on $\mathbb{P}^1$ at the origin. It is then natural to expect that the partition function with an insertion of Wilson loop operators will also be factorized into two partition functions of certain loop operators located at the poles in the blowup $\hat{\mathbb{C}}^2$. Under a blow-down transition $\hat{\mathbb{C}}^2 \rightarrow \mathbb{C}^2$, these loop operators on the $\mathbb{P}^1$ should reduce to a collection of Wilson loop operators at the origin of $\mathbb{C}^2$. Therefore we would expect to have after the transition a functional equation relating the Wilson loop partition function on an ordinary $\mathbb{C}^2$ and the factorized partition function with loop operators on a blowup $\hat{\mathbb{C}}^2$.

\begin{figure}[t]
\includegraphics[width=15cm]{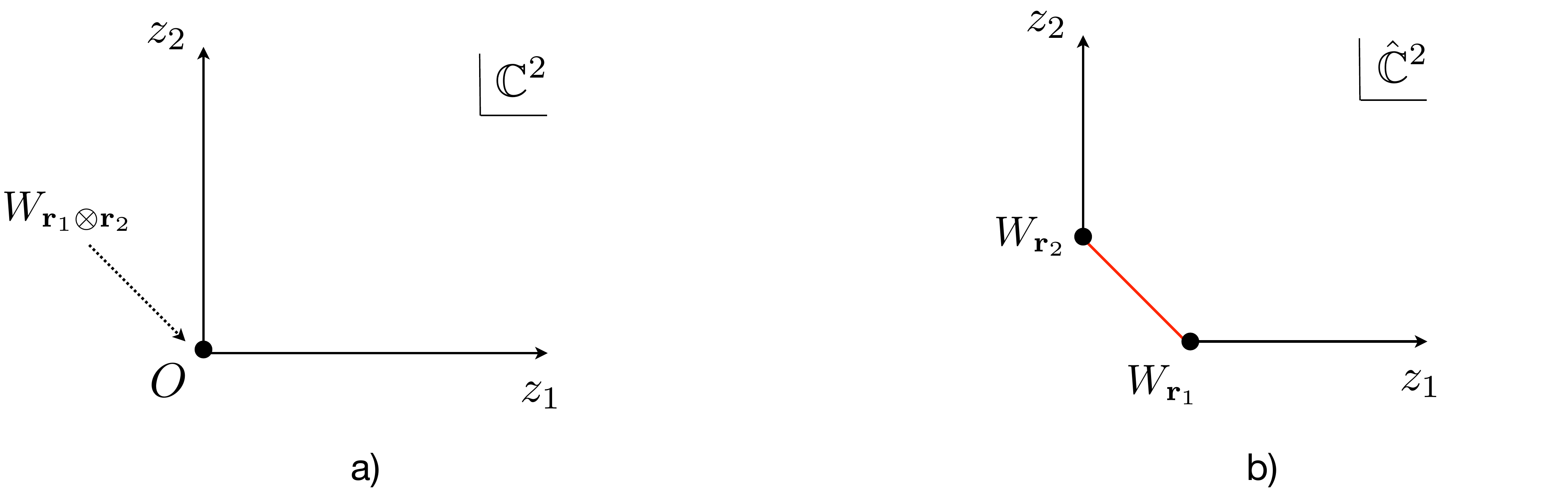}
\centering	
\caption{a) BPS Wilson loop $W_{{\bf r}_1\otimes{\bf r}_2}$ inserted at the origin of $\mathbb{C}^2$. b) The Wilson loop $W_{{\bf r}_1\otimes{\bf r}_2}$ in a) can be split into two Wilson loops $W_{{\bf r}_1}$ and $W_{{\bf r}_2}$ located at two poles on $\mathbb{P}^1$ (red line) in the blowup $\hat{\mathbb{C}}^2$.}
\label{fig:BlowupW}
\end{figure}

More concretely, we propose that, as shown in Figure \ref{fig:BlowupW}, Wilson loop operators $W_{{\bf r}_1}$ and $W_{{\bf r}_2}$ of the representations ${\bf r}_1$ and ${\bf r}_2$ at two poles on $\mathbb{P}^1$ will collapse and become a Wilson loop operator $W_{{\bf r}_1\otimes {\bf r}_2}$ of a product representation ${\bf r}_1\otimes {\bf r}_2$ at the origin of $\mathbb{C}^2$ after the transition. Therefore, we conjecture the blowup equations for the partition function $Z_{W_{\bf r}}$ in the presence of Wilson loop operators as
\begin{align}\label{eq:bleqW}
&\Lambda_0(m_j;\epsilon_1,\epsilon_2)\hat{Z}_{W_{{\bf r}_1\otimes{\bf r}_2}}(\phi_i, m_j; \epsilon_1,\epsilon_2)  \ (+\Lambda_1(m_j;\epsilon_1,\epsilon_2)\hat{Z}(\phi_i, m_j; \epsilon_1,\epsilon_2))  \\
&=\sum_{\vec{n}}(-1)^{|\vec{n}|}\hat{Z}^{(N)}_{W_{{\bf r}_1}}(\phi_i\!+\!n_i\epsilon_1,m_j\!+\!B_j\epsilon_1;\epsilon_1,\epsilon_2\!-\!\epsilon_1) \cdot \hat{Z}^{(S)}_{W_{{\bf r}_2}}(\phi_i\!+\!n_i\epsilon_2,m_j\!+\!B_j\epsilon_2;\epsilon_1\!-\!\epsilon_2,\epsilon_2) . \nonumber
\end{align} 
The representation ${\bf r_1}$ (or ${\bf r}_2$) here itself can be a product representation. The $\Lambda_0$ and $\Lambda_1$ factors are independent of the dynamical K\"ahler parameters $\phi_i$. The term including $\Lambda_1$ in the first line seems to be a bit unnatural from a localization point of view, so we put it in a parentheses. In the following discussions we mainly assume $\Lambda_1=0$. This term seems however necessary in certain cases with particular choices of magnetic fluxes, which will also be discussed for many examples.

By dividing both sides on \eqref{eq:bleqW} by $\hat{Z}$, the blowup equations can be rewritten in terms of VEVs of Wilson loop operators defined in \eqref{eq:VEV of BPS Wilson loop} as 
\begin{align}\label{eq:bleqW2}
&\Lambda_0\, \langle \hat{W}_{{\bf r}_1\otimes{\bf r}_2} \rangle \ (+\Lambda_1) = \sum_{\vec{n}}(-1)^{|\vec{n}|} \langle \hat{W}_{{\bf r}_1}^{(N)}\rangle(\vec{n},\vec{B}) \times \langle \hat{W}_{{\bf r}_2}^{(S)} \rangle(\vec{n},\vec{B}) \times \frac{\hat{Z}^{(N)}(\vec{n},\vec{B})\hat{Z}^{(S)}(\vec{n},\vec{B})}{\hat{Z}(\vec{n},\vec{B})}\, ,
\end{align}
where
\begin{align}
	\langle \hat{W}_{{\bf r}}^{(N)}\rangle(\vec{n},\vec{B}) & \equiv \langle \hat{W}_{{\bf r}} \rangle(\phi_i\!+\!n_i\epsilon_1,m_j\!+\!B_j\epsilon_1;\epsilon_1,\epsilon_2\!-\!\epsilon_1) \, , \nonumber \\
	\langle \hat{W}_{{\bf r}}^{(S)}\rangle(\vec{n},\vec{B}) & \equiv \langle \hat{W}_{{\bf r}} \rangle(\phi_i\!+\!n_i\epsilon_2,m_j\!+\!B_j\epsilon_2;\epsilon_1\!-\!\epsilon_2,\epsilon_2) \, .
\end{align}
In these equations, the hatted partition functions $\hat{Z}_{W_{\bf r}}$ and $\langle \hat{W}_{\bf r}\rangle$ are the same as the unhatted partition functions but with the shifted parameter $\epsilon_1 \rightarrow \epsilon_1+2\pi i$. 

It turns out that we can solve these blowup equations to calculate the partition function of Wilson loop operators as an expansion in the K\"ahler parameters of the theory. The blowup equation \eqref{eq:bleqW2} is a functional equation involving both the VEVs of the Wilson loops and the ordinary partition function $Z$. Therefore, before calculating the VEVs of the Wilson loops using these equations we must first know the partition function $Z$. As discussed, the partition function $Z$ without loop operators can be computed by solving the ordinary blowup equations from \eqref{eq:bleq}, for which we used a set of consistent magnetic fluxes $(\vec{n},\vec{B})$. Knowing the partition function $Z$, the remaining inputs in the blowup equation are the representations ${\bf r}_1,{\bf r}_2$ as well as the effective prepotential and the choice of magnetic fluxes. We propose that the blowup equations for Wilson loops in \eqref{eq:bleqW2} is valid only with the same consistent magnetic fluxes that we used for the ordinary blowup equations. Indeed, we can notice that equation \eqref{eq:bleqW2} for a trivial loop, i.e. ${\bf r}_1,{\bf r}_2 =\emptyset$, with $\Lambda_0 = \Lambda$ reduces to the ordinary blowup equation without loops.

By taking ${\bf r}_2=\emptyset$ for the Wilson loop $W^{(S)}_{{\bf r}_2}$ and plugging the partition function $Z$ into \eqref{eq:bleqW2}, we can construct a blowup equation containing Wilson loop operators of only one type in the representation ${\bf r}_1$. The resulting blowup equation is a linear functional equation for the VEV of the Wilson loop operator $\langle W_{{\bf r}_1}\rangle$ and it can be solved systematically. To solve the blowup equation with a given set of consistent magnetic fluxes $(\vec{n},\vec{B})$, we use the fact that the VEV of a Wilson loop operator is in fact the spectrum of 1d BPS states bound to the loop operator and thus it should take the form of a 1d index given in \eqref{eq:wilson-state}. The 1d index can be expressed as a Laurent expansion in terms of the K\"ahler parameters on the Coulomb branch. We put this Laurent expansion into the blowup equation \eqref{eq:bleqW2} and solve it order by order in the expansion of the K\"ahler parameters, in an iterative fashion, which will determine the degeneracies $\tilde{N}^{\bf d}_{j_l,j_r}$ of 1d bound states.

We remark that the blowup equation with ${\bf r}_2=\emptyset$ is a linear equation and thus any linear combination of solutions, $\sum_{\bf r} c_{\bf r} W_{\bf r}$ with constant coefficients $c_{\bf r}$, also becomes a solution of the blowup equation. There is hence an ambiguity in solving this equation for a given representation ${\bf r}$. Moreover, we can also consider the mass-dependent constants $c_{\bf r}(m_i)$ when the background magnetic flux for $m_i$ is not activated. In particular, this means in gauge theories that the constant $c_{\bf r}$ may depend on a gauge coupling when the background flux for the corresponding topological symmetry is switched off, which will spoil the non-perturbative instanton correction to the Wilson loop operator for a fixed classical VEV. Therefore, due to this ambiguity in general, the blowup equations of the form \eqref{eq:bleqW} or \eqref{eq:bleqW2} with ${\bf r}_2 =\emptyset$ do not have unique solution and cannot be use to correctly identify the Wilson loop partition functions.

We however claim that when ${\bf r}_1$ is a {\it minimal representation}, the blowup equation with ${\bf r}_2 =\emptyset$ has no such ambiguity. A {\it minimal representation} is defined as a representation whose primitive state carries electric charges $[e_1,e_2,\cdots, e_r]$ with $-1\le e_i\le 0$ (in Dynkin basis) and all other bound states have at least one electric charge with $e_i>0$. In geometry, this corresponds to an operator constructed by a non-compact primitive 2-cycle intersecting compact 4-cycles at most once. All (quasi-)minuscule representations in gauge algebras are also minimal representations. The reason for this claim is as follows.

Consider a non-compact primitive curve in a local CY 3-fold for a Wilson loop operator. Suppose that this curve intersects a surface $D_i$ at one generic point. Then exceptional curves with negative self-intersections in $D_i$ cannot form a bound state with the primitive curve because the exceptional curves are generically located at special points far from the location of the primitive curve. Hence the primitive curve can only be attached to curves with non-negative self-intersection having $e_i>1$, except for the elliptic fiber curve, for at least one surface $D_i$. This means all the bound states, but the primitive curve state, to the loop operator must have at least one electric charge with $e_i>0$. Consequently, we can find only a single BPS state of charge $[e_1,e_2,\cdots, e_r]$ with $-1\le e_i\le 0$, which is the primitive curve state itself, bound to the Wilson loop in a minimal representation in the 5d SCFT. So the ambiguity is not present in this case.

In a local CY 3-fold with an elliptic (or a genus-one) fibration for a 6d theory, the primitive curve can form bound states with the elliptic (or genus-one) fiber class which carries no gauge charge. The bound states provide a tower of KK momentum states with the same gauge charges as those of the primitive curve. However, these KK tower states are not genuine field theory states. They in fact decouple from the 6d field theory, so we need to exclude them from the spectrum of the loop operator. We therefore expect that there exists only a single BPS state of charge $[e_1,e_2,\cdots, e_r]$ with $-1\le e_i\le 0$ for the minimal representations in 6d as well. We will assume that such Wilson loop operators in the minimal representations in 6d do not involve in its spectrum any other states (but the primitive state) with $e_i\le0$ for all $i$'s.

The ambiguity for non-minimal representations can probably be removed by using the blowup equations with non-trivial ${\bf r}_1$ and ${\bf r}_2$. One can formulate blowup equations for the Wilson loop $W_{{\bf r}_1\otimes {\bf r}_2}$ in the product representation of two non-trivial representations ${\bf r}_1$ and ${\bf r}_2$ by inserting the known results for $\langle W_{{\bf r}_1^{(N)}} \rangle $ and $\langle W_{{\bf r}_2^{(S)}}\rangle $ into the RHS in the blowup equation. In this case, the blowup equation becomes a non-linear equation for Wilson loop operators in three different representations. Assuming we already know $\langle W_{{\bf r}_1^{(N)}} \rangle $ and $\langle W_{{\bf r}_2^{(S)}}\rangle $, the blowup equation can be solved in the same iterative manner to calculate $\langle W_{{\bf r}_1\otimes {\bf r}_2} \rangle$.

We conjecture that when the blowup equation for a Wilson loop operator is solved with the ansatz of \eqref{eq:wilson-state}, then the result correctly yields the BPS spectrum of 1d bound states to the Wilson loop operator up to the ambiguity for non-minimal representations explained above in this section. In the next section, we will explicitly illustrate how to solve the blowup equations with many non-trivial examples including Wilson loops in non-minimal representations in 5d $SU(2)$ gauge theories. We will verify the solutions by comparing them against the results from ADHM constructions and also by checking expected dualities.

As we will see, not all the blowup equations for Wilson loops with consistent magnetic fluxes are solvable with the ansatz of \eqref{eq:wilson-state}. In particular, we observe that the solutions to the blowup equations for Wilson loops in non-minimal representations generally involve 1d states which do not fit in the representations of Lorentz group $SU(2)_l\times SU(2)_r$ or have negative degeneracies in certain lowest orders in the K\"ahler parameter expansion. We do not expect that the solutions in these cases correctly capture the bound states to the loop operators. Some examples having this problem will be presented in the next section. Solving this problem for loop operators in non-minimal representations would be interesting future work.

So far, we have discussed blowup equations for Wilson loop (or Wilson surface) operators which are codimension-4 defect operators in 5d and in 6d. There is yet another type of codimension-4 defects that can be defined systematically by coupling 1d (or 2d) degrees of freedom to 5d (or 6d) supersymmetric field theories. These operators are often referred to as {\it qq-characters} as discussed in \cite{Nekrasov:2015wsu,Kim:2016qqs,Haouzi:2020yxy}. The spectrum of such operators can also be obtained by employing the blowup approach, which we will talk about briefly from now on.

For example, we can consider the 5d $SU(N)$ gauge theory coupled to a 1d fermion in the fundamental representation of the bulk $SU(N)$ gauge group. The partition function of this 1d/5d coupled system then becomes a generating function of the Wilson loops in the minuscule representations. On $S^1\times \mathbb{C}^2$ the partition function $Z^{1d/5d}$ can be written as, when normalized by the bare partition function $Z^{5d}$,
\begin{align}
	Z^{1d/5d}/Z^{5d} = \langle  e^{Nz/2}\prod_{w\in w_{\bf F}}(1-e^{-z} e^{-w\cdot \phi})\rangle = e^{Nz/2} \sum_{n=0}^N(-1)^{n}\langle W_{{\bf \Lambda}^n }\rangle e^{-nz} \, ,
\end{align}
where $ z $ is the $U(1)$ flavor mass parameter for the 1d fermion and $ w_\mathbf{F} $ is the fundamental weight, and $\langle W_{{\bf \Lambda}^n} \rangle$ denotes the expectation value of the Wilson loop in the rank-$n$ anti-symmetric tensor representation. This implies that this defect partition function can be obtained from the Wilson loop expectation values and vice versa. Likewise, we can define 1d defects of this type in different representations and also in other 5d gauge theories, and use the Wilson loop expectation values to calculate the partition functions of the 1d/5d coupled system.

In 6d, similar codimension-4 defects can be introduced by coupling 2d free fermions to the 6d bulk gauge fields. The partition function of the 2d/6d coupled system on the $\Omega$-background then can be written as
\begin{align}
	Z^{2d/6d}/Z^{6d} \sim \langle \prod_{w \in {\bf w}}\theta_1(\tau,z+w\cdot \phi) \rangle \, ,
\end{align}
where ${\bf w}$ denotes the weights for the representation of the 2d fermion. Now, since the 2d states are dressed by KK momenta, this partition function is not a usual generating function of Wilson loop operators. As we will see in the next section, one can also build blowup equations for the partition functions of this type of codimension-4 defects. Rather surprisingly, it turns out that these blowup equations appear to be the same as those of the Wilson loop operators we discussed in this section, which implies that the partition function $Z^{2d/6d}$ may also be written in terms of the expectation values of Wilson loop operators multiplied by certain elliptic functions. We will exhibit some concrete examples for this in the next section.

\section{Examples}\label{sec:examples}
In this section, we discuss various interesting 5d/6d theories and show how to compute the spectrum of BPS Wilson loops in these theories by solving the blowup equations.

\subsection{\texorpdfstring{$SU(2)$}{SU(2)} theories}

As a warm-up, let us start with the 5d $SU(2)$ gauge theories at the discrete theta angles $ \theta = 0 $ and $ \pi $. These two $SU(2)$ theories have the same effective prepotential on the $\Omega$-background given by
\begin{align}
\mathcal{E} = \frac{1}{\epsilon_1 \epsilon_2} \qty( \frac{4}{3}\phi^3 + m \phi^2 - \frac{\epsilon_1^2 + \epsilon_2^2}{12} \phi + \epsilon_+^2 \phi ) \, ,
\end{align}
where $ m = 1/g^2 $ is the inverse gauge coupling squared. The partition functions on the $\Omega$-background without a Wilson loop can be readily computed from the blowup equation with the following consistent magnetic fluxes \cite{Nakajima:2005fg,Kim:2020hhh}:
\begin{align}\label{eq:fluxes-su2}
&\theta = 0 \  : \ \ n \in \mathbb{Z} \, , \ B_m = 0 \, , \pm 1 \, , \nonumber \\
&\theta = \pi \ : \ \ n \in \mathbb{Z} \, , \ B_m = \pm 1/2 \, , 3/2 \, .
\end{align}

Now we insert a Wilson loop operator in a representation $ \mathbf{r} $ and compute its expectation value. The partition function $ Z_{W_\mathbf{r}} $ in the presence of the loop operator takes the form
\begin{align}
Z_{W_{\mathbf{r}}}(\phi, m; \epsilon_1, \epsilon_2) = Z_{\mathrm{pert}}(\phi; \epsilon_1, \epsilon_2) \sum_{k=0}^\infty e^{-k m} Z_{W_{\mathbf{r}}}^{(k)}(\phi; \epsilon_1, \epsilon_2) \, ,
\end{align}
where
\begin{align}
Z_{\mathrm{pert}}(\phi; \epsilon_1, \epsilon_2) = \PE\qty[-\frac{1+p_1 p_2}{(1-p_1)(1-p_2)} e^{-2\phi}]
\end{align}
is the perturbative contribution from the $SU(2)$ vector multiplet, which is the same as the case without loop operators, $ Z_{W_{\mathbf{r}}}^{(k)}$ is the $k$-th instanton contribution, and $ Z_{W_{\mathbf{r}}}^{(0)} = \langle W_{\mathbf{r}}^{\mathrm{cls}} \rangle $ is the classical VEV.

We will compute the Wilson loop partition function using the blowup formula with the same magnetic flux choice given in \eqref{eq:fluxes-su2}. As explained in the previous section, a Wilson loop operator in a product representation ${\bf r}_1\otimes {\bf r}_2$ will be factorized into a pair of Wilson loop operators in the representation $ \mathbf{r}_1 $ and $ \mathbf{r}_2 $ at two poles of $ \mathbb{P}^2 $ at the origin of a blowup $ \hat{\mathbb{C}}^2 $. Then the blowup equation \eqref{eq:bleqW} can be written as the form of an instanton expansion given by
\begin{align}\label{eq:blowup_qexpand}
&\sum_{k,k'=0}^\infty e^{-(k + k') m} \Lambda_{0,k'}(\epsilon_1, \epsilon_2) \hat{Z}_{W_{\mathbf{r}_1 \otimes \mathbf{r}_2}}^{(k)}(\phi; \epsilon_1, \epsilon_2) \nonumber \\
&= \sum_{n \in \mathbb{Z}} \sum_{k_1, k_2 = 0}^\infty (-1)^n e^{-V}\frac{\hat{Z}_{\mathrm{pert}}^{(N)} \hat{Z}_{\mathrm{pert}}^{(S)}}{\hat{Z}_{\mathrm{pert}}} e^{-(k_1 + k_2)m} p_1^{k_1 B_m} p_2^{k_2 B_m}  \\
&\qquad \qquad \quad \times \hat{Z}_{W_{\mathbf{r}_1}}^{(k_1)}(\phi + n \epsilon_1; \epsilon_1, \epsilon_2 - \epsilon_1) \hat{Z}_{W_{\mathbf{r}_2}}^{(k_2)}(\phi + n \epsilon_2; \epsilon_1 - \epsilon_2, \epsilon_2) \, . \nonumber
\end{align}
We now solve this equation for Wilson loops in various representations.

\paragraph{Fundamental Wilson loop}
We first consider the Wilson loop in the fundamental representation, which is a minimal representation, whose lowest weight is $ \mathbf{r} = [-1] $. Its classical VEV is
\begin{align}
Z_{W_{[-1]}}^{(0)} = \langle W_{[-1]}^{\mathrm{cls}} \rangle = e^\phi + e^{-\phi} \, .
\end{align}
The blowup equation for this Wilson loop operator can be formulated by considering the fundamental Wilson loop inserted at the North pole of $ \mathbb{P}^1 $ and a trivial operator at the South pole of $ \mathbb{P}^1 $, i.e. $ \mathbf{r}_1 = [-1] $ and $ \mathbf{r}_2 = \emptyset $ in \eqref{eq:blowup_qexpand}. We then easily see that the instanton expansion at the $k$-th order becomes
\begin{align}
\hat{Z}_{W_{[-1]}}^{(k)}(\phi; \epsilon_1, \epsilon_2)
&= p_1^{k B_m} \hat{Z}_{W_{[-1]}}^{(k)}(\phi; \epsilon_1, \epsilon_2 - \epsilon_1) + p_2^{k B_m} \hat{Z}^{(k)}(\phi; \epsilon_1 - \epsilon_2, \epsilon_2) \nonumber \\
& \quad + (\text{terms with } \hat{Z}_{W_{[-1]}}^{(n < k)} \text{ and } \hat{Z}^{(n < k)}) \, .
\end{align}
By noting that the partition function $ \hat{Z}^{(k)} $ without loops is already known, this equation becomes a functional equation for two unknown functions $\hat{Z}_{W_{[-1]}}^{(k)}(\phi; \epsilon_1, \epsilon_2)$ and $\hat{Z}_{W_{[-1]}}^{(k)}(\phi; \epsilon_1, \epsilon_2 - \epsilon_1)$ at $k$-th instanton order. The equation can be iteratively solved order by order in the instanton number expansion. Since the fundamental representation is a minimal representation, we expect the solution to this blowup equation with the consistent magnetic fluxes in \eqref{eq:fluxes-su2} uniquely determines the spectrum of the bound states to the fundamental Wilson loop operator.

For the two $SU(2)_\theta$ theories of two different theta angles $\theta$, we can use two distinct background fluxes as in \eqref{eq:fluxes-su2}. We take $ B_m = 0, 1 $ for the $ SU(2)_0 $ theory and $ B_m = \pm 1/2 $ for the $ SU(2)_\pi $ theory. It is then straightforward to obtain the closed expressions of instanton corrections to the Wilson loop expectation value at each instanton order. One finds that the closed expressions for the fundamental Wilson loop VEVs at $k=1$ are given by
\begin{align}
\langle W_{[-1]}^{(1)} \rangle
= Z_{W_{[-1]}}^{(1)} - Z_{W_{[-1]}}^{(0)} Z^{(1)}
= -\frac{p_1 p_2 (e^{-\phi} + e^{\phi}) }{(1-p_1 p_2 e^{-2\phi}) (1-p_1 p_2 e^{2\phi})} \ ,
\end{align}
for $ \theta = 0 $ and
\begin{align}
\langle W_{[-1]}^{(1)} \rangle
= Z_{W_{[-1]}}^{(1)} - Z_{W_{[-1]}}^{(0)} Z^{(1)}
= \frac{\sqrt{p_1 p_2} (1+p_1 p_2)}{(1-p_1 p_2 e^{-2\phi}) (1-p_1 p_2 e^{2\phi})} \ ,
\end{align}
for $ \theta = \pi $. The first case for $\theta=0$ perfectly matches the result from the ADHM calculation in \cite{Bullimore:2014upa,Gaiotto:2015una,Assel:2018rcw} and the case for $\theta=\pi$ agrees with the result in \cite{Haouzi:2020zls}. Higher order instanton corrections of these VEVs can be obtained in a similar fashion by solving the higher order equations. Instead of giving explicit expressions for the higher order results, we summarize some low order bound states to the Wilson loops in two $SU(2)$ theories in the K\"ahler parameter expansion in Table~\ref{table:SU2_0_fund} for $\theta=0$ and in Table \ref{table:SU2_pi_fund} for $\theta=\pi$. 

\begin{table}
	\centering
	\begin{tabular}{|c|C{30ex}||c|C{30ex}|} \hline
		$ \mathbf{d} $ & $ \oplus \tilde{N}_{j_l, j_r}^{\mathbf{d}} (j_l, j_r) $ & $ \mathbf{d} $ & $ \oplus \tilde{N}_{j_l, j_r}^{\mathbf{d}} (j_l, j_r) $ \\ \hline
		$ (1, 1) $ & $ (0, 0) $ & $ (1, 3) $ & $ (0, 1) $ \\ \hline
		$ (1, 5) $ & $ (0, 2) $ & $ (1, 7) $ & $ (0, 3) $ \\ \hline
		$ (1, 9) $ & $ (0, 4) $ & $ (2, 5) $ & $ (0, 2) $ \\ \hline
		$ (2, 7) $ & $ (0, 2) \oplus 2(0, 3) \oplus (\frac{1}{2}, \frac{7}{2}) $ & $ (2, 9) $ & $ (0, 2) \oplus 2(0, 3) \oplus 3(0, 4) \oplus (\frac{1}{2}, \frac{7}{2}) \oplus 2(\frac{1}{2}, \frac{9}{2}) \oplus (1, 5) $ \\ \hline
		$ (3, 7) $ & $ (0, 3) $ & $ (3, 9) $ & $ (0,2) \oplus 2(0,3) \oplus 3(0,4) \oplus (\frac{1}{2},\frac{7}{2}) \oplus 2(\frac{1}{2},\frac{9}{2}) \oplus (1,5) $ \\ \hline
	\end{tabular}
	\caption{Spectrum of BPS bound states to a fundamental Wilson loop in the $ SU(2)_0 $ theory for $ d_1 \leq 3 $ and $ d_2 \leq 10 $. Here, $ \mathbf{d} = (d_1, d_2) $ labels the bound states with charge $ d_1 m + d_2 \phi $.}\label{table:SU2_0_fund}
\end{table}

\begin{table}
	\centering
	\begin{tabular}{|c|C{30ex}||c|C{30ex}|} \hline
		$ \mathbf{d} $ & $ \oplus \tilde{N}_{j_l, j_r}^{\mathbf{d}} (j_l, j_r) $ & $ \mathbf{d} $ & $ \oplus \tilde{N}_{j_l, j_r}^{\mathbf{d}} (j_l, j_r) $ \\ \hline
		$ (1, 2) $ & $ (0, \frac{1}{2}) $ & $ (1, 4) $ & $ (0, \frac{3}{2}) $ \\ \hline
		$ (1, 6) $ & $ (0, \frac{5}{2}) $ & $ (1, 8) $ & $ (0, \frac{7}{2}) $ \\ \hline
		$ (1, 10) $ & $ (0, \frac{9}{2}) $ & $ (2, 5) $ & $ (0, 2) $ \\ \hline
		$ (2, 7) $ & $ (0, 2) \oplus 2(0, 3) \oplus (\frac{1}{2}, \frac{7}{2}) $ & $ (2, 9) $ & $ (0,2) \oplus 2(0,3) \oplus 3(0,4) \oplus (\frac{1}{2},\frac{7}{2}) \oplus 2(\frac{1}{2},\frac{9}{2}) \oplus (1,5) $ \\ \hline
		$ (3, 8) $ & $ (0,\frac{5}{2}) \oplus (0,\frac{7}{2}) \oplus (\frac{1}{2},4) $ & $ (3, 10) $ & $ (0,\frac{3}{2}) \oplus 2(0,\frac{5}{2}) \oplus 4(0,\frac{7}{2}) \oplus 3(0,\frac{9}{2}) \oplus (0,\frac{11}{2}) \oplus (\frac{1}{2},3) \oplus 3(\frac{1}{2},4) \oplus 4(\frac{1}{2},5) \oplus (1,\frac{9}{2}) \oplus 2(1,\frac{11}{2}) \oplus (\frac{3}{2},6) $ \\ \hline
	\end{tabular}
	\caption{Spectrum of BPS bound states to a fundamental Wilson loop in the $ SU(2)_\pi $ theory for $ d_1 \leq 3 $ and $ d_2 \leq 10 $. Here, $ \mathbf{d} = (d_1, d_2) $ labels the BPS states with charge $ d_1 m + d_2 \phi $.}\label{table:SU2_pi_fund}
\end{table}

\paragraph{Adjoint Wilson loop}

Let us now compute the partition function with a Wilson loop operator in the adjoint representation $ \mathbf{r} = [-2] $ whose classical VEV is
\begin{align}
Z_{W_{[-2]}}^{(0)} = \langle W_{[-2]}^{\mathrm{cls}} \rangle = e^{2\phi} + 1 + e^{-2\phi} \, .
\end{align}

The $SU(2)$ adjoint representation is a non-minimal representation with gauge charge $e_1 \le -2$. Thus, the blowup equation with ${\bf r}_2 = \emptyset$ has the ambiguity explained in section \ref{sec:blowup-loop}. To avoid this ambiguity, we instead utilize the blowup equation with ${\bf r}_1=[-1]$ and ${\bf r}_2=[-1]$. In this case the VEV of the fundamental Wilson loop computed above will be used as an input for the blowup equation. Since the product representation ${\bf r}_1\otimes {\bf r}_2$ involves an adjoint and a singlet representation, the LHS of the blowup equation in \eqref{eq:bleqW2} becomes the sum of an adjoint Wilson loop and a trivial loop, i.e. ${\bf 2}\otimes {\bf 2} = {\bf 3} + {\bf 1}$.

For the $SU(2)_0$ theory, we use magnetic fluxes as $n\in \mathbb{Z}$, $ B_m = 0 $ and solve the blowup equations. The result at $k=1$ is
\begin{align}
\langle W^{(1)}_{{\bf 2}\otimes {\bf 2}} \rangle = \frac{(1-p_1)(1-p_2)(1+p_1p_2)-2p_1p_2( e^{-2\phi}+2+e^{2\phi})}{(1-p_1p_2e^{2\phi})(1-p_1p_2e^{-2\phi})} \, ,
\end{align}
which exactly matches the ADHM result in \cite{Assel:2018rcw}. Some low order states of the adjoint Wilson loop operator at $\theta=0$ are given in Table~\ref{table:SU2_0_adj}, which also agrees with \cite{Assel:2018rcw}. Note that $W^{(k)}_{{\bf 2}\otimes {\bf 2}} = W^{(k)}_{{\bf 3}}$ for $k>0$ because the trivial loop receives no instanton correction.

\begin{table}
	\centering
	\begin{tabular}{|c|C{30ex}||c|C{30ex}|} \hline
		$ \mathbf{d} $ & $ \oplus \tilde{N}_{j_l, j_r}^{\mathbf{d}} (j_l, j_r) $ & $ \mathbf{d} $ & $ \oplus \tilde{N}_{j_l, j_r}^{\mathbf{d}} (j_l, j_r) $ \\ \hline
		$ (1, 2) $ & $ (0, 0) \oplus (0, 1) \oplus (\frac{1}{2}, \frac{1}{2}) $ & $ (1, 4) $ & $ (0, 1) \oplus (0, 2) \oplus (\frac{1}{2}, \frac{3}{2}) $ \\ \hline
		$ (1, 6) $ & $ (0, 2) \oplus (0, 3) \oplus (\frac{1}{2}, \frac{5}{2}) $ & $ (1, 8) $ & $ (0, 3) \oplus (0, 4) \oplus (\frac{1}{2}, \frac{7}{2}) $ \\ \hline
		$ (2, 2) $ & $ (0, 0) $ & $ (2, 4) $ & $ (0, 1) \oplus (0, 2) \oplus (\frac{1}{2}, \frac{3}{2}) $ \\ \hline
		$ (2, 6) $ & $ (0,0) \oplus 3(0,2) \oplus 2(0,3) \oplus (\frac{1}{2},\frac{3}{2}) \oplus 2(\frac{1}{2},\frac{5}{2}) \oplus (\frac{1}{2},\frac{7}{2}) \oplus (1,3) $ & $ (2, 8) $ & $ (0,1) \oplus 2(0,2) \oplus 5(0,3) \oplus 4(0,4) \oplus (\frac{1}{2},\frac{3}{2}) \oplus 2(\frac{1}{2},\frac{5}{2}) \oplus 5(\frac{1}{2},\frac{7}{2}) \oplus 2(\frac{1}{2},\frac{9}{2}) \oplus (1,3) \oplus 2(1,4) \oplus (1,5) \oplus (\frac{3}{2},\frac{9}{2}) $ \\ \hline
		$ (3, 6) $ & $ (0,2) \oplus (0,3) \oplus (\frac{1}{2},\frac{5}{2}) $ & $ (3, 8) $ & $ (0,1) \oplus 2(0,2) \oplus 5(0,3) \oplus 4(0,4) \oplus (\frac{1}{2},\frac{3}{2}) \oplus 2(\frac{1}{2},\frac{5}{2}) \oplus 5(\frac{1}{2},\frac{7}{2}) \oplus 2(\frac{1}{2},\frac{9}{2}) \oplus (1,3) \oplus 2(1,4) \oplus (1,5) \oplus (\frac{3}{2},\frac{9}{2}) $ \\ \hline
	\end{tabular}
	\caption{Spectrum of the adjoint Wilson loop operator in the $ SU(2)_0 $ theory for $ d_1 \leq 3 $ and $ d_2 \leq 8 $. }\label{table:SU2_0_adj}
\end{table}

For the $SU(2)_\pi$ theory, we solve the blowup equations with the magnetic fluxes $n\in\mathbb{Z}$, $ B_m=1/2 $. The result at $k=1$ is
\begin{align}
\langle W^{(1)}_{{\bf 2}\otimes {\bf 2}} \rangle = \frac{\sqrt{p_1 p_2} (1+p_1)(1+p_2) (e^\phi + e^{-\phi})}{(1-p_1 p_2 e^{2\phi})(1-p_1 p_2 e^{-2\phi})} \, .
\end{align}
We summarize in Table~\ref{table:SU2_pi_adj} some low order degeneracies of the adjoint Wilson loop states at $\theta=\pi$.

\begin{table}
	\centering
	\begin{tabular}{|c|C{30ex}||c|C{30ex}|} \hline
		$ \mathbf{d} $ & $ \oplus \tilde{N}_{j_l, j_r}^{\mathbf{d}} (j_l, j_r) $ & $ \mathbf{d} $ & $ \oplus \tilde{N}_{j_l, j_r}^{\mathbf{d}} (j_l, j_r) $ \\ \hline
		$ (1, 1) $ & $ (0, \frac{1}{2}) \oplus (\frac{1}{2}, 0) $ & $ (1, 3) $ & $ (0, \frac{1}{2}) \oplus (0, \frac{3}{2}) \oplus (\frac{1}{2}, 1) $ \\ \hline
		$ (1, 5) $ & $ (0, \frac{3}{2}) \oplus (0, \frac{5}{2}) \oplus (\frac{1}{2}, 2) $ & $ (1, 7) $ & $ (0, \frac{5}{2}) \oplus (0, \frac{7}{2}) \oplus (\frac{1}{2}, 3) $ \\ \hline
		$ (2, 4) $ & $ (0, 0) \oplus (0, 2) \oplus (\frac{1}{2}, \frac{3}{2}) $ & $ (2, 6) $ & $ (0,1) \oplus 2(0,2) \oplus 2(0,3) \oplus (\frac{1}{2},\frac{3}{2}) \oplus 2(\frac{1}{2},\frac{5}{2}) \oplus (\frac{1}{2},\frac{7}{2}) \oplus (1,3) $ \\ \hline
		$ (2, 8) $ & $ (0,0) \oplus 3(0,2) \oplus 4(0,3) \oplus 4(0,4) \oplus (\frac{1}{2},\frac{3}{2}) \oplus 2(\frac{1}{2},\frac{5}{2}) \oplus 5(\frac{1}{2},\frac{7}{2}) \oplus 2(\frac{1}{2},\frac{9}{2}) \oplus (1,3) \oplus 2(1,4) \oplus (1,5) \oplus (\frac{3}{2},\frac{9}{2}) $ & $ (3, 7) $ & $ (0,\frac{3}{2}) \oplus (0,\frac{5}{2}) \oplus (0,\frac{7}{2}) \oplus (\frac{1}{2},2) \oplus (\frac{1}{2},3) \oplus (\frac{1}{2},4) \oplus (1,\frac{7}{2}) $ \\ \hline
	\end{tabular}
	\caption{Spectrum of the adjoint Wilson loop operator in the $ SU(2)_\pi $ theory for $ d_1 \leq 3 $ and $ d_2 \leq 8 $.}\label{table:SU2_pi_adj}
\end{table}

Let us briefly discuss the ambiguity in the blowup equation for the adjoint Wilson loop with representations ${\bf r}_1 = [-2]$ and ${\bf r}_2 = \emptyset$. Consider first the blowup equation for the $\theta=0$ case with magnetic fluxes $n\in \mathbb{Z},B_m = 0$. One finds that the blowup equation leads to the Wilson loop spectrum given in Table~\ref{table:SU2_0_adj} which is consistent with the ADHM result in \cite{Assel:2012nf} in the instanton number expansion. However, since the blowup equation in this case is a linear functional equation for the adjoint Wilson loop expectation value, the solution is not unique. All the linear combinations of the form $W_{{\bf 2}\otimes {\bf 2}} + c\, e^{-n m} $ with a positive integer $n$ and a constant $c$ solve the same blowup equation,\footnote{We can also add a fundamental Wilson loop like $c'e^{- n' m} W_{\bf 2}$. However this factor can be excluded by using the fact that any compact 2-cycles in the geometry $\mathbb{F}_0$ which can be attached to the primitive non-compact 2-cycle of the Wilson loop $W_{{\bf 2}\otimes {\bf 2}}$ have volumes $k\phi$ with an even number $k$, and so all the BPS bound states of $W_{{\bf 2}\otimes {\bf 2}}$ carry even electric charges.} but they will have different instanton contributions while maintaining the same classical expectation value of the adjoint Wilson loop.

On the other hand, the ambiguity in the blowup equation with non-trivial background magnetic flux $B_m=1$ appears rather in a different way. In this case, the ADHM result or the spectrum in Table~\ref{table:SU2_0_adj} does not satisfy the usual blowup equation with $\Lambda_1=0$ in \eqref{eq:bleqW}. This issue is resolved by introducing an additional unnatural constant factor like $ \Lambda_1 = -e^{-m}(1-p_1)^2 $ in the blowup equation. Once this constant factor $\Lambda_1$ is allowed, however, the blowup equation possesses the same ambiguity as the previous $B_m=0$ case, e.g., any combinations such as $W_{{\bf 2}\otimes {\bf 2}} + c\,e^{-n m} $ can solve the blowup equation by properly setting the factor $\Lambda_1$. This hence affects the instanton contributions. 

We note that blowup equations with a non-minimal representation ${\bf r}_1$ and a trivial one ${\bf r}_2=\emptyset$ generically possess similar ambiguities, requiring other additional data to uniquely determine the expectation value of the loop operator. For this reason, the blowup equations with both non-trivial representations ${\bf r}_1$ and ${\bf r}_2$ would be more favored for the Wilson loops in non-minimal representations. In practice, we can first formulate a blowup equation for a product representation ${\bf r}_1\otimes {\bf r}_2$ by embedding the non-minimal representation into the product representation and solve it. The desired Wilson loop expectation value for the non-minimal representation can then be extracted from the solution by decomposing the product representation into irreducible representations, as we demonstrated in this section for the adjoint Wilson loop.

\paragraph{Wilson loop in the rank-3 symmetric representation} The classical VEV of the Wilson loop in the rank-3 symmetric representation $ \mathbf{r} = [-3] $ is
\begin{align}
\ev*{W_{[-3]}^{\mathrm{cls}}} = e^{3\phi} + e^{\phi} + e^{-\phi} + e^{-3\phi} \, .
\end{align}
One may try to use the blowup equations with $ \mathbf{r}_1 = [-3] $, $ \mathbf{r}_2 = \emptyset $ to compute the instanton contributions to $ \langle W_{[-3]} \rangle$. However, we find that this method does not work well: the blowup equations are linear and therefore suffer from the ambiguity explained above, and moreover, the solutions contain unexpected terms breaking $ \epsilon_1 \leftrightarrow \epsilon_2 $ symmetry which cannot be absorbed into the $ \Lambda_1 $. 

We can instead consider the blowup equations with $ \mathbf{r}_1 = \mathbf{2} \otimes \mathbf{2} $, $ \mathbf{r}_2 = \mathbf{2} $ for the product representation ${\bf r}={\bf 2}^{\otimes 3}$ whose classical VEV reads\footnote{For $SU(2)$, $\mathbf{2}^{\otimes 3}= \mathbf{4}\oplus \mathbf{2}\oplus \mathbf{2}$ and $\mathbf{2}^{\otimes 4}= \mathbf{5}\oplus \mathbf{3}\oplus \mathbf{3}\oplus \mathbf{3}\oplus\mathbf{1}\oplus\mathbf{1}$.}
\begin{align}\label{eq:su2_rank3_W0}
\langle W_{\mathbf{2}^{\otimes 3}}^{\rm cls} \rangle
= e^{3\phi} + 3e^{\phi} + 3e^{-\phi} + e^{-3\phi} \, .
\end{align}

For the $SU(2)_0$ gauge theory, we use the blowup equation with background magnetic flux $ B_m = 0 $ to compute $\langle W_{\mathbf{2}^{\otimes 3}}\rangle$ in the instanton expansion. At $k=1$, the solution is
\begin{align}
\langle W_{\mathbf{2}^{\otimes 3}}^{(1)} \rangle
&= \frac{(1-p_1)(1-p_2)(2+p_1+p_2+2p_1p_2)(e^{\phi} + e^{-\phi})}{(1-p_1 p_2 e^{2\phi})(1-p_1 p_2 e^{-2\phi})} \nonumber \\
& \quad - \frac{3p_1 p_2 (e^{3\phi} + 3e^{\phi} + 3e^{-\phi} + e^{-3\phi})}{(1-p_1 p_2 e^{2\phi})(1-p_1 p_2 e^{-2\phi})} \, ,
\end{align}
which precisely matches the ADHM result in \cite{Assel:2018rcw}. One can then extract the expectation value of the Wilson loop $ W_{[-3]}$ from $\langle W_{\mathbf{2}^{\otimes 3}}\rangle$ by using the decomposition $ W_{[-3]} = W_{\mathbf{2}^{\otimes 3}}-2W_{\mathbf{2}}$. We summarize the spectrum of the Wilson loop $ W_{[-3]}$ in the $SU(2)_0$ theory in Table~\ref{table:SU2_0_rank3}.

\begin{table}
	\centering
	\begin{tabular}{|c|C{30ex}||c|C{30ex}|} \hline
		$ \mathbf{d} $ & $ \oplus \tilde{N}_{j_l, j_r}^{\mathbf{d}} (j_l, j_r) $ & $ \mathbf{d} $ & $ \oplus \tilde{N}_{j_l, j_r}^{\mathbf{d}} (j_l, j_r) $ \\ \hline
		$ (1, -1) $ & $ 3(0, 0) $ & $ (1, 1) $ & $ 3(0, 0) \oplus (0, 1) \oplus (\frac{1}{2}, \frac{1}{2}) \oplus (1, 0) $ \\ \hline
		$ (1, 3) $ & $ (0, 0) \oplus (0, 1) \oplus (0, 2) \oplus (\frac{1}{2}, \frac{1}{2}) \oplus (\frac{1}{2}, \frac{3}{2}) \oplus (1, 1) $ & $ (1, 5) $ & $ (0, 1) \oplus (0, 2) \oplus (0, 3) \oplus (\frac{1}{2}, \frac{3}{2}) \oplus (\frac{1}{2}, \frac{5}{2}) \oplus (1, 2) $ \\ \hline
		$ (2, 1) $ & $ 3(0, 0) $ & $ (2, 3) $ & $ (0, 0) \oplus 3(0, 1) \oplus (0, 2) \oplus (\frac{1}{2}, \frac{1}{2}) \oplus (\frac{1}{2}, \frac{3}{2}) \oplus (1, 1) $ \\ \hline
		$ (2, 5) $ & $ (0, 0) \oplus (0, 1) \oplus 6(0, 2) \oplus 2(0, 3) \oplus (\frac{1}{2}, \frac{1}{2}) \oplus 3(\frac{1}{2}, \frac{3}{2}) \oplus 2(\frac{1}{2}, \frac{5}{2}) \oplus (\frac{1}{2}, \frac{7}{2}) \oplus (1, 1) \oplus 2(1, 2) \oplus (1, 3) \oplus (\frac{3}{2}, \frac{5}{2}) $ & $ (3, 3) $ & $ (0, 0) $ \\ \hline
		$ (3, 5) $ & \multicolumn{3}{c|}{$ (0, 1) \oplus 3(0, 2) \oplus (0, 3) \oplus (\frac{1}{2}, \frac{3}{2}) \oplus (\frac{1}{2}, \frac{5}{2}) \oplus (1, 2) $} \\ \hline
	\end{tabular}
	\caption{Spectrum of the Wilson loop operator in the rank-3 symmetric representation in the $ SU(2)_0 $ theory for $ d_1 \leq 3 $ and $ d_2 \leq 6 $.}\label{table:SU2_0_rank3}
\end{table}

Next, for the $ SU(2)_\pi$ theory, we use the blowup equation with background magnetic flux $ B_m = 1/2 $ to compute $\langle W_{\mathbf{2}^{\otimes 3}}\rangle$ in instanton expansion. The result at $k=1$ is then given by
\begin{align}\label{eq:su2_pi_rank3_W0}
\ev*{W_{\mathbf{2}^{\otimes 3}}^{(1)}}
&= \frac{(1-p_1)^2 (1-p_2)^2 (1+p_1p_2)}{\sqrt{p_1 p_2} (1-p_1 p_2 e^{2\phi}) (1-p_1 p_2 e^{-2\phi})} + \frac{3\sqrt{p_1 p_2}(p_1+p_2) (e^{2\phi}+2+e^{-2\phi})}{(1-p_1p_2 e^{2\phi})(1-p_1p_2 e^{-2\phi})} \nonumber \\
&\quad - \frac{(1-p_1)^2 (1-p_2)}{\sqrt{p_1p_2}} \, .
\end{align}
Notice that the first two terms on the RHS of \eqref{eq:su2_pi_rank3_W0} are symmetric under $\epsilon_1 \leftrightarrow \epsilon_2$, while the last term is not. We checked that the first two terms agree with the ADHM result computed in a similar manner as \cite{Assel:2018rcw}. The last term breaking $ \epsilon_1 \leftrightarrow \epsilon_2 $ symmetry is absorbed into $\Lambda_1$. We checked that there are no other terms breaking $ \epsilon_1 \leftrightarrow \epsilon_2 $ symmetry up to 3-instantons. We summarize the spectrum of the Wilson loop $ W_{[-3]}$ in the $SU(2)_\pi$ theory in Table~\ref{table:SU2_pi_rank3}.

\begin{table}
	\centering
	\begin{tabular}{|c|C{30ex}||c|C{30ex}|} \hline
		$ \mathbf{d} $ & $ \oplus \tilde{N}_{j_l, j_r}^{\mathbf{d}} (j_l, j_r) $ & $ \mathbf{d} $ & $ \oplus \tilde{N}_{j_l, j_r}^{\mathbf{d}} (j_l, j_r) $ \\ \hline
		$ (1, 2) $ & $ (0,\frac{1}{2}) \oplus (0,\frac{3}{2}) \oplus (\frac{1}{2},0) \oplus (\frac{1}{2},1) \oplus (1,\frac{1}{2}) $ & $ (1, 4) $ & $ (0,\frac{1}{2}) \oplus (0,\frac{3}{2}) \oplus (0,\frac{5}{2}) \oplus (\frac{1}{2},1) \oplus (\frac{1}{2},2) \oplus (1,\frac{3}{2}) $ \\ \hline
		$ (1, 6) $ & $ (0,\frac{3}{2}) \oplus (0,\frac{5}{2}) \oplus (0,\frac{7}{2}) \oplus (\frac{1}{2},2) \oplus (\frac{1}{2},3) \oplus (1,\frac{5}{2}) $ & $ (2, 3) $ & $ (0,0) \oplus (0,2) \oplus (\frac{1}{2},\frac{1}{2}) \oplus (\frac{1}{2},\frac{3}{2}) \oplus (1,1) $ \\ \hline
		$ (2, 5) $ & $ (0,0) \oplus (0,1) \oplus 3(0,2) \oplus 2(0,3) \oplus (\frac{1}{2},\frac{1}{2}) \oplus 3(\frac{1}{2},\frac{3}{2}) \oplus 2(\frac{1}{2},\frac{5}{2}) \oplus (\frac{1}{2},\frac{7}{2}) \oplus (1,1) \oplus 2 (1,2) \oplus (1,3) \oplus (\frac{3}{2},\frac{5}{2}) $ & $ (3, 6) $ & $ (0,\frac{3}{2}) \oplus 2(0,\frac{5}{2}) \oplus (0,\frac{7}{2}) \oplus (\frac{1}{2},1) \oplus 2 (\frac{1}{2},2) \oplus (\frac{1}{2},3) \oplus (\frac{1}{2},4) \oplus (1,\frac{3}{2}) \oplus (1,\frac{5}{2}) \oplus (1,\frac{7}{2}) \oplus (\frac{3}{2},3) $ \\ \hline
	\end{tabular}
	\caption{Spectrum of the Wilson loop operator in the rank-3 symmetric representation in the $ SU(2)_\pi $ theory for $ d_1 \leq 3 $ and $ d_2 \leq 6 $.}\label{table:SU2_pi_rank3}
\end{table}

\paragraph{Wilson loop in the rank-4 symmetric representation}

The classical VEV of the Wilson loop $W_{[-4]}$ in the rank-4 symmetric representation ${\bf r}=[-4]$ is
\begin{align}
\langle W_{[-4]}^{\rm cls} \rangle = e^{4\phi} + e^{2\phi} + 1 + e^{-2\phi} + e^{-4\phi} \, .
\end{align}
This Wilson loop can be embedded into a Wilson loop in the product representation ${\bf r}={\bf 2}^{\otimes 4}$ whose classical VEV is
\begin{align}
\langle W_{\mathbf{2}^{\otimes 4}}^{ \rm cls} \rangle = e^{4\phi} + 4e^{2\phi} + 6 + 4e^{-2\phi} + e^{-4\phi} \, .
\end{align}

The blowup equation for the loop operator $W_{\mathbf{2}^{\otimes 4}}^{ \rm cls}$ in the $SU(2)_0$ gauge theory can be formulated with $ \mathbf{r}_1 = \mathbf{2} \otimes \mathbf{2} $, $ \mathbf{r}_2 = \mathbf{2} \otimes \mathbf{2} $ and $ B_m = 0 $. At $k=1$, we find 
\begin{align}
\langle W_{\mathbf{2}^{\otimes 4}}^{(1)} \rangle
&= -\frac{(1-p_1)^2 (1-p_2)^2 (1+p_1) (1+p_2) (1+p_1 p_2)}{p_1 p_2 (1-p_1 p_2 e^{2\phi}) (1-p_1 p_2 e^{-2\phi})} \nonumber \\
&\quad + \frac{2(1-p_1) (1-p_2) (2+p_1+p_2+2p_1 p_2) (e^{2\phi}+2+e^{-2\phi})}{(1-p_1 p_2 e^{2\phi}) (1-p_1 p_2 e^{-2\phi})} \nonumber \\
&\quad - \frac{4p_1p_2 (e^{4\phi} + 4e^{-2\phi} + 6 + 4e^{-2\phi} + e^{-4\phi})}{(1-p_1 p_2 e^{2\phi}) (1-p_1 p_2 e^{-2\phi})} \, .
\end{align}
Here, we use the results for $W_{\mathbf{2} \otimes \mathbf{2}}$. We checked that the solution agrees up to 2-instantons with the ADHM result (modulo the constant term at 1-instanton order which can be again absorbed into $\Lambda_1$) and also with the solution to another blowup equation with $ \mathbf{r}_1 = \mathbf{2} \otimes \mathbf{2} \otimes \mathbf{2} $, $ \mathbf{r}_2 = \mathbf{2} $, $ B_m = 0 $. We summarize the spectrum of the Wilson loop operator $W_{[-4]}$, which can be extracted from the result of $\ev*{W_{\mathbf{2}^{\otimes 4}}}$, in Table~\ref{table:SU2_0_rank4}.

\begin{table}
	\centering
	\begin{tabular}{|c|C{30ex}||c|C{30ex}|} \hline
		$ \mathbf{d} $ & $ \oplus \tilde{N}_{j_l, j_r}^{\mathbf{d}} (j_l, j_r) $ & $ \mathbf{d} $ & $ \oplus \tilde{N}_{j_l, j_r}^{\mathbf{d}} (j_l, j_r) $ \\ \hline
		$ (1, -2) $ & $ 4(0, 0) $ & $ (1, 2) $ & $ \! 4(0,0) \oplus (0,1) \oplus (0,2) \oplus (\frac{1}{2},\frac{1}{2}) \oplus (\frac{1}{2},\frac{3}{2}) \oplus (1,0) \oplus (1,1) \oplus \
		(\frac{3}{2},\frac{1}{2}) \! $ \\ \hline
		$ (1, 4) $ & $ (0,0) \oplus (0,1) \oplus (0,2) \oplus (0,3) \oplus (\frac{1}{2},\frac{1}{2}) \oplus (\frac{1}{2},\frac{3}{2}) \oplus (\frac{1}{2},\frac{5}{2}) \oplus (1,1) \oplus (1,2) \oplus (\frac{3}{2},\frac{3}{2}) $ & $ (2, 2) $ & $ 4(0,0) \oplus 4(0,1) \oplus (0,2) \oplus 4(\frac{1}{2},\frac{1}{2}) \oplus (\frac{1}{2},\frac{3}{2}) \oplus (1,0) \oplus (1,1) \oplus (\frac{3}{2},\frac{1}{2}) $ \\ \hline
		$ (2, 4) $ & $ 3(0,0) \oplus 4(0,1) \oplus 7(0,2) \oplus 2(0,3) \oplus 2(\frac{1}{2},\frac{1}{2}) \oplus 7(\frac{1}{2},\frac{3}{2}) \oplus 2(\frac{1}{2},\frac{5}{2}) \oplus (\frac{1}{2},\frac{7}{2}) \oplus 4(1,1) \oplus 2 (1,2) \oplus (1,3) \oplus (\frac{3}{2},\frac{1}{2}) \oplus 2(\frac{3}{2},\frac{3}{2}) \oplus (\frac{3}{2},\frac{5}{2}) \oplus (2,2) $ & $ (3, 2) $ & $ 4(0, 0) $ \\ \hline
		$ (3, 4) $ & \multicolumn{3}{C{70ex}|}{$ (0,0) \oplus 4(0,1) \oplus 4(0,2) \oplus (0,3) \oplus (\frac{1}{2},\frac{1}{2}) \oplus 4(\frac{1}{2},\frac{3}{2}) \oplus (\frac{1}{2},\frac{5}{2}) \oplus (1,1) \oplus (1,2) \oplus (\frac{3}{2},\frac{3}{2}) $} \\ \hline
	\end{tabular}
	\caption{Spectrum of the Wilson loop operator in the rank-4 symmetric representation in the $ SU(2)_0 $ theory for $ d_1 \leq 3 $ and $ d_2 \leq 5 $.}\label{table:SU2_0_rank4}
\end{table}

Lastly, one may try to construct blowup equations for the Wilson loop operator $W_{{\bf 2}^{\otimes 4}}$ in the $SU(2)_\pi$ theory. We find that any choice of the representations ${\bf r}_1$ and ${\bf r}_2$ and of the magnetic fluxes could not give a consistent blowup equation whose solution yields a physically sensible spectrum of the loop operator. The solutions to the blowup equations appear to involve unphysical states which either do not form representations under the $SO(4)$ Lorentz rotation or have negative degeneracies. So our blowup approach fails to determine the degeneracies of BPS bound states to the loop operator $W_{{\bf 2}^{\otimes 4}}$ in the $SU(2)_\pi$ theory.

\subsection{Local \texorpdfstring{$\mathbb{P}^2$}{P2}}

The minimal rank-one SCFT with no mass parameter is engineered by M-theory compactified on a local $\mathbb{P}^2$ embedded in a Calabi-Yau 3-fold. The effective prepotential on the $\Omega$-background of this theory can be calculated using the geometric data as
\begin{align}
\mathcal{E} = \frac{1}{\epsilon_1 \epsilon_2} \qty(\frac{3}{2}\phi^3 - \frac{\epsilon_1^2 + \epsilon_2^2}{8}\phi + \epsilon_+^2 \phi ) \, .
\end{align}
The partition function for this theory on the $\Omega$-background was computed in \cite{Huang:2017mis, Kim:2020hhh}, based on the blowup equations with consistent magnetic flux given by 
\begin{align}\label{eq:P2_flux}
n \in \mathbb{Z} + 1/6 \, .
\end{align}

This theory does not have mass deformations leading to a gauge theory description and hence it is a non-Lagrangian theory. It is therefore difficult to define a loop operator in a conventional gauge theoretic way. We instead introduce a loop operator in a geometric sense, which is to insert a non-compact primitive curve intersecting with the $\mathbb{P}^2$ at one point, as depicted in Figure \ref{fig:P2}. This enables us to consider the Wilson loop operator whose expectation value takes the form of K\"ahler parameter expansion as
\begin{align}
\langle W_{[-1]} \rangle = e^\phi + \mathcal{O}(e^{-2\phi}) \, ,
\end{align}
where terms of higher orders in this expectation value can be obtained by solving the blowup equation. For this, we formulate the blowup equation with ${\bf r}_1=[-1]$ at the North pole and an empty operator ${\bf r}_2=\emptyset$ at the South pole. Then the blowup equation \eqref{eq:bleqW2} can be written as
\begin{align}
\Lambda_0 \langle \hat{W}_{[-1]} \rangle
= \sum_{n \in \mathbb{Z} + 1/6} (-1)^n \langle\hat{W}_{[-1]}^{(N)}\rangle \frac{\hat{Z}^{(N)} \hat{Z}^{(S)}}{\hat{Z}} \, .
\end{align}
Here, $\Lambda_0 $ is fixed at the first order to be $ (-1)^{1/6} p_1^{-2/9} p_2^{-1/18} $. This blowup equation is solvable and the solution is summarized in Table~\ref{table:P2}. 

We note that this Wilson loop operator can also be obtained by an RG-flow from the fundamental Wilson loop in the $SU(2)_\pi$ gauge theory by integrating out an instantonic hypermultiplet state. We checked that the result in Table~\ref{table:P2} as well as other higher order degeneracies in the K\"ahler parameter expansion indeed match the spectrum of the fundamental Wilson loop in the $SU(2)_\pi$ gauge theory in Table~\ref{table:SU2_pi_fund} after the RG-flow.

\begin{table}
	\centering
	\begin{tabular}{|c|C{33ex}||c|C{33ex}|} \hline
		$ d $ & $ \oplus \tilde{N}_{j_l, j_r}^{d} (j_l, j_r) $ & $ d $ & $ \oplus \tilde{N}_{j_l, j_r}^{d} (j_l, j_r) $ \\ \hline
		$ -1 $ & $ (0, 0) $ & $ 2 $ & $ (0, \frac{1}{2}) $ \\ \hline
		$ 5 $ & $ (0, 2) $ & $ 8 $ & $ (0, \frac{5}{2}) \oplus (0, \frac{7}{2}) \oplus (\frac{1}{2}, 4) $ \\ \hline
		$ 11 $ & \multicolumn{3}{C{72.5ex}|}{$ (0, 2) \oplus (0, 3) \oplus 2(0, 4) \oplus (0, 5) \oplus (0, 6) \oplus (\frac{1}{2}, \frac{7}{2}) \oplus 2(\frac{1}{2}, \frac{9}{2}) \oplus 2(\frac{1}{2}, \frac{11}{2}) \oplus (1, 5) \oplus (1, 6) \oplus (\frac{3}{2}, \frac{13}{2}) $} \\ \hline
	\end{tabular}
	\caption{Spectrum of the SCFT of a local $ \mathbb{P}^2 $ in the presence of a Wilson loop $ \ev*{W_{[-1]}} $ for $ d \leq 13 $. Here, $ d $ labels BPS states with charge $ d\phi $.}\label{table:P2}
\end{table}

\subsection{\texorpdfstring{$SU(3)_7, \, G_2$}{SU(3)7 and G2} theory}
Next, we consider theories that are UV dual to each other and discuss how Wilson loop operators in such a theory are mapped to those in dual theories under the duality. As a representative example, we consider the loop operators in the 5d SCFT described by the pure $ SU(3) $ gauge theory at the CS-level $ 7 $ which is dual to the pure $ G_2 $ gauge theory. These theories are realized geometrically by gluing two Hirzebruch surfaces as
\begin{align}\label{eq:su3_7}
\begin{tikzpicture}
\draw[thick](-3,0)--(0,0);	
\node at(-3.6,0) {$\mathbb{F}_{8}{}\big|_1$};
\node at(-2.8,0.3) {${}_e$};
\node at(0.5,0) {$\mathbb{F}_{0}{}\big|_2$\ .};
\node at(-0.6,0.3) {${}_{h+3f}$};
\end{tikzpicture}
\end{align}
The fiber-base duality in this geometry relates the $SU(3)_7$ and the $G_2$ gauge theories \cite{Jefferson:2018irk, Hayashi:2018lyv}.

We shall now compute the VEVs of Wilson loop operators in these two dual gauge theory descriptions using the blowup equations. As we will see, the duality predicts quite non-trivial relationships between Wilson loop operators in the two different gauge descriptions. We will check the relationship by explicitly computing partition functions of the loop operators.

For computational ease, we will use the basis $ \{Q_1, Q_2, Q_3\} $ for the K\"ahler parameters which are related to the parameters in the $SU(3)$ frame as
\begin{align}
Q_1 = e^{-(m - 3\phi_1 + 2\phi_2)} \, , \quad
Q_2 = e^{-(2\phi_1 - \phi_2)} \, , \quad
Q_3 = e^{-(-\phi_1 + 2\phi_2)} \, ,
\end{align}
and those in the $G_2$ frame as
\begin{align}
Q_1 = e^{-(- 3\phi'_1 + 2\phi'_2)} \, , \quad
Q_2 = e^{-(2\phi'_1 - \phi'_2)} \, , \quad
Q_3 = e^{-(m' -\phi'_1 + 2\phi'_2)} \, .
\end{align}
Here, the unprimed and the primed parameters are the K\"ahler parameters in the $SU(3)_7$ and in the $G_2$ gauge theory descriptions, respectively.

In terms of the $SU(3)_7$ gauge theory parameters, the effective prepotential of this theory takes the form
\begin{align}
\mathcal{E} &= \frac{1}{\epsilon_1 \epsilon_2} \qty(\mathcal{F} - \frac{\epsilon_1^2 + \epsilon_2^2}{12}(\phi_1 + \phi_2) + \epsilon_+^2 (\phi_1 + \phi_2))\ , \nonumber \\
6\mathcal{F} &= 8\phi_1^3 + 18\phi_1^2 \phi_2 - 24\phi_1 \phi_2^2 + 8\phi_2^3 + 6m(\phi_1^2 - \phi_1 \phi_2 + \phi_2^2) \, ,
\end{align}
while that in terms of the $ G_2 $ gauge theory parameters is given by
\begin{align}
\mathcal{E} &= \frac{1}{\epsilon_1 \epsilon_2} \qty(\mathcal{F}' - \frac{\epsilon_1^2 + \epsilon_2^2}{12}(\phi'_1 + \phi'_2) + \epsilon_+^2 (\phi'_1 + \phi'_2))\ , \nonumber \\
6\mathcal{F}' &= 8\phi'_1{}^3 + 18\phi'_1{}^2 \phi'_2 - 24\phi'_1 \phi'_2{}^2 + 8\phi'_2{}^3 + 6m'(3\phi'_1{}^2 - 3\phi_1 \phi_2 + \phi'_2{}^2) \, .
\end{align}

The partition functions for these theories without loop operators can be calculated using the blowup equation with a set of consistent magnetic fluxes 
\begin{align}
	\left\{\begin{array}{lll} n_i \in \mathbb{Z} \, , & \ 
B_m = 0, \pm 1 \, & \ {\rm for} \ SU(3)_7 \, ,\\
n_i' \in \mathbb{Z} \, , & \ 
B_{m'} = 0, \pm 1 \, & \ {\rm for} \ G_2 \, .
 \end{array}\right.
\end{align}
See \cite{Kim:2020hhh} for the detailed calculation.

\paragraph{Representation $ \mathbf{r} = [0, -1] $}

Consider now a Wilson loop in the representation $ \mathbf{r} = [0, -1] $ which is geometrically realized by a heavy M2-brane state wrapping a non-compact curve intersecting the second surface $\mathbb{F}_0$ at one point. This loop operator amounts to the fundamental Wilson loop in the $ SU(3)_7 $ theory whose classical VEV is given by
\begin{align}\label{eq:classical-su3-7}
\ev*{W_{[0, -1]}^{\mathrm{cls}}}_{SU(3)_7} = e^{\phi_2} + e^{\phi_1 - \phi_2} + e^{-\phi_1} \, .
\end{align}
On the other hand, in the $G_2$ gauge theory this operator corresponds to the adjoint representation with the classical VEV as
\begin{align}
\ev*{W_{[0, -1]}^{\mathrm{cls}}}_{G_2} &= e^{\phi'_2} + e^{3\phi'_1-\phi'_2} + e^{\phi'_1} + e^{-\phi'_1 + \phi'_2} + e^{-3\phi'_1 + 2\phi'_2} + e^{2\phi'_1 - \phi'_2} + 2 \nonumber \\
& \quad + e^{-2\phi'_1 + \phi'_2} + e^{3\phi'_1 - 2\phi'_2} + e^{\phi'_1 - \phi'_2} + e^{-\phi'_1} + e^{-3\phi'_1 + \phi'_2} + e^{-\phi'_2} \, .
\end{align}
Quite interestingly, the duality predicts an interesting identification of the fundamental Wilson loop in the $SU(3)_7$ gauge theory with the adjoint Wilson loop in the $G_2$ gauge theory. We can check this duality of loop operators by comparing their partition functions.

By plugging the classical VEV \eqref{eq:classical-su3-7} into the blowup equation \eqref{eq:bleqW2} with fluxes $ B_m = 0, 1 $, we compute the VEV of the fundamental Wilson loop in the $SU(3)_7$ theory as
\begin{align}\label{eq:su3_7_fund}
Q_2^{1/3} Q_3^{2/3} \langle W_{[0, -1]} \rangle_{SU(3)_7}
&= 1 + Q_1 + Q_3 +  Q_1 Q_2 +  Q_2 Q_3 +\chi_1(p_1 p_2) Q_1 Q_3 + Q_1 Q_2^2 \nonumber \\
& \quad + (1 + \chi_1(p_1 p_2)) Q_1 Q_2 Q_3 + \chi_2(p_1 p_2) Q_1 Q_3^2  \nonumber \\
& \quad + \chi_2(p_1 p_2) Q_1^2 Q_3 + \cdots \, ,
\end{align}
where $\chi_n(p)$ is the $SU(2)$ character of spin $n$. Here, for convenience, we have multiplied the factor $Q_2^{1/3}Q_3^{2/3}=e^{-\phi_2}$ in the LHS so that the ground state of the loop operator with electric charge $[0,-1]$ (so labeled by the chemical potential $e^{\phi_2}$) corresponds to the state `1' in the RHS. Similarly, we compute the VEV of the adjoint Wilson loop in the $G_2$ gauge theory by solving the blowup equation with the fluxes $ B_{m'} = 0, 1 $. The solution is
\begin{align}\label{eq:g2_adj}
Q_1^2 Q_2^3 \langle W_{[0, -1]}\rangle_{G_2}
&= 1 + Q_1 + Q_3 +  Q_1 Q_2 +  Q_2 Q_3 +\chi_1(p_1 p_2) Q_1 Q_3 + Q_1 Q_2^2 \nonumber \\
& \quad + (1 + \chi_1(p_1 p_2)) Q_1 Q_2 Q_3 + \chi_2(p_1 p_2) Q_1 Q_3^2  \nonumber \\
& \quad + \chi_2(p_1 p_2) Q_1^2 Q_3 + \cdots \, .
\end{align}
Here we again normalized the result by multiplying the factor $Q_1^{2}Q_2^{3}=e^{-\phi'_2}$.
As expected from the duality we proposed above, the result shows perfect agreement between these two Wilson loop operators. Table~\ref{table:SU3_7_fund} contains more BPS spectrum from the perspective of the $ SU(3)_7 $ gauge theory.

\begin{table}[t]
	\centering
	\begin{tabular}{|c|C{28ex}||c|C{28ex}|} \hline
		$ \mathbf{d} $ & $ \oplus \tilde{N}_{j_l, j_r}^{\mathbf{d}} (j_l, j_r) $ & $ \mathbf{d} $ & $ \oplus \tilde{N}_{j_l, j_r}^{\mathbf{d}} (j_l, j_r) $ \\ \hline
		$ (1, 0, 0) $ & $ (0, 0) $ & $ (1, 0, 1) $ & $ (0, 1) $ \\ \hline
		$ (1, 0, 2) $ & $ (0, 2) $ & $ (1, 0, 3) $ & $ (0, 3) $ \\ \hline
		$ (1, 1, 0) $ & $ (0, 0) $ & $ (1, 1, 1) $ & $ (0, 0) \oplus (0, 1) $ \\ \hline
		$ (1, 1, 2) $ & $ (0, 1) \oplus (0, 2) $ & $ (1, 1, 3) $ & $ (0, 2) \oplus (0, 3) $ \\ \hline
		$ (1, 2, 0) $ & $ (0, 0) $ & $ (1, 2, 1) $ & $ (0, 0) \oplus (0, 1) $ \\ \hline
		$ (1, 2, 2) $ & $ (0, 0) \oplus (0, 1) \oplus (0, 2) $ & $ (1, 2, 3) $ & $ (0, 1) \oplus (0, 2) \oplus (0, 3) $ \\ \hline
		$ (1, 3, 0) $ & $ (0, 0) $ & $ (1, 3, 1) $ & $ (0, 0) \oplus (0, 1) $ \\ \hline
		$ (1, 3, 2) $ & $ (0, 0) \oplus (0, 1) \oplus (0, 2) $ & $ (1, 3, 3) $ & $ (0, 0) \oplus (0, 1) \oplus (0, 2) \oplus (0, 3) $ \\ \hline
		$ (2, 0, 1) $ & $ (0, 2) $ & $ (2, 0, 2) $ & $ (0, 2) \oplus 2(0, 3) \oplus (\frac{1}{2}, \frac{7}{2}) $ \\ \hline
		$ (2, 0, 3) $ & $ (0,2) \oplus 2(0,3) \oplus 3(0,4) \oplus (\frac{1}{2},\frac{7}{2}) \oplus 2(\frac{1}{2},\frac{9}{2}) \oplus (1,5) $ & $ (2, 1, 1) $ & $ (0, 1) \oplus (0, 2) $ \\ \hline
		$ (2, 1, 2) $ & $ (0,1) \oplus 4(0,2) \oplus 3(0,3) \oplus (\frac{1}{2},\frac{5}{2}) \oplus (\frac{1}{2},\frac{7}{2}) $ & $ (2, 1, 3) $ & $ (0,1) \oplus 4(0,2) \oplus 8(0,3) \oplus 5(0,4) \oplus (\frac{1}{2},\frac{5}{2}) \oplus 4(\frac{1}{2},\frac{7}{2}) \oplus 3(\frac{1}{2},\frac{9}{2}) \oplus (1,4) \oplus (1,5) $ \\ \hline
		$ (2, 2, 0) $ & $ (0, 0) $ & $ (2, 2, 1) $ & $ (0, 0) \oplus 2(0, 1) \oplus (0, 2) $ \\ \hline
		$ (2, 2, 2) $ & $ (0,0) \oplus 4(0,1) \oplus 6(0,2) \oplus 3(0,3) \oplus (\frac{1}{2},\frac{3}{2}) \oplus (\frac{1}{2},\frac{5}{2}) \oplus (\frac{1}{2},\frac{7}{2}) $ & $ (2, 2, 3) $ & $ (0,0) \oplus 4(0,1) \oplus 11(0,2) \oplus 12(0,3) \oplus 6(0,4) \oplus
		(\frac{1}{2},\frac{3}{2})\oplus 4(\frac{1}{2},\frac{5}{2})\oplus 5(\frac{1}{2},\frac{7}{2})\oplus 3(\frac{1}{2},\frac{9}{2})\oplus (1,3) \oplus (1,4) \oplus (1,5) $ \\ \hline
		$ (2, 3, 1) $ & $ 3(0,0) \oplus 4(0,1) \oplus (0,2) \oplus (\frac{1}{2},\frac{1}{2}) $ & $ (2, 3, 2) $ & $ 3(0,0) \oplus 8(0,1) \oplus 8(0,2) \oplus 3(0,3) \oplus (\frac{1}{2},\frac{1}{2})\oplus 2(\frac{1}{2},\frac{3}{2})\oplus (\frac{1}{2},\frac{5}{2})\oplus (\frac{1}{2},\frac{7}{2}) $ \\ \hline
		$ (2, 3, 3) $ & \multicolumn{3}{C{68.5ex}|}{$ 3(0,0) \oplus 11(0,1) \oplus 17(0,2) \oplus 15(0,3) \oplus 6(0,4) \oplus (\frac{1}{2},\frac{1}{2}) \oplus 4(\frac{1}{2},\frac{3}{2}) \oplus 6(\frac{1}{2},\frac{5}{2}) \oplus 5(\frac{1}{2},\frac{7}{2}) \oplus 3(\frac{1}{2},\frac{9}{2}) \oplus (1,2) \oplus (1,3) \oplus (1,4) \oplus (1,5) $} \\ \hline
	\end{tabular}
	\caption{Spectrum of the bound states to the fundamental Wilson loop in the $SU(3)_7$ theory up to $ d_1 \leq 2 $ and $ d_2, d_3 \leq 3 $. Here, $ \mathbf{d} = (d_1, d_2, d_3) $ labels the BPS states with fugacity $ Q_2^{-1/3} Q_3^{-2/3} \prod Q_i^{d_i} $. For spectrum of the $G_2$ theory, one takes the fugacity to be $ Q_1^{-2} Q_2^{-3} \prod Q_i^{d_i} $.}\label{table:SU3_7_fund}
\end{table}

\paragraph{Representation $ \mathbf{r} = [-1, 0] $}

Now we consider the Wilson loop operator in the representation $ \mathbf{r} = [-1, 0] $ corresponding to an M2-brane state wrapping a non-compact curve intersecting the surface $\mathbb{F}_8$ at one point in the geometry \eqref{eq:su3_7}.

In the $SU(3)_7$ gauge theory description, this operator is the Wilson loop in the anti-fundamental representation. In the dual $G_2$ gauge theory description, it is realized by the Wilson loop in the fundamental representation. This duality map between the Wilson loop operators is rather interesting. As representations, the fundamental and the anti-fundamental Wilson loops in the $SU(3)$ gauge theory are complex conjugate to each other. However, they are respectively mapped to the adjoint and the fundamental Wilson loops in the dual $G_2$ gauge theory in which two loops are not conjugate to each other. As we will see shortly, the fundamental and the anti-fundamental loop operators in the $SU(3)$ theory remain completely distinct even after 
 the non-perturbative corrections are included. Indeed, this was expected because the charge conjugation symmetry of the $SU(3)_7$ gauge theory was broken already by the Chern-Simons term at level $7$. In 5d gauge theories, the naive relation between Wilson loop expectations can be significantly modified by non-perturbative instanton effects. 


Let us first verify the duality for the Wilson loop in the representation $ \mathbf{r} = [-1, 0] $ by explicitly computing its partition function. The classical VEV of this loop operator in two gauge theory descriptions are
\begin{align}
\ev*{W_{[-1, 0]}^{\mathrm{cls}}}_{SU(3)_7} = e^{\phi_1} + e^{-\phi_1 + \phi_2} + e^{-\phi_2} \, ,
\end{align}
in the $ SU(3)_7$ frame and
\begin{align}
\ev*{W_{[-1, 0]}^{\mathrm{cls}}}_{G_2} = e^{\phi'_1} + e^{-\phi'_1 + \phi'_2} + e^{2\phi'_1 - \phi'_2} + 1 + e^{-2\phi'_1 + \phi'_2} + e^{\phi'_1 - \phi'_2} + e^{-\phi'_1} \, ,
\end{align}
in the $G_2$ frame. We then compute the non-perturbative instanton corrections by solving the blowup equations with the magnetic fluxes $ B_m ,B_{m'} = 0, 1 $. The solution for the anti-fundamental Wilson loop in the $ SU(3)_7 $ theory is
\begin{align}
Q_2^{2/3} Q_3^{1/3} \langle W_{[-1, 0]}\rangle_{SU(3)_7}
&= 1 + Q_2 + Q_2 Q_3 + Q_1 Q_2 + Q_1 Q_2^2 + \chi_1(p_1 p_2) Q_1 Q_2 Q_3 \nonumber \\
&\quad + Q_1 Q_2^3 + (1 + \chi_1(p_1 p_2)) Q_1 Q_2^2 Q_3 + \chi_2(p_1 p_2) Q_1 Q_2 Q_3^2 \nonumber \\
&\quad + \chi_2(p_1 p_2) Q_1^2 Q_2 Q_3 + \cdots \ ,
\end{align}
and the solution for the fundamental Wilson loop in the $ G_2 $ theory is
\begin{align}
Q_1 Q_2^{2} \langle W_{[-1, 0]}\rangle_{G_2}
&= 1 + Q_2 + Q_2 Q_3 + Q_1 Q_2 + Q_1 Q_2^2 + \chi_1(p_1 p_2) Q_1 Q_2 Q_3 \nonumber \\
&\quad + Q_1 Q_2^3 + (1 + \chi_1(p_1 p_2)) Q_1 Q_2^2 Q_3 + \chi_2(p_1 p_2) Q_1 Q_2 Q_3^2 \nonumber \\
&\quad + \chi_2(p_1 p_2) Q_1^2 Q_2 Q_3 + \cdots \ .
\end{align}
These two results completely agree, which is consistent with the expected duality. We summarize in Table~\ref{table:SU3_7_antifund} some leading BPS states bound to the Wilson loop in the perspective of the $ SU(3) $ gauge theory. Comparing Tables~\ref{table:SU3_7_fund} and \ref{table:SU3_7_antifund}, one can see that spectra of two Wilson loop operators are not conjugate to each other.

\begin{table}[t]
	\centering
	\begin{tabular}{|c|C{28ex}||c|C{28ex}|} \hline
		$ \mathbf{d} $ & $ \oplus \tilde{N}_{j_l, j_r}^{\mathbf{d}} (j_l, j_r) $ & $ \mathbf{d} $ & $ \oplus \tilde{N}_{j_l, j_r}^{\mathbf{d}} (j_l, j_r) $ \\ \hline
		$ (1, 1, 0) $ & $ (0, 0) $ & $ (1, 1, 1) $ & $ (0, 1) $ \\ \hline
		$ (1, 1, 2) $ & $ (0, 2) $ & $ (1, 1, 3) $ & $ (0, 3) $ \\ \hline
		$ (1, 2, 0) $ & $ (0, 0) $ & $ (1, 2, 1) $ & $ (0, 0) \oplus (0, 1) $ \\ \hline
		$ (1, 2, 2) $ & $ (0, 1) \oplus (0, 2) $ & $ (1, 2, 3) $ & $ (0, 2) \oplus (0, 3) $ \\ \hline
		$ (1, 3, 0) $ & $ (0, 0) $ & $ (1, 3, 1) $ & $ (0, 0) \oplus (0, 1) $ \\ \hline
		$ (1, 3, 2) $ & $ (0, 0) \oplus (0, 1) \oplus (0, 2) $ & $ (1, 3, 3) $ & $ (0, 1) \oplus (0, 2) \oplus (0, 3) $ \\ \hline
		$ (1, 4, 1) $ & $ (0, 1) $ & $ (1, 4, 2) $ & $ (0, 1) \oplus (0, 2) $ \\ \hline
		$ (1, 4, 3) $ & $ (0, 0) \oplus (0, 1) \oplus (0, 2) \oplus (0, 3) $ & $ (2, 1, 1) $ & $ (0, 2) $ \\ \hline
		$ (2, 1, 2) $ & $ (0, 2) \oplus 2(0, 3) \oplus (\frac{1}{2}, \frac{7}{2}) $ & $ (2, 1, 3) $ & $ (0,2) \oplus 2(0,3) \oplus 3(0,4) \oplus (\frac{1}{2},\frac{7}{2}) \oplus 2(\frac{1}{2},\frac{9}{2}) \oplus (1,5) $ \\ \hline
		$ (2, 2, 1) $ & $ (0, 1) \oplus (0, 2) $ & $ (2, 2, 2) $ & $ (0,1) \oplus 4(0,2) \oplus 3(0,3) \oplus (\frac{1}{2},\frac{5}{2}) \oplus (\frac{1}{2},\frac{7}{2}) $ \\ \hline
		$ (2, 2, 3) $ & $ (0,1) \oplus 4(0,2) \oplus 8(0,3) \oplus 5(0,4) \oplus (\frac{1}{2},\frac{5}{2}) \oplus 4(\frac{1}{2},\frac{7}{2}) \oplus 3(\frac{1}{2},\frac{9}{2}) \oplus (1,4) \oplus (1,5) $ & $ (2, 3, 0) $ & $ (0, 0) $ \\ \hline
		$ (2, 3, 1) $ & $ (0,0) \oplus 2(0,1) \oplus (0,2) $ & $ (2, 3, 2) $ & $ (0,0) \oplus 4(0,1) \oplus 6(0,2) \oplus 3(0,3) \oplus (\frac{1}{2},\frac{3}{2})\oplus (\frac{1}{2},\frac{5}{2})\oplus (\frac{1}{2},\frac{7}{2}) $ \\ \hline
		$ (2, 3, 3) $ & $ (0,0) \oplus 4(0,1) \oplus 11(0,2) \oplus 12(0,3) \oplus 6(0,4) \oplus (\frac{1}{2},\frac{3}{2}) \oplus 4(\frac{1}{2},\frac{5}{2}) \oplus 5(\frac{1}{2},\frac{7}{2}) \oplus 3(\frac{1}{2},\frac{9}{2}) \oplus (1,3) \oplus (1,4) \oplus (1,5) $ & $ (2, 4, 0) $ & $ (0,0) $ \\ \hline
		$ (2, 4, 1) $ & $ 2(0,0) \oplus 3(0,1) \oplus (0,2) $ & $ (2, 4, 2) $ & $ 3(0,0) \oplus 7(0,1) \oplus 7(0,2) \oplus 3(0,3) \oplus (\frac{1}{2},\frac{1}{2}) \oplus (\frac{1}{2},\frac{3}{2}) \oplus (\frac{1}{2},\frac{5}{2}) \oplus (\frac{1}{2},\frac{7}{2}) $ \\ \hline
		$ (2, 4, 3) $ & \multicolumn{3}{C{68.5ex}|}{$ 3(0,0) \oplus 11(0,1) \oplus 16(0,2) \oplus 14(0,3) \oplus 6(0,4) \oplus (\frac{1}{2},\frac{1}{2}) \oplus 4(\frac{1}{2},\frac{3}{2}) \oplus 5(\frac{1}{2},\frac{5}{2}) \oplus 5(\frac{1}{2},\frac{7}{2}) \oplus 3(\frac{1}{2},\frac{9}{2}) \oplus (1,2) \oplus (1,3) \oplus (1,4) \oplus (1,5) $} \\ \hline
	\end{tabular}
	\caption{Spectrum of the anti-fundamental Wilson loop in the $ SU(3)_7 $ theory for $ (d_1, d_2, d_3) \leq (2, 4, 3) $. Here, $ \mathbf{d} = (d_1, d_2, d_3) $ labels the BPS states with fugacity $ Q_2^{-2/3} Q_3^{1/3} \prod Q_i^{d_i} $.}\label{table:SU3_7_antifund}
\end{table}

\subsection{\texorpdfstring{$SU(4)_6, \ Sp(3)_0$}{SU(4)6 and Sp(3)0} theory}

As another interesting example of the loop operators in UV dual theories, we consider Wilson loops in the $ SU(4) $ gauge theory at the CS-level 6 and those in the dual $ Sp(3) $ gauge theory at $ \theta = 0 $. The geometric construction of these theories is given by
\begin{align}\label{eq:geo-su4-7}
\begin{tikzpicture}
\draw (0, 0) node {$ \mathbb{F}_8 $}
(3, 0) node {$ \mathbb{F}_6 $}
(6, 0) node {$ \mathbb{F}_0 $};
\draw[thick] (0.5, 0) -- (2.5, 0)
(3.5, 0) -- (5.5, 0);
\draw (0.7, 0.2) node {$ _e $}
(2.3, 0.2) node {$ _h $}
(3.7, 0.2) node {$ _e $}
(5.1, 0.2) node {$ _{e+2f} $};
\end{tikzpicture}
\end{align}
The duality between the $SU(4)_6$ theory and the $ Sp(3)_0 $ theory proposed in \cite{Gaiotto:2015una} is geometrically realized by exchanging $ e $ and $ f $ curves in $ \mathbb{F}_0 $.

We shall consider the minimal Wilson loops $W_{\bf r}$ with ${\bf r}=[-1,0,0], [0,-1,0],$ and $[0,0,-1]$ which intersect one of three Hirzebruch surfaces at one point. These loop operators in the $SU(4)_6$ theory correspond to Wilson loops in the anti-fundamental, anti-symmetric and fundamental representations, respectively. In the $Sp(3)_0$ theory, on the other hand, they correspond to Wilson loops in the fundamental, rank-2 and rank-3 anti-symmetric representations, respectively. 
We note that the  map between the Wilson loops in two dual gauge theories is also consistent with the $\mathbb{Z}_2$ 1-form symmetry which acts non-trivially on the (anti-)fundamentals of $SU(4)$ and also on the fundamental and the rank-3 anti-symmetric representations of $Sp(3)$,    whereas it  acts trivially on the rank-2 anti-symmetric representations of both $SU(4)$ and $Sp(3)$. Note  also  that the Wilson loop expectation values of ${\bf r}=[-1,0,0]$ and ${\bf r}=[0,0,-1]$ are classically complex conjugate to each other in the $SU(4)$ theory; however their duals in the $Sp(3)$ theory are completely independent representations. Hence the duality implies that the classical relations between Wilson loop expectation values in the $SU(4)_6$ theory are not maintained under the non-perturbative corrections, which was already anticipated since the non-vanishing Chern-Simons term breaks charge conjugation symmetry. We will verify these dualities between the Wilson loops explicitly by evaluating their expectation values.



Let $ Q_1 $, $ Q_2 $, $ Q_3, Q_4 $ be K\"ahler parameters for four basis curves in the geometry \eqref{eq:geo-su4-7} which are the fiber curves in three Hirzebruch surfaces and the $ e $ curve in $ \mathbb{F}_0 $. They can be written in terms of the Coulomb branch parameters as well as the mass parameter in the $ SU(4) $ theory as
\begin{align}\label{eq:SU4_basis}
Q_1 = e^{-(m - 2\phi_2 + 2\phi_3)} \, , \quad
Q_2 = e^{-(2\phi_1 - \phi_2)} \, , \quad
Q_3 = e^{-(-\phi_1 + 2\phi_2 - \phi_3)} \, , \quad
Q_4 = e^{-(-\phi_2 + 2\phi_3)} \, .
\end{align}
In the $ Sp(3) $ theory, the K\"ahler parameters become
\begin{align}\label{eq:Sp3_basis}
Q_1 = e^{-(-2\phi'_2 + 2\phi'_3)} \, , \quad
Q_2 = e^{-(2\phi'_1 - \phi'_2)} \, , \quad
Q_3 = e^{-(-\phi'_1 + 2\phi'_2 - \phi'_3)} \, , \quad
Q_4 = e^{-(m' -\phi'_2 + 2\phi'_3)} \, .
\end{align}

We first compute the partition function without loops by solving the blowup equations.
The effective prepotential in the $ SU(4)_6 $ frame is
\begin{align}
\mathcal{E} &= \frac{1}{\epsilon_1 \epsilon_2} \qty(\mathcal{F} - \frac{\epsilon_1^2 + \epsilon_2^2}{12}(\phi_1 + \phi_2 + \phi_3) + \epsilon_+^2 (\phi_1 + \phi_2 + \phi_3)) \\
6\mathcal{F} &= 8\phi_1^3 + 18\phi_1^2 \phi_2 - 24\phi_1 \phi_2^2 + 8\phi_2^3 + 12\phi_2 \phi_3^2 - 18\phi_2 \phi_3^2 + 8\phi_3^3 \nonumber \\
&\quad + 6m (\phi_1^2 - \phi_1 \phi_2 + \phi_2^2 - \phi_2 \phi_3 + \phi_3^2) \, ,
\end{align}
while in the $ Sp(3)_0 $ frame, it is given by
\begin{align}
\mathcal{E} &= \frac{1}{\epsilon_1 \epsilon_2} \qty(\mathcal{F}'' - \frac{\epsilon_1^2 + \epsilon_2^2}{12}(\phi'_1 + \phi'_2 + \phi'_3) + \epsilon_+^2 (\phi'_1 + \phi'_2 + \phi'_3)) \\
6\mathcal{F}'' &= 8\phi'_1{}^3 + 18\phi'_1{}^2 \phi'_2 - 24\phi'_1 \phi'_2{}^2 + 8\phi'_2{}^3 + 12\phi'_2 \phi'_3{}^2 - 18\phi'_2 \phi'_3{}^2 + 8\phi'_3{}^3 \nonumber \\
&\quad + 6m' (2\phi'_1{}^2 - 2\phi'_1 \phi'_2 + 2\phi'_2{}^2 - 2\phi'_2 \phi'_3 + \phi'_3{}^2) \, .
\end{align}
Using these effective prepotentials and the following magnetic fluxes
\begin{align}
\left\{\begin{array}{lll} n_i \in \mathbb{Z} \, , & \ 
B_m = 0, \pm 1 \, & \ {\rm for} \ SU(4)_6 \ ,\\
n_i' \in \mathbb{Z} \, , & \ 
B_{m'} = 0, \pm 1 \, & \ {\rm for} \ Sp(3)_0 \, ,
\end{array}\right.
\end{align}
we can formulate the blowup equations, and we checked that the solutions to the blowup equations in two dual gauge theories completely agree with each other in the K\"ahler parameter expansion. The result is listed in Table~\ref{table:SU4_6}.

\begin{table}
	\centering
	\begin{tabular}{|c|C{26ex}||c|C{28ex}|} \hline
		$ \mathbf{d} $ & $ \oplus N_{j_l, j_r}^{\mathbf{d}} (j_l, j_r) $ & $ \mathbf{d} $ & $ \oplus N_{j_l, j_r}^{\mathbf{d}} (j_l, j_r) $ \\ \hline
		$ (1, 0, 0, 0) $ & $ (0, \frac{1}{2}) $ & $ (1, 0, 0, 1) $ & $ (0, \frac{3}{2}) $ \\ \hline
		$ (1, 0, 0, 2) $ & $ (0, \frac{5}{2}) $ & $ (1, 0, 1, 0) $ & $ (0, \frac{1}{2}) $ \\ \hline
		$ (1, 0, 1, 1) $ & $ (0, \frac{1}{2}) \oplus (0, \frac{3}{2}) $ & $ (1, 0, 1, 2) $ & $ (0, \frac{3}{2}) \oplus (0, \frac{5}{2}) $ \\ \hline
		$ (1, 0, 2, 0) $ & $ (0, \frac{1}{2}) $ & $ (1, 0, 2, 1) $ & $ (0, \frac{1}{2}) \oplus (0, \frac{3}{2}) $ \\ \hline
		$ (1, 0, 2, 2) $ & $ (0, \frac{1}{2}) \oplus (0, \frac{3}{2}) \oplus (0, \frac{5}{2}) $ & $ (1, 1, 1, 0) $ & $ (0, \frac{1}{2}) $ \\ \hline
		$ (1, 1, 1, 1) $ & $ (0, \frac{1}{2}) \oplus (0, \frac{3}{2}) $ & $ (1, 1, 1, 2) $ & $ (0, \frac{3}{2}) \oplus (0, \frac{5}{2}) $ \\ \hline
		$ (1, 1, 2, 0) $ & $ (0, \frac{1}{2}) $ & $ (1, 1, 2, 1) $ & $ 2(0, \frac{1}{2}) \oplus (0, \frac{3}{2}) $ \\ \hline
		$ (1, 1, 2, 2) $ & $ (0, \frac{1}{2}) \oplus 2(0, \frac{3}{2}) \oplus (0, \frac{5}{2}) $ & $ (1, 2, 2, 0) $ & $ (0, \frac{1}{2}) $ \\ \hline
		$ (1, 2, 2, 1) $ & $ (0, \frac{1}{2}) \oplus (0, \frac{3}{2}) $ & $ (1, 2, 2, 2) $ & $ (0, \frac{1}{2}) \oplus (0, \frac{3}{2}) \oplus (0, \frac{5}{2}) $ \\ \hline
		$ (2, 0, 0, 1) $ & $ (0, \frac{5}{2}) $ & $ (2, 0, 0, 2) $ & $ (0, \frac{5}{2}) \oplus (0, \frac{7}{2}) \oplus (\frac{1}{2}, 4) $ \\ \hline
		$ (2, 0, 1, 1) $ & $ (0, \frac{3}{2}) \oplus (0, \frac{5}{2}) $ & $ (2, 0, 1, 2) $ & $ (0,\frac{3}{2}) \oplus 3(0,\frac{5}{2}) \oplus 2(0,\frac{7}{2}) \oplus (\frac{1}{2},3) \oplus (\frac{1}{2},4) $ \\ \hline
		$ (2, 0, 2, 1) $ & $ (0, \frac{1}{2}) \oplus (0, \frac{3}{2}) \oplus (0, \frac{5}{2}) $ & $ (2, 0, 2, 2) $ & $ (0,\frac{1}{2}) \oplus 3(0,\frac{3}{2}) \oplus 4(0,\frac{5}{2}) \oplus 2(0,\frac{7}{2}) \oplus (\frac{1}{2},2) \oplus (\frac{1}{2},3) \oplus (\frac{1}{2},4) $ \\ \hline
		$ (2, 1, 1, 1) $ & $ (0, \frac{3}{2}) \oplus (0, \frac{5}{2}) $ & $ (2, 1, 1, 2) $ & $ (0,\frac{3}{2}) \oplus 3(0,\frac{5}{2}) \oplus 2(0,\frac{7}{2}) \oplus (\frac{1}{2},3) \oplus (\frac{1}{2},4) $ \\ \hline
		$ (2, 1, 2, 1) $ & $ (0, \frac{1}{2}) \oplus 2(0, \frac{3}{2}) \oplus (0, \frac{5}{2}) $ & $ (2, 1, 2, 2) $ & $ (0,\frac{1}{2}) \oplus 5(0,\frac{3}{2}) \oplus 7(0,\frac{5}{2}) \oplus 3(0,\frac{7}{2}) \oplus (\frac{1}{2},2) \oplus 2(\frac{1}{2},3) \oplus (\frac{1}{2},4) $ \\ \hline
		$ (2, 2, 2, 1) $ & $ (0, \frac{1}{2}) \oplus (0, \frac{3}{2}) \oplus (0, \frac{5}{2}) $ & $ (2, 2, 2, 2) $ & $ (0,\frac{1}{2}) \oplus 3(0,\frac{3}{2}) \oplus 4(0,\frac{5}{2}) \oplus 2(0,\frac{7}{2}) \oplus (\frac{1}{2},2) \oplus (\frac{1}{2},3) \oplus (\frac{1}{2},4) $ \\ \hline
	\end{tabular}
	\caption{BPS spectrum of $ SU(4)_6 $ theory for $ d_i \leq 2 $. Here, $ \mathbf{d} = (d_1, d_2, d_3, d_4) $ labels the BPS states with fugacity $ \prod Q_i^{d_i} $.}\label{table:SU4_6}
\end{table}

\paragraph{Representation $ \mathbf{r} = [0, 0, -1] $}

We now consider the Wilson loop operator in the representation $ \mathbf{r} = [0, 0, -1] $ corresponding to the fundamental representation $ \mathbf{4} $ of $ SU(4) $ and also to the rank-3 anti-symmetric representation $ \mathbf{14'} $ of $ Sp(3) $. The classical VEVs take the form
\begin{align}
\langle W_{[0, 0, -1]}^{\mathrm{cls}} \rangle_{SU(4)_6}
= \sum_{w \in \mathbf{4}} e^{-w \cdot \phi} \, , \quad
\langle W_{[0, 0, -1]}^{\mathrm{cls}}\rangle_{Sp(3)_0}
= \sum_{w \in \mathbf{14}'} e^{-w \cdot \phi'} \, .
\end{align}

We can formulate blowup equations for this loop operator by choosing ${\bf r}_1 = [0, 0, -1]$ and ${\bf r}_2=\emptyset$ and with the background magnetic fluxes $ B_m,B_{m'} = 0, 1 $. In the $ SU(4)_6 $ frame, we find the solution to the blowup equation written as
\begin{align}
Q_2^{1/4} Q_3^{1/2} Q_4^{3/4} \langle W_{[0, 0, -1]}\rangle_{SU(4)_6}
= \sum_{j_l, j_r, \mathbf{d}} (-1)^{2(j_l + j_r)} \tilde{N}_{j_l, j_r}^{\mathbf{d}} \chi_{j_l}(p_1/p_2) \chi_{j_r}(p_1 p_2) \prod_i Q_i^{d_i} \, ,
\end{align}
where the BPS degeneracies $ \tilde{N}_{j_l, j_r}^{\mathbf{d}} $ are given in Table~\ref{table:SU4_6_fund}. Similarly, we solve the blowup equations in the $Sp(3)$ frame and the result is written as
\begin{align}
Q_1^{3/2} Q_2 Q_3^2 \langle W_{[0, 0, -1]}\rangle_{Sp(3)_6}
= \sum_{j_l, j_r, \mathbf{d}} (-1)^{2(j_l + j_r)} \tilde{N}_{j_l, j_r}^{\mathbf{d}} \chi_{j_l}(p_1/p_2) \chi_{j_r}(p_1 p_2) \prod_i Q_i^{d_i} \, ,
\end{align}
with the same $ \tilde{N}_{j_l, j_r}^{\mathbf{d}} $ in Table~\ref{table:SU4_6_fund}. Two BPS spectra from two gauge theory descriptions indeed yield the same BPS spectrum $ \tilde{N}_{j_l, j_r}^{\mathbf{d}} $ as a consequence of the duality.

\begin{table}
	\centering
	\begin{tabular}{|c|C{26ex}||c|C{26ex}|} \hline
		$ \mathbf{d} $ & $ \oplus \tilde{N}_{j_l, j_r}^{\mathbf{d}} (j_l, j_r) $ & $ \mathbf{d} $ & $ \oplus \tilde{N}_{j_l, j_r}^{\mathbf{d}} (j_l, j_r) $ \\ \hline
		$ (1, 0, 0, 0) $ & $ (0, 0) $ & $ (1, 0, 0, 1) $ & $ (0, 1) $ \\ \hline
		$ (1, 0, 0, 2) $ & $ (0, 2) $ & $ (1, 0, 1, 0) $ & $ (0, 0) $ \\ \hline
		$ (1, 0, 1, 1) $ & $ (0, 0) \oplus (0, 1) $ & $ (1, 0, 1, 2) $ & $ (0, 1) \oplus (0, 2) $ \\ \hline
		$ (1, 0, 2, 0) $ & $ (0, 0) $ & $ (1, 0, 2, 1) $ & $ (0, 0) \oplus (0, 1) $ \\ \hline
		$ (1, 0, 2, 2) $ & $ (0, 0) \oplus (0, 1) \oplus (0, 2) $ & $ (1, 1, 1, 0) $ & $ (0, 0) $ \\ \hline
		$ (1, 1, 1, 1) $ & $ (0, 0) \oplus (0, 1) $ & $ (1, 1, 1, 2) $ & $ (0, 1) \oplus (0, 2) $ \\ \hline
		$ (1, 1, 2, 0) $ & $ (0, 0) $ & $ (1, 1, 2, 1) $ & $ 2(0, 0) \oplus (0, 1) $ \\ \hline
		$ (1, 1, 2, 2) $ & $ (0, 0) \oplus 2(0, 1) \oplus (0, 2) $ & $ (1, 2, 2, 0) $ & $ (0, 0) $ \\ \hline
		$ (1, 2, 2, 1) $ & $ (0, 0) \oplus (0, 1) $ & $ (1, 2, 2, 2) $ & $ (0, 0) \oplus (0, 1) \oplus (0, 2) $ \\ \hline
		$ (2, 0, 0, 1) $ & $ (0, 2) $ & $ (2, 0, 0, 2) $ & $ (0, 2) \oplus 2(0, 3) \oplus (\frac{1}{2}, \frac{7}{2}) $ \\ \hline
		$ (2, 0, 1, 1) $ & $ (0, 1) \oplus (0, 2) $ & $ (2, 0, 1, 2) $ & $ (0,1) \oplus 4(0,2) \oplus 3(0,3) \oplus (\frac{1}{2},\frac{5}{2}) \oplus (\frac{1}{2},\frac{7}{2}) $ \\ \hline
		$ (2, 0, 2, 0) $ & $ (0, 0) $ & $ (2, 0, 2, 1) $ & $ (0, 0) \oplus 2(0, 1) \oplus (0, 2) $ \\ \hline
		$ (2, 0, 2, 2) $ & $ (0,0) \oplus 4(0,1) \oplus 6(0,2) \oplus 3(0,3) \oplus (\frac{1}{2},\frac{3}{2}) \oplus (\frac{1}{2},\frac{5}{2}) \oplus (\frac{1}{2},\frac{7}{2}) $ & $ (2, 1, 1, 1) $ & $ (0, 1) \oplus (0, 2) $ \\ \hline
		$ (2, 1, 1, 2) $ & $ (0,1) \oplus 4(0,2) \oplus 3(0,3) \oplus (\frac{1}{2},\frac{5}{2}) \oplus (\frac{1}{2},\frac{7}{2}) $ & $ (2, 1, 2, 0) $ & $ (0, 0) $ \\ \hline
		$ (2, 1, 2, 1) $ & $ (0, 0) \oplus 3(0, 1) \oplus (0, 2) $ & $ (2, 1, 2, 2) $ & $ (0,0) \oplus 6(0,1) \oplus 10(0,2) \oplus 4(0,3) \oplus (\frac{1}{2},\frac{3}{2}) \oplus 2(\frac{1}{2},\frac{5}{2}) \oplus (\frac{1}{2},\frac{7}{2}) $ \\ \hline
		$ (2, 2, 2, 0) $ & $ (0, 0) $ & $ (2, 2, 2, 1) $ & $ (0, 0) \oplus 2(0, 1) \oplus (0, 2) $ \\ \hline
		$ (2, 2, 2, 2) $ & \multicolumn{3}{c|}{$ (0,0) \oplus 4(0,1)\oplus 6(0,2) \oplus 3(0,3) \oplus (\frac{1}{2},\frac{3}{2}) \oplus (\frac{1}{2},\frac{5}{2}) \oplus (\frac{1}{2},\frac{7}{2}) $} \\ \hline
	\end{tabular}
	\caption{Spectrum of the fundamental Wilson loop states in the $ SU(4)_6 $ theory for $ d_i \leq 2 $. Here, $ \mathbf{d} = (d_1, d_2, d_3, d_4) $ labels the BPS states with fugacity $ Q_2^{-1/4} Q_3^{-1/2} Q_4^{-3/4} \prod Q_i^{d_i} $.}\label{table:SU4_6_fund}
\end{table}

\paragraph{Representation $ \mathbf{r} = [0, -1, 0] $}

This Wilson loop corresponds to the Wilson loop in the rank-2 anti-symmetric representation $ \mathbf{6} $ of $ SU(4) $ and to the rank 2 anti-symmetric representation $ \mathbf{14} $ of $ Sp(3) $. Their classical VEVs are
\begin{align}
\langle W_{[0, 0, -1]}^{\mathrm{cls}}\rangle_{SU(4)_6}
= \sum_{w \in \mathbf{6}} e^{-w \cdot \phi} \, , \qquad
\langle W_{[0, 0, -1]}^{\mathrm{cls}}\rangle_{Sp(3)_0}
= \sum_{w \in \mathbf{14}} e^{-w \cdot \phi'} \, .
\end{align}

We can formulate the blowup equations for this loop operator by choosing ${\bf r}_1=[0,-1,0]$ and ${\bf r}_2=\emptyset$ and with the background fluxes $ B_m,B_{m'} = 0, 1 $. The solution to the blowup equations can be written as
\begin{align}
Q_2^{1/2} Q_3 Q_4^{1/2} \langle W_{[0, -1, 0]}\rangle_{SU(4)_6}
= \sum_{j_l, j_r, \mathbf{d}} (-1)^{2(j_l + j_r)} \tilde{N}_{j_l, j_r}^{\mathbf{d}} \chi_{j_l}(p_1/p_2) \chi_{j_r}(p_1 p_2) \prod_i Q_i^{d_i} \, ,
\end{align}
in the $SU(4)_6$ frame and written as
\begin{align}
Q_1 Q_2 Q_3^2 \langle W_{[0, -1, 0]}\rangle_{Sp(3)_0}
= \sum_{j_l, j_r, \mathbf{d}} (-1)^{2(j_l + j_r)} \tilde{N}_{j_l, j_r}^{\mathbf{d}} \chi_{j_l}(p_1/p_2) \chi_{j_r}(p_1 p_2) \prod_i Q_i^{d_i}
\end{align}
in the $Sp(3)_0$ frame. We checked that their BPS spectra yield the same BPS degeneracies $ \tilde{N}_{j_l, j_r}^{\mathbf{d}} $ given in Table~\ref{table:SU4_6_antisymm}. This confirms the duality for the Wilson loop operators of ${\bf r}=[0,-1,0]$.

\begin{table}
	\centering
	\begin{tabular}{|c|C{26ex}||c|C{26ex}|} \hline
		$ \mathbf{d} $ & $ \oplus \tilde{N}_{j_l, j_r}^{\mathbf{d}} (j_l, j_r) $ & $ \mathbf{d} $ & $ \oplus \tilde{N}_{j_l, j_r}^{\mathbf{d}} (j_l, j_r) $ \\ \hline
		$ (1, 0, 1, 0) $ & $ (0, 0) $ & $ (1, 0, 1, 1) $ & $ (0, 1) $ \\ \hline
		$ (1, 0, 1, 2) $ & $ (0, 2) $ & $ (1, 0, 2, 0) $ & $ (0, 0) $ \\ \hline
		$ (1, 0, 2, 1) $ & $ (0, 0) \oplus (0, 1) $ & $ (1, 0, 2, 2) $ & $ (0, 1) \oplus (0, 2) $ \\ \hline
		$ (1, 1, 1, 0) $ & $ (0, 0) $ & $ (1, 1, 1, 1) $ & $ (0, 1) $ \\ \hline
		$ (1, 1, 1, 2) $ & $ (0, 2) $ & $ (1, 1, 2, 0) $ & $ 2(0, 0) $ \\ \hline
		$ (1, 1, 2, 1) $ & $ 2(0, 0) \oplus 2(0, 1) $ & $ (1, 1, 2, 2) $ & $ 2(0, 1) \oplus 2(0, 2) $ \\ \hline
		$ (1, 2, 2, 0) $ & $ (0, 0) $ & $ (1, 2, 2, 1) $ & $ (0, 0) \oplus (0, 1) $ \\ \hline
		$ (1, 2, 2, 2) $ & $ (0, 1) \oplus (0, 2) $ & $ (2, 0, 1, 1) $ & $ (0, 2) $ \\ \hline
		$ (2, 0, 1, 2) $ & $ (0,2) \oplus 2(0,3) \oplus (\frac{1}{2},\frac{7}{2}) $ & $ (2, 0, 2, 1) $ & $ (0, 1) \oplus (0, 2) $ \\ \hline
		$ (2, 0, 2, 2) $ & $ (0,1) \oplus 4(0,2) \oplus 3(0,3) \oplus (\frac{1}{2},\frac{5}{2}) \oplus (\frac{1}{2},\frac{7}{2}) $ & $ (2, 1, 1, 1) $ & $ (0, 2) $ \\ \hline
		$ (2, 1, 1, 2) $ & $ (0,2) \oplus 2(0,3) \oplus (\frac{1}{2},\frac{7}{2}) $ & $ (2, 1, 2, 1) $ & $ 2(0, 1) \oplus 2(0, 2) $ \\ \hline
		$ (2, 1, 2, 2) $ & $ 2(0,1) \oplus 8(0,2) \oplus 6(0,3) \oplus 2(\frac{1}{2},\frac{5}{2}) \oplus 2(\frac{1}{2},\frac{7}{2}) $ & $ (2, 2, 2, 1) $ & $ (0, 1) \oplus (0, 2) $ \\ \hline
		$ (2, 2, 2, 2) $ & \multicolumn{3}{c|}{$ (0,1) \oplus 4(0,2) \oplus 3(0,3) \oplus (\frac{1}{2},\frac{5}{2}) \oplus (\frac{1}{2},\frac{7}{2}) $} \\ \hline
	\end{tabular}
	\caption{Spectrum of the Wilson loop states in the anti-symmetric representation in the $SU(4)_6$ theory for $ d_i \leq 2 $. Here, $ \mathbf{d} = (d_1, d_2, d_3, d_4) $ labels the BPS states with fugacity $ Q_2^{-1/2} Q_3^{-1} Q_4^{-1/2} \prod Q_i^{d_i} $.}\label{table:SU4_6_antisymm}
\end{table}

\paragraph{Representation $ \mathbf{r} = [-1, 0, 0] $}

The last example is the Wilson loop in the representation $ \mathbf{r} = [-1, 0, 0] $ corresponding to the anti-fundamental representation $ \bar{\mathbf{4}} $ in $ SU(4) $ and to the fundamental representation $ \mathbf{6} $ in $ Sp(3) $. Their classical VEVs are
\begin{align}
\langle W_{[0, 0, -1]}^{\mathrm{cls}}\rangle_{SU(4)_6}
= \sum_{w \in \bar{\mathbf{4}}} e^{-w \cdot \phi} \, , \qquad
\langle W_{[0, 0, -1]}^{\mathrm{cls}}\rangle_{Sp(3)_0}
= \sum_{w \in \mathbf{6}} e^{-w \cdot \phi'} \, .
\end{align}

The blowup equations for this Wilson loop can be formulated by choosing ${\bf r}_1=[-1,0,0], {\bf r}_2=\emptyset$ and with the background fluxes $ B_m,B_{m'} = 0, 1 $. The solution can be written as
\begin{align}
Q_2^{3/4} Q_3^{1/2} Q_4^{1/4} \langle W_{[0, -1, 0]}\rangle_{SU(4)_6}
=\sum_{j_l, j_r, \mathbf{d}}\! (-1)^{2(j_l + j_r)} \tilde{N}_{j_l, j_r}^{\mathbf{d}} \chi_{j_l}(p_1/p_2) \chi_{j_r}(p_1 p_2)\! \prod_i Q_i^{d_i}\!
\end{align}
in the $SU(4)_6$ frame and written as
\begin{align}
Q_1^{1/2} Q_2 Q_3 \ev*{W_{[0, -1, 0]}}_{Sp(3)_0}
= \sum_{j_l, j_r, \mathbf{d}} (-1)^{2(j_l + j_r)} \tilde{N}_{j_l, j_r}^{\mathbf{d}} \chi_{j_l}(p_1/p_2) \chi_{j_r}(p_1 p_2) \prod_i Q_i^{d_i}
\end{align}
in the $Sp(3)_0$ frame. As expected, they have the same $ \tilde{N}_{j_l, j_r}^{\mathbf{d}} $ given in Table~\ref{table:SU4_6_antifund}. This result confirms the duality of the loop operator of ${\bf r}_1=[-1,0,0]$ and also shows that the fundamental Wilson loop is no longer complex conjugate to the anti-fundamental Wilson loop in the $SU(4)_6$ gauge theory when the instanton corrections are taken into account.

\begin{table}
	\centering
	\begin{tabular}{|c|C{26ex}||c|C{26ex}|} \hline
		$ \mathbf{d} $ & $ \oplus \tilde{N}_{j_l, j_r}^{\mathbf{d}} (j_l, j_r) $ & $ \mathbf{d} $ & $ \oplus \tilde{N}_{j_l, j_r}^{\mathbf{d}} (j_l, j_r) $ \\ \hline
		$ (1, 1, 1, 0) $ & $ (0, 0) $ & $ (1, 1, 1, 1) $ & $ (0, 1) $ \\ \hline
		$ (1, 1, 1, 2) $ & $ (0, 2) $ & $ (1, 1, 2, 0) $ & $ (0, 0) $ \\ \hline
		$ (1, 1, 2, 1) $ & $ (0, 0) \oplus (0, 1) $ & $ (1, 1, 2, 2) $ & $ (0, 1) \oplus (0, 2) $ \\ \hline
		$ (1, 1, 3, 1) $ & $ (0, 1) $ & $ (1, 1, 3, 2) $ & $ (0, 0) \oplus (0, 1) \oplus (0, 2) $ \\ \hline
		$ (1, 2, 2, 0) $ & $ (0, 0) $ & $ (1, 2, 2, 1) $ & $ (0, 0) \oplus (0, 1) $ \\ \hline
		$ (1, 2, 2, 2) $ & $ (0, 1) \oplus (0, 2) $ & $ (1, 2, 3, 1) $ & $ (0, 0) \oplus (0, 1) $ \\ \hline
		$ (1, 2, 3, 2) $ & $ (0,0) \oplus 2(0,1) \oplus (0,2) $ & $ (1, 3, 3, 1) $ & $ (0, 1) $ \\ \hline
		$ (1, 3, 3, 2) $ & $ (0,0) \oplus (0,1) \oplus (0,2) $ & $ (2, 1, 1, 1) $ & $ (0, 2) $ \\ \hline
		$ (2, 1, 1, 2) $ & $ (0,2) \oplus 2(0,3) \oplus (\frac{1}{2},\frac{7}{2}) $ & $ (2, 1, 2, 1) $ & $ (0, 1) \oplus (0, 2) $ \\ \hline
		$ (2, 1, 2, 2) $ & $ (0,1) \oplus 4(0,2) \oplus 3(0,3) \oplus (\frac{1}{2},\frac{5}{2}) \oplus (\frac{1}{2},\frac{7}{2}) $ & $ (2, 1, 3, 1) $ & $ (0,0) \oplus (0,1) \oplus (0,2) $ \\ \hline
		$ (2, 1, 3, 2) $ & $ (0,0) \oplus 4(0,1) \oplus 5(0,2) \oplus 3(0,3) \oplus (\frac{1}{2},\frac{3}{2}) \oplus (\frac{1}{2},\frac{5}{2}) \oplus (\frac{1}{2},\frac{7}{2}) $ & $ (2, 2, 2, 1) $ & $ (0, 1) \oplus (0, 2) $ \\ \hline
		$ (2, 2, 2, 2) $ & $ (0,1) \oplus 4(0,2) \oplus 3(0,3) \oplus (\frac{1}{2},\frac{5}{2}) \oplus (\frac{1}{2},\frac{7}{2}) $ & $ (2, 2, 3, 1) $ & $ (0,0) \oplus 2(0,1) \oplus (0,2) $ \\ \hline
		$ (2, 2, 3, 2) $ & $ (0,0) \oplus 6(0,1) \oplus 9(0,2) \oplus 4(0,3) \oplus (\frac{1}{2},\frac{3}{2}) \oplus 2(\frac{1}{2},\frac{5}{2}) \oplus (\frac{1}{2},\frac{7}{2}) $ & $ (2, 3, 3, 1) $ & $ (0,0) \oplus (0,1) \oplus (0,2) $ \\ \hline
		$ (2, 3, 3, 2) $ & \multicolumn{3}{c|}{$ (0,0) \oplus 4(0,1) \oplus 5(0,2) \oplus 3(0,3) \oplus (\frac{1}{2},\frac{3}{2})\oplus (\frac{1}{2},\frac{5}{2})\oplus (\frac{1}{2},\frac{7}{2}) $} \\ \hline
	\end{tabular}
	\caption{Spectrum of the anti-fundamental Wilson loop states in the $ SU(4)_6 $ theory for $ (d_1, d_2, d_3, d_4) \leq (2, 3, 3, 2) $. Here, $ \mathbf{d} = (d_1, d_2, d_3, d_4) $ labels the BPS states with fugacity $ Q_2^{-3/4} Q_3^{-1/2} Q_4^{-1/4} \prod Q_i^{d_i} $.}\label{table:SU4_6_antifund}
\end{table}

\subsection{\texorpdfstring{$F_4$}{F4} theory}

Let us discuss the fundamental Wilson loop in the 5d pure $ F_4 $ gauge theory which has a geometric construction as \cite{DelZotto:2017pti,Esole:2017rgz}
\begin{align}\label{eq:geo-F4}
\begin{tikzpicture}
\draw (0, 0) node {$ \mathbb{F}_1 $}
(3, 0) node {$ \mathbb{F}_1 $}
(6, 0) node {$ \mathbb{F}_6 $}
(9, 0) node {$ \mathbb{F}_8 $};
\draw[thick] (0.5, 0) -- (2.5, 0)
(3.5, 0) -- (5.5, 0)
(6.5, 0) -- (8.5, 0);
\draw (0.7, 0.2) node {$ _e $}
(2.3, 0.2) node {$ _e $}
(3.7, 0.2) node {$ _{2h} $}
(5.3, 0.2) node {$ _e $}
(6.7, 0.2) node {$ _h $}
(8.3, 0.2) node {$ _e $};
\end{tikzpicture} \, .
\end{align}
For later convenience, we define the K\"ahler parameters of the base and the fiber curves in this geometry as
\begin{align}
&Q_1 = e^{-(-\phi_1 + \phi_2 + m)} \, , \qquad
Q_2 = e^{-(2\phi_1 - \phi_2)} \, , \qquad
Q_3 = e^{-(-\phi_1 + 2\phi_2 - 2\phi_3)} \, , \nonumber \\
&Q_4 = e^{-(-\phi_2 + 2\phi_3 - \phi_4)} \, , \qquad
Q_5 = e^{-(-\phi_3 + 2\phi_4)} \, .
\end{align}
The effective prepotential of the $ F_4 $ gauge theory in the phase of the geometry is
\begin{align}
\mathcal{E} &= \frac{1}{\epsilon_1 \epsilon_2} \qty(\mathcal{F} - \frac{\epsilon_1^2 + \epsilon_2^2}{12}(\phi_1 + \phi_2 + \phi_3 + \phi_4) + \epsilon_+^2 (\phi_1 + \phi_2 + \phi_3 + \phi_4))\ , \nonumber \\
6\mathcal{F} &= 8\phi_1^3 - 3\phi_1^2 \phi_2 - 3\phi_1 \phi_2^2 + 8\phi_2^3 - 18\phi_2^2 \phi_3 + 12\phi_2 \phi_3^2 + 8\phi_3^3 - 24\phi_3^2 \phi_4 + 18\phi_3 \phi_4^2 \nonumber \\
& \quad + 8\phi_4^3 + 6m (\phi_1^2 - \phi_1 \phi_2 + \phi_2^2 - 2\phi_2 \phi_3 + 2\phi_3^2 - 2\phi_3 \phi_4 + 2\phi_4^2) \, .
\end{align}

One can formulate the blowup equations with the consistent magnetic fluxes
\begin{align}
n_i \in \mathbb{Z} \, , \quad
B_m = \pm 1/2, 3/2 \ ,
\end{align}
and solve them to compute the BPS spectrum of the $F_4$ theory. We list some BPS degeneracies in Table~\ref{table:F4}. See also \cite{Keller:2012da,Kim:2019uqw} for similar computations in the $F_4$ gauge theories with/without matters.

\begin{table}[h]
	\centering
	\begin{tabular}{|c|C{22.5ex}||c|C{22.5ex}|} \hline
		$ \mathbf{d} $ & $ \oplus {N}_{j_l, j_r}^{\mathbf{d}} (j_l, j_r) $ & $ \mathbf{d} $ & $ \oplus {N}_{j_l, j_r}^{\mathbf{d}} (j_l, j_r) $ \\ \hline
		$ (1, 4, 6, 8, 4) $ & $ (0, 0) $ & $ (1, 4, 7, 8, 4) $ & $ (0, 1) $ \\ \hline
		$ (1, 4, 7, 9, 4) $ & $ (0, 0) \oplus (0, 1) $ & $ (1, 3, 7, 9, 5) $ & $ (0, 0) \oplus (0, 1) $ \\ \hline
		$ (1, 4, 7, 10, 4) $ & $ (0, 1) $ & $ (1, 4, 7, 10, 5) $ & $ (0,0) \oplus (0,1) $ \\ \hline
		$ (1, 4, 7, 10, 6) $ & $ (0, 1) $ & $ (1, 4, 8, 8, 4) $ & $ (0, 2) $ \\ \hline
		$ (1, 4, 8, 9, 4) $ & $ (0, 1) \oplus (0, 2) $ & $ (1, 4, 8, 9, 5) $ & $ (0, 1) \oplus (0, 2) $ \\ \hline
		$ (1, 4, 8, 10, 4) $ & $ (0,0) \oplus (0,1) \oplus (0,2) $ & $ (1, 4, 8, 10, 5) $ & $ (0,0) \oplus 2(0,1) \oplus (0,2) $ \\ \hline
		$ (1, 4, 8, 10, 6) $ & $ (0,0) \oplus (0,1) \oplus (0,2) $ & $ (1, 5, 6, 8, 4) $ & $ (0, 1) $ \\ \hline
		$ (1, 5, 7, 8, 4) $ & $ (0, 0) \oplus (0, 1) $ & $ (1, 5, 7, 9, 4) $ & $ (0, 0) \oplus (0, 1) $ \\ \hline
		$ (1, 5, 7, 9, 5) $ & $ (0, 0) \oplus (0, 1) $ & $ (1, 5, 7, 10, 4) $ & $ (0, 0) \oplus (0, 1) $ \\ \hline
		$ (1, 5, 7, 10, 5) $ & $ (0, 0) \oplus (0, 1) $ & $ (1, 5, 7, 10, 6) $ & $ (0, 0) \oplus (0, 1) $ \\ \hline
		$ (1, 5, 8, 8, 4) $ & $ (0, 1) \oplus (0, 2) $ & $ (1, 5, 8, 9, 4) $ & $ (0,0) \oplus 2(0,1) \oplus (0,2) $ \\ \hline
		$ (1, 5, 8, 9, 5) $ & $ (0,0) \oplus 2(0,1) \oplus (0,2) $ & $ (1, 5, 8, 10, 4) $ & $ 2(0,0) \oplus 3(0,1) \oplus (0,2) $ \\ \hline
		$ (1, 5, 8, 10, 5) $ & $ 3(0,0) \oplus 4(0,1) \oplus (0,2) $ & $ (1, 5, 8, 10, 6) $ & $ 2(0,0) \oplus 3(0,1) \oplus (0,2) $ \\ \hline
		$ (2, 8, 14, 16, 8) $ & $ (0, \frac{5}{2}) $ & $ (2, 8, 14, 17, 8) $ & $ (0, \frac{3}{2}) \oplus (0, \frac{5}{2}) $ \\ \hline
		$ (2, 8, 14, 17, 9) $ & $ (0, \frac{3}{2}) \oplus (0, \frac{5}{2}) $ & $ (2, 8, 15, 16, 8) $ & $ (0,\frac{5}{2}) \oplus (0,\frac{7}{2}) \oplus (\frac{1}{2},4) $ \\ \hline
		$ (2, 8, 15, 17, 8) $ & $ (0,\frac{3}{2}) \oplus 3(0,\frac{5}{2}) \oplus 2(0,\frac{7}{2}) \oplus (\frac{1}{2},3) \oplus (\frac{1}{2},4) $ & $ (2, 8, 15, 17, 9) $ & $ (0,\frac{3}{2}) \oplus 3(0,\frac{5}{2}) \oplus 2(0,\frac{7}{2}) \oplus (\frac{1}{2},3) \oplus (\frac{1}{2},4) $ \\ \hline
		$ (2, 9, 14, 16, 8) $ & $ (0,\frac{3}{2}) \oplus (0,\frac{5}{2}) $ & $ (2, 9, 14, 17, 8) $ & $ (0,\frac{1}{2}) \oplus 2(0,\frac{3}{2}) \oplus (0,\frac{5}{2}) $ \\ \hline
		$ (2, 9, 14, 17, 9) $ & $ (0,\frac{1}{2}) \oplus 2(0,\frac{3}{2}) \oplus (0,\frac{5}{2}) $ & $ (2, 9, 15, 16, 8) $ & $ (0,\frac{3}{2}) \oplus 3(0,\frac{5}{2}) \oplus 2(0,\frac{7}{2}) \oplus (\frac{1}{2},3) \oplus (\frac{1}{2},4) $ \\ \hline
		$ (2, 9, 15, 17, 8) $ & $ (0,\frac{1}{2}) \oplus 5(0,\frac{3}{2}) \oplus 7(0,\frac{5}{2}) \oplus 3(0,\frac{7}{2}) \oplus (\frac{1}{2},2) \oplus 2(\frac{1}{2},3) \oplus (\frac{1}{2},4) $ & $ (2, 9, 15, 17, 9) $ & $ (0,\frac{1}{2}) \oplus 5(0,\frac{3}{2}) \oplus 7(0,\frac{5}{2}) \oplus 3(0,\frac{7}{2}) \oplus (\frac{1}{2},2) \oplus 2(\frac{1}{2},3) \oplus (\frac{1}{2},4) $ \\ \hline
	\end{tabular}
	\caption{BPS spectrum of the $ F_4 $ theory for $ d_1 = 1 $, $ (d_2, d_3, d_4, d_5) \leq (5, 8, 10, 6) $ and $ d_1 = 2 $, $ (d_2, d_3, d_4, d_5) \leq (9, 15, 17, 9) $. Here, $ \mathbf{d} = (d_1, d_2, d_3, d_4, d_5) $ labels the BPS states with fugacity $\prod Q_i^{d_i} $. }\label{table:F4}
\end{table}

Now consider a Wilson loop operator in the fundamental representation, which is a minimal representation denoted by $ \mathbf{r} = [0, 0, 0, -1] $ or ${\bf r}={\bf 26}$ of $F_4$. In the geometry, this loop operator is associated to a non-compact 2-cycle intersecting the $ \mathbb{F}_8 $ surface at one point. The classical expectation value of this loop operator is given by
\begin{align}\label{eq:F4-cls}
\langle W_{[0,0,0,-1]}^{\mathrm{cls}}\rangle = \sum_{w \in \mathbf{26}} e^{-w \cdot \phi} \, .
\end{align}

The blowup equations for this Wilson loop can be formulated by choosing $ \mathbf{r}_1 = [0, 0, 0, -1] $, $ \mathbf{r}_2 = \emptyset $. One can try to solve these blowup equations with a minimal assumption that the Wilson loop partition function begins with the primitive state such as
\begin{align}
\langle W_{[0,0,0,-1]} \rangle = e^{\phi_4} + \cdots = (Q_2Q_3^2Q_4^3Q_5^2)^{-1}\left(1+ \mathcal{O}(Q_i)\right)\, ,
\end{align}
where we used $e^{\phi_4 }= (Q_2Q_3^2Q_4^3Q_5^2)^{-1}$. We insert this into the blowup equations and solve them iteratively, up to the order ${\bf d}=(1, 9, 10, 17, 13)$. We find that the classical expectation value is then fixed as
\begin{align}
&\langle W_{[0,0,0,-1]}^{\mathrm{cls}}\rangle= Q_4+Q_3Q_4+Q_2Q_3Q_4+Q_2Q_3^2Q_4^3Q_5^2+\tilde{N}^{(0,0,0,0,0)}_{0,0} \\
&\qquad +\tilde{N}^{(0,-1,-2,-2,-1)}_{0,0}\left(Q_5+Q_3Q_4^2Q_5+Q_2Q_3Q_4^2Q_5+Q_2Q_3^2Q_4^2Q_5\right) \nonumber \\
&\qquad+ \tilde{N}^{(0,-1,-2,-3,-1)}_{0,0}(Q_4Q_5+Q_3Q_4Q_5+Q_2Q_3Q_4Q_5+Q_2Q_3^2Q_4^3Q_5) + (Q_i\rightarrow Q_i^{-1}) \nonumber \, ,
\end{align}
with three undetermined degeneracies $\tilde{N}^{(0,0,0,0,0)}_{0,0}$\footnote{The degeneracy $\tilde{N}^{(0,0,0,0,0)}_{0,0}$ for the classical gauge singlet states cannot be fixed by using the blowup equations since the singlet states trivially satisfy the blowup equations. We should fix it by requiring the states form representations of $F_4$ gauge algebra.}, $\tilde{N}^{(0,-1,-2,-2,-1)}_{0,0}$, and $\tilde{N}^{(0,-1,-2,-3,-1)}_{0,0}$. All other terms in the perturbative sector (at 0-instanton) vanish. We expect that the undetermined degeneracies, but $\tilde{N}^{(0,0,0,0,0)}_{0,0}$, can also be fixed once we perform higher order calculations. One may notice that the result at 0-instanton sector precisely reproduces the classical Wilson loop VEV in \eqref{eq:F4-cls} if we set
\begin{align}\label{eq:F4-deg}
	\tilde{N}^{(0,0,0,0,0)}_{0,0} = 2 \, , \quad \tilde{N}^{(0,-1,-2,-2,-1)}_{0,0} = 1 \, , \quad \tilde{N}^{(0,-1,-2,-3,-1)}_{0,0} = 1 \, .
\end{align}
This is a non-trivial evidence that the blowup equations can correctly capture the Wilson loop spectrum in the $F_4$ gauge theory.

Assuming the classical degeneracies in \eqref{eq:F4-deg}, we solve the blowup equations to determine all other degeneracies of the fundamental Wilson loop states in the K\"ahler parameter expansion. The result is summarized in Table~\ref{table:F4_fund}.

\begin{table}[t]
	\centering
	\begin{tabular}{|c|C{22.5ex}||c|C{22.5ex}|} \hline
		$ \mathbf{d} $ & $ \oplus \tilde{N}_{j_l, j_r}^{\mathbf{d}} (j_l, j_r) $ & $ \mathbf{d} $ & $ \oplus \tilde{N}_{j_l, j_r}^{\mathbf{d}} (j_l, j_r) $ \\ \hline
		$ (1, 3, 5, 6, 3) $ & $ (0, \frac{1}{2}) $ & $ (1, 3, 5, 7, 3) $ & $ (0, \frac{1}{2}) $ \\ \hline
		$ (1, 3, 5, 7, 4) $ & $ (0, \frac{1}{2}) $ & $ (1, 3, 6, 6, 3) $ & $ (0, \frac{3}{2}) $ \\ \hline
		$ (1, 3, 6, 7, 3) $ & $ (0, \frac{1}{2}) \oplus (0, \frac{3}{2}) $ & $ (1, 3, 6, 7, 4) $ & $ (0, \frac{1}{2}) \oplus (0, \frac{3}{2}) $ \\ \hline
		$ (1, 3, 6, 8, 3) $ & $ (0, \frac{1}{2}) \oplus (0, \frac{3}{2}) $ & $ (1, 3, 6, 8, 4) $ & $ 2(0,\frac{1}{2}) \oplus (0,\frac{3}{2}) $ \\ \hline
		$ (1, 3, 6, 8, 5) $ & $ (0,\frac{1}{2}) \oplus (0,\frac{3}{2}) $ & $ (1, 3, 7, 6, 3) $ & $ (0, \frac{5}{2}) $ \\ \hline
		$ (1, 3, 7, 7, 3) $ & $ (0,\frac{3}{2}) \oplus (0,\frac{5}{2}) $ & $ (1, 3, 7, 7, 4) $ & $ (0,\frac{3}{2}) \oplus (0,\frac{5}{2}) $ \\ \hline
		$ (1, 3, 7, 8, 3) $ & $ (0,\frac{1}{2}) \oplus (0,\frac{3}{2}) \oplus (0,\frac{5}{2}) $ & $ (1, 3, 7, 8, 4) $ & $ (0,\frac{1}{2}) \oplus 2(0,\frac{3}{2}) \oplus (0,\frac{5}{2}) $ \\ \hline
		$ (1, 3, 7, 8, 5) $ & $ (0,\frac{1}{2}) \oplus (0,\frac{3}{2}) \oplus (0,\frac{5}{2}) $ & $ (1, 4, 5, 6, 3) $ & $ (0, \frac{1}{2}) $ \\ \hline
		$ (1, 4, 5, 7, 3) $ & $ (0, \frac{1}{2}) $ & $ (1, 4, 5, 7, 4) $ & $ (0, \frac{1}{2}) $ \\ \hline
		$ (1, 4, 6, 6, 3) $ & $ (0,\frac{1}{2}) \oplus (0,\frac{3}{2}) $ & $ (1, 4, 6, 7, 3) $ & $ 3(0,\frac{1}{2}) \oplus (0,\frac{3}{2}) $ \\ \hline
		$ (1, 4, 6, 7, 4) $ & $ 3(0,\frac{1}{2}) \oplus (0,\frac{3}{2}) $ & $ (1, 4, 6, 8, 3) $ & $ 3(0,\frac{1}{2}) \oplus (0,\frac{3}{2}) $ \\ \hline
		$ (1, 4, 6, 8, 4) $ & $ 5(0,\frac{1}{2}) \oplus (0,\frac{3}{2}) $ & $ (1, 4, 6, 8, 5) $ & $ 3(0,\frac{1}{2}) \oplus (0,\frac{3}{2}) $ \\ \hline
		$ (1, 4, 7, 6, 3) $ & $ (0,\frac{3}{2}) \oplus (0,\frac{5}{2}) $ & $ (1, 4, 7, 7, 3) $ & $ (0,\frac{1}{2}) \oplus 3(0,\frac{3}{2}) \oplus (0,\frac{5}{2}) $ \\ \hline
		$ (1, 4, 7, 7, 4) $ & $ (0,\frac{1}{2}) \oplus 3(0,\frac{3}{2}) \oplus (0,\frac{5}{2}) $ & $ (1, 4, 7, 8, 3) $ & $ \! 4(0,\frac{1}{2}) \oplus 4(0,\frac{3}{2}) \oplus (0,\frac{5}{2}) \! $ \\ \hline
		$ (1, 4, 7, 8, 4) $ & $ 6(0,\frac{1}{2}) \oplus 6(0,\frac{3}{2}) \oplus (0,\frac{5}{2}) $ & $ (1, 4, 7, 8, 5) $ & $ \! 4(0,\frac{1}{2}) \oplus 4(0,\frac{3}{2}) \oplus (0,\frac{5}{2}) \! $ \\ \hline
		$ (2, 7, 12, 14, 7) $ & $ (0, 2) $ & $ (2, 7, 12, 15, 7) $ & $ (0, 1) \oplus (0, 2) $ \\ \hline
		$ (2, 7, 12, 15, 8) $ & $ (0, 1) \oplus (0, 2) $ & $ (2, 7, 13, 14, 7) $ & $ (0,2) \oplus 2(0,3) \oplus (\frac{1}{2},\frac{7}{2}) $ \\ \hline
		$ (2, 7, 13, 15, 7) $ & $ (0,1) \oplus 4(0,2) \oplus 3(0,3) \oplus (\frac{1}{2},\frac{5}{2}) \oplus (\frac{1}{2},\frac{7}{2}) $ & $ (2, 7, 13, 15, 8) $ & $ (0,1) \oplus 4(0,2) \oplus 3(0,3) \oplus (\frac{1}{2},\frac{5}{2}) \oplus (\frac{1}{2},\frac{7}{2}) $ \\ \hline
		$ (2, 8, 12, 14, 7) $ & $ (0,1)\oplus (0,2) $ & $ (2, 8, 12, 15, 7) $ & $ (0,0) \oplus 2(0,1) \oplus (0,2) $ \\ \hline
		$ (2, 8, 12, 15, 8) $ & $ (0,0) \oplus 2(0,1) \oplus (0,2) $ & $ (2, 8, 13, 14, 7) $ & $ (0,1) \oplus 4(0,2) \oplus 3(0,3) \oplus (\frac{1}{2},\frac{5}{2}) \oplus (\frac{1}{2},\frac{7}{2}) $ \\ \hline
		$ (2, 8, 13, 15, 7) $ & $ (0,0) \oplus 6(0,1) \oplus 10(0,2) \oplus 4(0,3) \oplus (\frac{1}{2},\frac{3}{2}) \oplus 2(\frac{1}{2},\frac{5}{2}) \oplus (\frac{1}{2},\frac{7}{2}) $ & $ (2, 8, 13, 15, 8) $ & $ (0,0) \oplus 6(0,1) \oplus 10(0,2) \oplus 4(0,3) \oplus (\frac{1}{2},\frac{3}{2}) \oplus 2(\frac{1}{2},\frac{5}{2}) \oplus (\frac{1}{2},\frac{7}{2}) $ \\ \hline
	\end{tabular}
	\caption{Spectrum of the fundamental Wilson loop in the $ F_4 $ gauge theory for $ d_1 = 1 $, $ (d_2, d_3, d_4, d_5) \leq (4, 7, 8, 5) $ and $ d_1 = 2 $, $ (d_2, d_3, d_4, d_5) \leq (8, 13, 15, 8) $. Here, $ \mathbf{d} = (d_1, d_2, d_3, d_4, d_5) $ labels the BPS states with fugacity $ \prod Q_i^{d_i} $.}\label{table:F4_fund}
\end{table}


\subsection{6d \texorpdfstring{$\mathcal{N}=(2,0)$}{N=(2,0)} \texorpdfstring{$A_1$}{A1} theory}

We now move on to Wilson loop (or Wilson surface) operators in the 5d KK theories coming from circle compactifications of 6d SCFTs. The first example for this is the fundamental Wilson loop in the 5d $ SU(2)$ gauge theory at $\theta=0$ with one adjoint hypermultiplet that arises from a circle compactification of the 6d $\mathcal{N}=(2,0)$ $A_1$ theory.

The BPS spectrum of this theory without loop operators has been computed based on the blowup method in \cite{Gu:2019pqj,Kim:2020hhh}. To solve the blowup equation, one uses the effective prepotential
\begin{align}\label{eq:su2_adj_E}
\mathcal{E} = \frac{1}{\epsilon_1 \epsilon_2} \qty(m_0 \phi^2 - m_1^2 \phi + \epsilon_+^2 \phi) \, ,
\end{align}
and the magnetic fluxes
\begin{align}\label{eq:su2_adj_flux}
n \in \mathbb{Z} \, , \quad
B_{m_0} = 0 \, , \quad
B_{m_1} = 1/2 \, .
\end{align}
where $ m_0 $ is the inverse gauge coupling squared and $ m_1 $ is the adjoint mass parameter. 
The solution to the blowup equation matches the result in \cite{Hwang:2014uwa} from the ADHM calculation. 

We now introduce a fundamental Wilson loop of $ \mathbf{r} = [-1] $ into this 5d KK theory.
The classical VEV of the Wilson loop in the 5d gauge theory is given by
\begin{align}
\ev*{W_{[-1]}^{\mathrm{cls}}}_{\mathrm{5d}} = e^{\phi} + e^{-\phi} \, .
\end{align}
In 6d, this loop operator corresponds to the Wilson surface operator carrying a unit 2-form tensor charge studied in \cite{Chen:2007ir,Agarwal:2018tso}. 

The blowup equation for this Wilson loop operator can be constructed by choosing ${\bf r}_1=[-1]$ and ${\bf r}_2=\emptyset$ and with the magnetic fluxes in \eqref{eq:su2_adj_flux}. Note here that since this theory is a 5d KK theory, the spectrum may involve a KK tower state for the primitive curve of the loop operator, i.e. the KK states of fugacity $e^{\phi- nm_0}$ with $n\in \mathbb{Z}_+$. However, these states will decouple from the field theory. So we will assume that such a KK tower of the primitive loop state is absent in the genuine field theory spectrum. Under this assumption, we can solve the blowup equation and uniquely determine the BPS spectrum of the fundamental Wilson loop operator. We summarize the BPS spectrum in Table~\ref{table:A1}. We checked in the K\"ahler parameter expansion that our solution agrees with the result of the ADHM calculation in \cite{Agarwal:2018tso}.

\begin{table}
	\centering
	\begin{tabular}{|c|C{27ex}||c|C{27ex}|} \hline
		$ \mathbf{d} $ & $ \oplus \tilde{N}_{j_l, j_r}^{\mathbf{d}} (j_l, j_r) $ & $ \mathbf{d} $ & $ \oplus \tilde{N}_{j_l, j_r}^{\mathbf{d}} (j_l, j_r) $ \\ \hline
		$ (1, 1, -2) $ & $ (0, 0) $ & $ (1, 1, -1) $ & $ (0, \frac{1}{2}) \oplus (\frac{1}{2}, 0) $ \\ \hline
		$ (1, 1, 0) $ & $ 2(0, 0) \oplus (\frac{1}{2}, \frac{1}{2}) $ & $ (1, 3, -2) $ & $ (0, 1) $ \\ \hline
		$ (1, 3, -1) $ & $ (0, \frac{1}{2}) \oplus (0, \frac{3}{2}) \oplus (\frac{1}{2}, 1) $ & $ (1, 3, 0) $ & $ 2(0, 1) \oplus (\frac{1}{2}, \frac{1}{2}) \oplus (\frac{1}{2}, \frac{3}{2}) $ \\ \hline
		$ (2, 1, -2) $ & $ 2(0,0) \oplus (\frac{1}{2},\frac{1}{2}) $ & $ (2, 1, -1) $ & $ 3(0,\frac{1}{2}) \oplus 3(\frac{1}{2},0) \oplus (\frac{1}{2},1) \oplus (1,\frac{1}{2}) $ \\ \hline
		$ (2, 1, 0) $ & $ 5(0,0) \oplus (0,1) \oplus 4(\frac{1}{2},\frac{1}{2}) \oplus (1,0) \oplus (1,1) $ & $ (2, 3, -3) $ & $ (0,\frac{1}{2}) \oplus (0,\frac{3}{2}) \oplus (\frac{1}{2},1) $ \\ \hline
		$ (2, 3, -2) $ & $ 2(0,0) \oplus 5(0,1) \oplus (0,2) \oplus 3(\frac{1}{2},\frac{1}{2}) \oplus 4(\frac{1}{2},\frac{3}{2}) \oplus (1,1) \oplus (1,2) $ & $ (2, 3, -1) $ & $ 8(0,\frac{1}{2}) \oplus 8(0,\frac{3}{2}) \oplus 3(\frac{1}{2},0) \oplus 10(\frac{1}{2},1) \oplus 4(\frac{1}{2},2) \oplus 2(1,\frac{1}{2}) \oplus 4(1,\frac{3}{2}) \oplus (1,\frac{5}{2}) \oplus (\frac{3}{2},2) $ \\ \hline
		$ (2, 3, 0) $ & \multicolumn{3}{C{68ex}|}{$ 5(0,0) \oplus 13(0,1) \oplus 4(0,2) \oplus 9(\frac{1}{2},\frac{1}{2}) \oplus 11(\frac{1}{2},\frac{3}{2}) \oplus (\frac{1}{2},\frac{5}{2}) \oplus (1,0) \oplus 5(1,1) \oplus 4(1,2) \oplus (\frac{3}{2},\frac{3}{2}) \oplus (\frac{3}{2},\frac{5}{2}) $} \\ \hline
	\end{tabular}
	\caption{Spectrum of the Wilson surface of $ \mathbf{r} = [-1] $ representation in the 6d $ \mathcal{N} = (2, 0) $ $ A_1 $ theory for $ d_1 \leq 2 $ and $ d_2 \leq 3 $. Here, $ \mathbf{d} = (d_1, d_2, d_3) $ labels the BPS states with charge $ d_1 m_0 + d_2 \phi + d_3 m_1 $. The states related by the symmetry $ d_3 \leftrightarrow -d_3 $ are omitted.}\label{table:A1}
\end{table}

It is also possible to solve the blowup equation from the perspective of 6d theory. In this case, we first expand the partition function with/without the Wilson loop operator in the self-dual string number and solve it order by order in the expansion. The fundamental Wilson loop expectation value $\langle W_{[-1]}\rangle$ is expanded as
\begin{align}
	\langle W_{[-1]}\rangle \equiv Z_{W_{[-1]}}/Z = e^{\phi} \sum_{k=0}^\infty e^{-2k\phi} \, W_{[-1]}^{(k)} \, ,
\end{align}
where $Z$ is the partition function without the Wilson loop operator, $\phi$ is the 6d tensor parameter and $k$ is the string number with $W^{(0)}_{[-1]}=1$. We can obtain a closed expression of the Wilson loop expectation value $W_{[-1]}^{(k)}$ at each string number by solving two blowup equations with magnetic fluxes
\begin{align}
\left\{\begin{array}{ll}    n \in \mathbb{Z}\,, &  B_{\tau} = 0 \, , \ B_{m_1} = 1/2 \\  n\in \mathbb{Z}+1/2 \,,& B_{\tau} = 0 \,, \ B_{m_1} = 1/2\quad  \end{array} \right.
\end{align}
and the effective prepotential \eqref{eq:su2_adj_E}, where $n$ represents the magnetic fluxes for the tensor symmetry, $m_0 = \tau\equiv 1/R$ is the inverse radius of the 6d circle and $m_1$ is the mass parameter for the R-symmetry $SU(2)\subset SO(5)_R$ commuting with the supercharge $Q$ used to define the Witten index.

At 0-string sector, the blowup equations can be written as
\begin{align}
\Lambda_0 e^{\phi} &= e^{\phi}\sum_{n}(-1)^n e^{-V(n)} p_1^{-n} \, , 
\end{align}
with $V(n)=V(m_1,n,B_{m_1};\epsilon_1,\epsilon_2)$ defined in \eqref{eq:GV-V}. After summing over $n$, we can compute
\begin{align}
\Lambda_0 = \left\{
\begin{array}{ll}
\theta_4(2\tau, m_1+\epsilon_++\epsilon_1) &\ \text{for} \ n \in \mathbb{Z} \\
\theta_1(2\tau, m_1 + \epsilon_+ + \epsilon_1) &\  \text{for} \ n \in \mathbb{Z}+1/2
\end{array}\right. \, .
\end{align}
See Appendix~\ref{appendix:elliptic} for the definition of the elliptic functions $\theta_1$ and $\theta_4$. Next, the blowup equations at 1-string order are given by
\begin{align}
\Lambda_0 e^{-\phi} \hat{Z}_{W_{[-1]}}^{(1)} = \sum_n (-1)^n e^{-V(n)} \qty(e^{-\phi} p_1^n \hat{Z}_{W_{[-1]}}^{(N), (1)} + e^{-\phi} p_1^{-n} p_2^{2n} \hat{Z}^{(S),(1)}) \, ,
\end{align}
for both $n\in\mathbb{Z}$ and $n\in \mathbb{Z}+1/2$. Using these two blowup equations, we can compute $\hat{Z}_{W_{[-1]}}^{(1)}$ and $\hat{Z}_{W_{[-1]}}^{(N),(1)}$ in terms of $Z_{1}$ for which we can use the result from the ADHM computation \cite{Haghighat:2013gba}
\begin{align}
Z^{(1)} = \frac{\theta_1(m_1 \pm \epsilon_+)}{\theta_1(\epsilon_{1,2})} \, ,
\end{align}
where we used the following short hand notation: $\theta(a\pm b)\equiv\theta(a+b)\,\theta(a-b)$ and $\theta(\epsilon_{1,2}) \equiv\theta(\epsilon_1)\theta(\epsilon_2).$ By plugging this into the solution, we obtain the fundamental Wilson loop VEV at 1-string order as
\begin{align}\label{eq:A1_W1_blowup}
\langle W_{[-1]}^{(1)} \rangle \equiv Z_{W_{[-1]}}^{(1)} - Z^{(1)} = \frac{\theta_1(m_1 \pm \epsilon_-) - \theta_1(m_1 \pm \epsilon_+)}{\theta_1(\epsilon_{1,2})} \, .
\end{align}
This result when expanded in terms of $e^{-\tau}$ and $e^{-m_1}$ is of course in agreement with the spectrum in Table \ref{table:A1}.

We can compare our result from the blowup equations against the result of the 6d ADHM calculation in \cite{Agarwal:2018tso}. The 5d fundamental Wilson loop operator is mapped to a codimension-4 defect introduced by coupling a 2d fundamental fermion to the 6d theory. The expectation value of the codimension-4 defect at 1-string order is given by
\begin{align}\label{eq:A1_W1}
\ev*{W^{(1)}_{[-1]}} = \frac{\theta_1(z + \epsilon_{1,2}) \theta_1(m_1 \pm \epsilon_+)}{\theta_1(\epsilon_{1,2}) \theta_1(z) \theta_1(z + 2\epsilon_+)} + \frac{\theta_1(z + \epsilon_+ \pm m_1)}{\theta_1(z) \theta_1(z + 2\epsilon_+)} - \frac{\theta_1(m_1 \pm \epsilon_+)}{\theta_1(\epsilon_{1,2})} \, ,
\end{align}
where $z$ is the 2d fermion mass parameter. This looks quite different from the above result \eqref{eq:A1_W1_blowup} from the blowup equations. However, we checked that two results \eqref{eq:A1_W1_blowup} and \eqref{eq:A1_W1} perfectly agree with each other up to order $e^{-10\tau}$, which may imply a non-trivial identity between the elliptic functions. It is worthy of noting that although the RHS of \eqref{eq:A1_W1} looks as if it is a function of $z$, the codimension-4 partition function \eqref{eq:A1_W1} is independent of the chemical potential $z$, as explicitly checked in the expansion of $e^{-\tau}$. Our result \eqref{eq:A1_W1_blowup}, on the other hand, provides an alternative expression that is manifestly independent of $z$. We also checked that the Wilson loop VEV from the blowup equations at 2-string order matches the ADHM result given in \cite{Agarwal:2018tso}.

\subsection{\texorpdfstring{E-string theory}{E-string theory}}

The E-string theory on a circle reduces to the 5d $ SU(2) $ gauge theory with 8 fundamental hypermultiplets. The effective prepotential is given by
\begin{align}
\mathcal{E} = \frac{1}{\epsilon_1 \epsilon_2} \qty(m_0 \phi^2 - \sum_{i=1}^8 m_i^2 \phi + \frac{\epsilon_1^2 + \epsilon_2^2}{4}\phi + \epsilon_+^2 \phi ) \, ,
\end{align}
where $ m_0$ is the inverse of the 5d gauge coupling and $ m_{i=1, \cdots, 8} $ are the flavor mass parameters. The partition function without a defect can be calculated using the blowup equations with consistent magnetic fluxes \cite{Gu:2019pqj,Kim:2020hhh}
\begin{align}\label{eq:e-string-flux}
n \in \mathbb{Z} \, , \quad
B_{m_0} = 0 \, , \quad
B_{m_i} = 1/2 \ (1 \leq i \leq 8) \, .
\end{align}

We now introduce a fundamental Wilson loop in the 5d theory and compute its expectation value. The classical expectation value of the loop operator is given by
\begin{align}
\langle W_{[-1]}^{\mathrm{cls}} \rangle = e^{\phi} + e^{-\phi} \, .
\end{align}
One can formulate the blowup equation for this loop operator by choosing $ \mathbf{r}_1 = [-1] $, $ \mathbf{r}_2 = \emptyset $ and with the same consistent magnetic fluxes in \eqref{eq:e-string-flux}. The solution of the blowup equation is summarized in Table~\ref{table:E-string}. We checked that the solution agrees with the result in \cite{Chen:2021ivd} based on the ADHM construction up to 2-instanton order.

\begin{table}
	\centering
	\begin{tabular}{|c|C{30ex}||c|C{30ex}|} \hline
		$ \mathbf{d} $ & $ \oplus \tilde{N}_{j_l, j_r}^{\mathbf{d}} (j_l, j_r) $ & $ \mathbf{d} $ & $ \oplus \tilde{N}_{j_l, j_r}^{\mathbf{d}} (j_l, j_r) $ \\ \hline
		$ (1, 1) $ & $ 128(0, 0) $ & $ (1, 2) $ & $ 128(0, \frac{1}{2}) $ \\ \hline
		$ (1, 3) $ & $ 128(0, 1) $ & $ (1, 4) $ & $ 128(0, \frac{3}{2}) $ \\ \hline
		$ (2, 1) $ & $ 2063(0, 0) \oplus 122(\frac{1}{2}, \frac{1}{2}) \oplus (1, 1) $ & $ (2, 2) $ & $ 5536(0, \frac{1}{2}) \oplus 16(\frac{1}{2}, 0) \oplus 592(\frac{1}{2}, 1) \oplus 16(1, \frac{3}{2}) $ \\ \hline
		$ (2, 3) $ & $ 1941(0,0) \oplus 12012(0,1) \oplus 122(\frac{1}{2},\frac{1}{2}) \oplus 2063(\frac{1}{2},\frac{3}{2}) \oplus (1,1) \oplus 122(1,2) \oplus (\frac{3}{2},\frac{5}{2}) $ & $ (2, 4) $ & $ 5520(0,\frac{1}{2}) \oplus 21920(0,\frac{3}{2}) \oplus 16(\frac{1}{2},0) \oplus 592(\frac{1}{2},1) \oplus 5536(\frac{1}{2},2) \oplus 16(1,\frac{3}{2}) \oplus 592(1,\frac{5}{2}) \oplus 16(\frac{3}{2},3) $ \\ \hline
	\end{tabular}
	\caption{Spectrum of the fundamental Wilson loop operator in the E-string theory for $ (d_1, d_2) \leq (2, 4) $. Here, $ \mathbf{d} = (d_1, d_2) $ labels the BPS states with charge $ d_1 m_0 + d_2 \phi $. For convenience, we turn off all flavor mass parameters $ m_{i=1, \cdots, 8} $.}\label{table:E-string}
\end{table}

\subsection{\texorpdfstring{6d $ SU(2) $ theory on a $ -2 $ curve}{6d SU(2) theory on a -2 curve}}

We next study Wilson loop operators in the 6d $ SU(2) $ gauge theory with 4 fundamental hypermultiplets which, on a circle, reduces to the 5d $ SU(2) \times SU(2) $ quiver gauge theory with the discrete theta angles $\theta_1=\theta_2=0$ coupled to two bi-fundamental hypermultiplets. 

The effective prepotential of the theory in the 6d frame is
\begin{align}\label{eq:6d_su2_E}
\mathcal{E} &= \frac{1}{\epsilon_1 \epsilon_2} \qty( \mathcal{E}_{\mathrm{tree}} + \mathcal{F}_{\text{1-loop}} + \frac{\epsilon_1^2 + \epsilon_2^2}{12} \phi_1 + \epsilon_+^2 \phi_1 ) \ , \nonumber \\
\mathcal{E}_{\mathrm{tree}} &= \tau \phi_0^2 + 2\phi_0 \qty(\phi_1^2 - \frac{m_1^2}{2} - \frac{m_2^2}{2} - b^2 + \epsilon_+^2) \ , \nonumber \\
\mathcal{F}_{\text{1-loop}}
&= \frac{1}{12} \sum_{n \in \mathbb{Z}} \qty(\abs{n\tau \pm 2\phi_1}^3 - \sum_{i=1}^2 \abs{n\tau \pm \phi_1 + (m_i \pm b)}^3) \nonumber \\
&= \frac{2}{3}\phi_1^3 - (m_1^2 + m_2^2 + 2b^2) \phi_1 + \frac{\tau}{6} (m_1^2 + m_2^2 + 2b^2) \, ,
\end{align}
where $\phi_0$ is the tensor scalar VEV and $\phi_1$ is the $SU(2)$ gauge holonomy, and $\tau$ is complex structure of a torus, $m_{1,2} $ and $ b $ are the chemical potentials for the flavor symmetry. The consistent magnetic fluxes are given by
\begin{align}\label{eq:6d_su2_B}
n_0 \in \mathbb{Z} \ \ \text{or}\ \ \mathbb{Z} + 1/2 \, , \quad
n_1 \in \mathbb{Z} \, , \quad
B_\tau = 0 \, , \quad
B_{m_{1,2}} = 1/2 \, , \quad
B_{b} = 0 \, .
\end{align}
The elliptic genera of self-dual strings in this theory can be computed by solving the blowup equations with these fluxes. For this, we first write the index part of the partition function as
\begin{align}
Z & = Z_{\mathrm{pert}}^{\mathrm{vec}} \times Z_{\mathrm{pert}}^{\mathrm{hyper}} \times Z_{\mathrm{str}} \, , \nonumber \\
Z_{\mathrm{pert}}^{\mathrm{vec}}
&= \PE\qty[ -\frac{1+p_1 p_2}{(1-p_1)(1-p_2)} (e^{-2\phi_1} + q e^{2\phi_1}) \frac{1}{1-q} ] \, , \nonumber \\
Z_{\mathrm{pert}}^{\mathrm{hyper}}
&= \PE\qty[ \frac{\sqrt{p_1 p_2}}{(1-p_1)(1-p_2)} (e^{-\phi_1} + q e^{\phi_1}) \sum_{i=1}^2 (e^{\pm (m_i+b)} + e^{\pm (m_i-b)} ) \frac{1}{1-q} ] \, , \nonumber \\
Z_{\mathrm{str}} & = 1+\sum_{k=1}^\infty e^{-2 k\phi_0} Z_k \, ,
\end{align}
with $ q = e^{-\tau}$. Here, $Z_{\rm pert}^{\rm vec}$ and $Z_{\rm pert}^{\rm hyper}$ are the perturbative contributions from the vector and the hypermultiplets, respectively, and $Z_k$ is the $k$-string elliptic genus. The tensor multiplet contribution is omitted as it is independent of the dynamical K\"ahler parameters. One can insert this into the blowup equation and solve the equation to determine the elliptic genera of the self-dual strings $Z_k$.

The solution of the blowup equations in 5d perspective is given in \cite{Kim:2020hhh} which matches the 6d elliptic genus calculation in \cite{Haghighat:2013gba,Haghighat:2013tka}. For the comparison between the 5d and the 6d results, one should use the parameter maps
\begin{align}\label{eq:su2-su2-params}
&\phi_0^{6d} = \phi_1^{5d} \, , \quad
\phi_1^{6d} = -\phi_1^{5d} + \phi_2^{5d} + \frac{1}{2}m_2^{5d} \, , \quad
\tau = m_1^{5d} + m_2^{5d} \, , \nonumber \\
& m_1^{6d} = m_3^{5d} \, , \quad
m_2^{6d} = m_4^{5d} \, , \quad
b = \frac{1}{2}m_2^{5d} \, ,
\end{align}
where $ (\phi_1^{5d}, \phi_2^{5d}) $ are the 5d Coulomb branch parameters, $ (m_1^{5d}, m_2^{5d}) $ are the inverse couplings of the $ SU(2) \times SU(2) $ gauge symmetry, and $ (m_3^{5d}, m_4^{5d}) $ are two bi-fundamental masses.

\paragraph{Codimension-4 defect: 2d fermion} We can consider a codimension-4 defect introduced by coupling a 2d free fermion to the 6d theory. We first compute the expectation value of this 6d/2d coupled system and then relate it to the Wilson loop operator. The codimension-4 defect in the 6d $SU(2)$ theory with 4 fundamentals has been recently studied in \cite{Chen:2020jla}. The defect is constructed by coupling a 2d fundamental fermion to the 6d bulk $SU(2)$ gauge field. In presence of the codimension-4 defect, we write the partition function as
\begin{align}\label{eq:codim4-su2-su2}
Z^{6d/2d} &= Z^{\rm vec}_{\rm pert} \times Z^{\rm hyper}_{\rm pert} \times e^{\phi_0} \sum_{k=0}^\infty e^{-2k \phi_0} Z_k^{6d/2d} \, , \nonumber \\
Z^{2d}_{\mathrm{pert}} &= e^{\phi_0} Z^{6d/2d}_{k=0} = e^{\phi_0} \frac{\theta_1(z \pm \phi_1)}{(i\eta)^2} \, .
\end{align}
Here $Z^{2d}_{\mathrm{pert}}$ is the perturbative contribution from the 2d fundamental fermion with mass $z$. The factor $ e^{\phi_0} $ is multiplied because we find otherwise the solution to the blowup equations does not fit into representations of the $SO(4)$ transverse rotation group which is necessary so that the solution becomes a correct spectrum for the defect.

One can formulate the blowup equations for this codimension-4 defect with two sets of magnetic fluxes in \eqref{eq:6d_su2_B} together with $ B_z = 0 $. Inserting the partition function of the form \eqref{eq:codim4-su2-su2} into the blowup equation, we can solve the equation to calculate $Z_k^{6d/2d}$ order by order in the string number $k$-expansion. At $k=0$ order, we can fix the constant prefactor $\Lambda_0$ in the blowup equation as\footnote{Note here that the factor $(-1)^{|\vec{n}^{5d}|} = (-1)^{n^{5d}_1+n^{5d}_2}$ in the blowup equations in the 5d Dynkin basis, which agrees with the geometric basis, becomes $(-1)^{n_1}$ in the 6d field theory basis.}
\begin{align}
\Lambda_0
= \sum_{\vec{n}=(n_0, 0)} e^{-V(\vec{n})} p_1^{-n_0} = \left\{
\begin{array}{ll}
q^{1/12} \theta_3(2\tau, m_1 + m_2 + 2\epsilon_1 + \epsilon_2) & \ \text{for} \ n_0 \in \mathbb{Z} \\
q^{1/12} \theta_2(2\tau, m_1 + m_2 + 2\epsilon_1 + \epsilon_2) & \ \text{for} \ n_0 \in \mathbb{Z} + 1/2 \, .
\end{array} \right.
\end{align}
At 1-string order, the blowup equations are given by
\begin{align}
\Lambda e^{-\phi_0} \hat{Z}_{1}^{6d/2d}
&= \sum_{\vec{n}=(n_0, 0)} e^{-V(\vec{n})} e^{-\phi_0} \qty( p_1^{n_0} \hat{Z}_{1}^{6d/2d, (N)} +  p_1^{-n_0} p_2^{2n_0} \hat{Z}_{0}^{6d/2d, (N)} \hat{Z}_1^{(S)} ) \nonumber \\
& \quad - \sum_{\vec{n} = (n_0, \pm 1)} e^{-V(\vec{n})} e^{\phi_0} p_1^{-n_0} \frac{\hat{Z}_{\mathrm{pert}}^{(N)} \hat{Z}_{\mathrm{pert}}^{(S)}}{\hat{Z}_{\mathrm{pert}}} \hat{Z}_{0}^{6d/2d, (N)} \, .
\end{align}
The term involving $ \hat{Z}_{\mathrm{pert}} $ can be simplified using \eqref{eq:PE_vec_theta} and \eqref{eq:PE_mat_theta}. We then compute
\begin{align}
Z_{1}^{6d/2d} &= \frac{\theta_1(m_1+m_2+\epsilon_1) \theta_1(2\epsilon_-) Z_{0}^{6d/2d}(\phi_1, \tau, z) Z_1(\phi_1, m_i + \epsilon_2/2, \tau; \epsilon_1 - \epsilon_2, \epsilon_2)}{\theta_1(m_1 + m_2 + 2\epsilon_+) \theta_1(\epsilon_1)} \nonumber \\
& \quad - \Bigg( \frac{\theta_1(2\phi_1 -m_1 - m_2) \theta_1(\phi_1 \pm b + m_{1,2} + \epsilon_+) Z_{0}^{6d/2d}(\phi_1 + \epsilon_1, \tau, z)}{\theta_1(m_1 + m_2 + 2\epsilon_+) \theta_1(\epsilon_1) \theta_1(2\phi_1) \theta_1(2\phi_1 + \epsilon_{1}) \theta_1(2\phi_1 + 2\epsilon_+)} \nonumber \\
& \qquad \qquad + (m_{1,2} \to -m_{1,2} \, , \ \epsilon_{1,2} \to -\epsilon_{1,2}) \Bigg)\, ,
\end{align}
where we used the identities \eqref{eq:theta_shift} and \eqref{eq:theta_identity} to simplify the solution. We compared this solution against the result from the ADHM calculation in \cite{Chen:2020jla} in the $ e^{-\tau} $ expansion and verified that two results, although they look completely different, perfectly match up to $ e^{-5\tau} $ order. We also checked that the solution at 2-string also matches the 2-string elliptic genus from the ADHM calculation up to $ e^{-5\tau} $ order.

\paragraph{Fundamental Wilson loops} Now we will relate the partition function of the 6d/2d coupled system to the Wilson loop expectation values in the 5d $SU(2)\times SU(2)$ gauge theory. We first note that using the identity \eqref{eq:theta_identity}, we can recast $Z_{\rm pert}^{2d}$ as
\begin{align}
Z^{2d}_{\rm pert} = \frac{\theta_2(2\tau, 2z)}{\eta(\tau)^2} \mathcal{W}_1 - q^{1/4} e^{b} \frac{\theta_3(2\tau, 2z)}{\eta(\tau)^2} \mathcal{W}_2 \ ,
\end{align}
where
\begin{align}
\mathcal{W}_1 = e^{\phi_0} \theta_3(2\tau, 2\phi_1) \, , \qquad
\mathcal{W}_2 = q^{-1/4} e^{-b} e^{\phi_0} \theta_2(2\tau, 2\phi_1) \, .
\end{align}
Notice that the terms depending on the dynamical K\"ahler parameters $\phi_0$ and $\phi_1$ are separated from the $z$-dependent terms. Hence when we insert this expression into the blowup equations, the terms other than $\mathcal{W}_1$ and $\mathcal{W}_2$ can be factored out, since the magnetic fluxes for $ \tau $, $ b $ and $ z $ are switched off and the blowup equations are linear equations in the 6d/2d partition function. This implies that the blowup equations for the 6d/2d partition function are in fact satisfied by the expectation value $\langle\mathcal{W}_1\rangle$ and independently by $\langle\mathcal{W}_2\rangle$ defined as 
\begin{align}
\langle \mathcal{W}_{1,2}\rangle = \mathcal{W}_{1,2}+ \sum_{k=1}^\infty e^{-2k\phi_0}\mathcal{W}^{(k)}_{1,2} \ ,
\end{align}
where $\mathcal{W}_{1,2}^{(k)}$ stand for the $k$-th string corrections to the expectation values.

It turns out that $\langle\mathcal{W}_1\rangle$ and $\langle\mathcal{W}_2\rangle$ are the 6d dual expressions of the fundamental Wilson loop expectation values in the 5d $SU(2)\times SU(2)$ gauge theory. In terms of the 5d parameters given in \eqref{eq:su2-su2-params}, one reads that these functions are VEVs of the fundamental Wilson loops for the first and the second $SU(2)$ gauge groups, respectively, expanded as
\begin{align}
\langle\mathcal{W}_1\rangle &= \langle W_{[-1, 0]}\rangle = e^{\phi_1^{5d}} +e^{-\phi_1^{5d}} + \cdots \, , \nonumber \\
\langle\mathcal{W}_2\rangle &= \langle W_{[0, -1]}\rangle = e^{\phi_2^{5d}} +e^{-\phi_2^{5d}}+ \cdots \, ,
\end{align}
where $\cdots$ involve the 5d instanton corrections. Also, as expected, there is a symmetry exchanging two Wilson loop VEVs when we swap the 5d parameters as $ \phi_1^{5d} \leftrightarrow \phi_2^{5d} $ and $ m_1^{5d} \leftrightarrow m_2^{5d} $.

The instanton corrections to the Wilson loop VEVs in the 5d $SU(2)\times SU(2)$ gauge theory can also be calculated using their ADHM constructions by generalizing the study in \cite{Assel:2012nf}. For example, we compute the $(1,0)$ and $(0,1)$-instanton corrections as
\begin{align}
\langle W_{[-1, 0]}^{(1, 0)} \rangle
&= \frac{4 \sqrt{p_1 p_2}}{(1-p_1 p_2 e^{-2\phi_1})(1-p_1 p_2 e^{2\phi_1})} \big[(1+p_1 p_2) (\cosh m_1 + \cosh m_2) \cosh \phi_2 \nonumber \\
&\hspace{15ex} - 2\sqrt{p_1 p_2} (\cosh m_1 \cosh m_2 + \cosh^2 \phi_2) \cosh \phi_1 \big]\ 
\end{align}
and $ \langle W_{[-1, 0]}^{(0, 1)} \rangle = 0 $. The same result can be obtained by solving the blowup equations from the perspective of 5d KK theory, which we checked up to $ (1, 1) $-instanton order. We also checked that the 5d Wilson loop VEVs agree with $\langle \mathcal{W}_{1,2}\rangle$ extracted above from our solutions for the 6d/2d partition function up to 2-string order.

\subsection{6d \texorpdfstring{$SU(3)$}{SU(3)} gauge theory}

The 6d minimal SCFT with $ SU(3) $ gauge group is realized in F-theory compactified on a local elliptic 3-fold embedding three $ \mathbb{F}_1 $ surfaces glued together as follows \cite{DelZotto:2017pti}:
\begin{align}\label{eq:6d_su3_geo}
\begin{tikzpicture}
\draw (0, 0) node {$ \mathbb{F}_1 $}
(-1.2, -1.7) node {$ \mathbb{F}_1 $}
(1.2, -1.7) node {$ \mathbb{F}_1 $};
\draw[thick] (-0.3, -0.3) -- (-1.1, -1.4)
(-0.8, -1.7) -- (0.8, -1.7)
(0.3, -0.3) -- (1.1, -1.4);
\draw (-0.6, -0.3) node {$ _e $}
(-1.2, -1.1) node {$ _e $}
(-0.6, -1.5) node {$ _e $}
(0.6, -1.5) node {$ _e $}
(0.6, -0.3) node {$ _e $}
(1.2, -1.1) node {$ _e $};
\end{tikzpicture}
\end{align}
Let us introduce K\"ahler parameters $ Q_1, Q_2,  Q_3,  Q_4$ for the $ e $ curve in an $ \mathbb{F}_1 $ and the fiber curves in other three $ \mathbb{F}_1$'s respectively. These parameters can be written in terms of the K\"ahler parameters $\phi_0,\phi_1,\phi_2$ for three $\mathbb{F}_1$ surfaces and the K\"ahler parameter $\tau$ for the elliptic curve as
\begin{align}
Q_1 &= e^{-(\phi_0 + \phi_1 + \phi_2)} \, , 
&Q_2 &= e^{-(2\phi_0 - \phi_1 - \phi_2 + \tau)} \, , \cr
Q_3 &= e^{-(-\phi_0 + 2\phi_1 - \phi_2)} \, , 
&Q_4 &= e^{-(-\phi_0 - \phi_1 + 2\phi_2)} \, .
\end{align}

The BPS spectrum of this theory without loop operators has been computed using the blowup method in \cite{Gu:2018gmy,Kim:2020hhh}. To solve the blowup equations, one uses the effective prepotential given by
\begin{align}
\mathcal{E} = \frac{1}{\epsilon_1 \epsilon_2} \bigg( &\frac{1}{6} \qty(9\phi_0^3 + 9\phi_1^3 + 9\phi_2^3 - (\phi_0 + \phi_1 + \phi_2)^3 + 9\tau \phi_0^2 + 3\tau^2 \phi_0) \nonumber \\
& - \frac{\epsilon_1^2 + \epsilon_2^2}{12}(\phi_0 + \phi_1 + \phi_2) + \epsilon_+^2 (\phi_0 + \phi_1 + \phi_2) \bigg) \, ,
\end{align}
and magnetic fluxes given by
\begin{align}\label{eq:6d_su3_flux}
n_i \in \mathbb{Z} \pm 1/6 \, , \quad
B_\tau = 0 \, \qquad
\text{or} \qquad
n_i \in \mathbb{Z} + 1/2 \, , \quad
B_\tau = 0 \, .
\end{align}
The solution to the blowup equations matches the elliptic genera of self-dual strings in the 6d $SU(3)$ gauge theory computed based on the ADHM construction in \cite{Kim:2016foj}. 

\paragraph{Fundamental Wilson loops} Next, consider the Wilson loop operator in the representation $ \mathbf{r} = [-1, 0, 0] $ corresponding to an M2-brane wrapping a non-compact curve intersecting the first $\mathbb{F}_1$ at one point. We call this operator as a fundamental Wilson loop. Its expectation value can be expanded as
\begin{align}\label{eq:6d_su3_cls}
\ev*{W_{[-1, 0, 0]}} = e^{\phi_0} + \cdots = Q_1^{-1/3} Q_3^{1/3} Q_4^{1/3} \qty(1 + \mathcal{O}(Q_{i})) \, .
\end{align}
We can compute the spectrum of this loop operator by solving the blowup equation of $ \mathbf{r}_1 = [-1, 0, 0] $, $ \mathbf{r}_2 = \emptyset $ formulated with the effective prepotential and the magnetic fluxes given above. We assume in the computation, as we did for the $ \mathcal{N} = (2, 0) $ $ A_1 $ theory, that the KK tower of the primitive state with charge $ -\phi_0 + n \tau $ decouples from the 6d field theory and thus they are not involved in the spectrum in \eqref{eq:6d_su3_cls}. This removes any ambiguity in solving the blowup equations. One then obtains the spectrum of the Wilson loop operator expanded as
\begin{align}\label{eq:6d_su3_Wcls}
\ev*{W_{[-1, 0, 0]}}
= Q_1^{-1/3} Q_3^{1/3} Q_4^{1/3} (& 1+Q_2 + Q_2 Q_4 + Q_2 Q_3 + Q_2 Q_3 Q_4^2 \nonumber \\
& + Q_2 Q_3^2 Q_4 + Q_2 Q_3^2 Q_4^2 + \mathcal{O}(Q_1,Q_2^3, Q_3^3, Q_4^3) ) \, .
\end{align}
The higher order spectrum is summarized in Table~\ref{table:6d_SU3}.

\begin{table}[t]
	\centering
	\begin{tabular}{|c|C{27ex}||c|C{27ex}|} \hline
		$ \mathbf{d} $ & $ \oplus \tilde{N}_{j_l, j_r}^{\mathbf{d}} (j_l, j_r) $ & $ \mathbf{d} $ & $ \oplus \tilde{N}_{j_l, j_r}^{\mathbf{d}} (j_l, j_r) $ \\ \hline
		$ (1, 1, 0, 0) $ & $ (0, \frac{1}{2}) $ & $ (1, 1, 0, 1) $ & $ (0, \frac{1}{2}) $ \\ \hline
		$ (1, 1, 0, 2) $ & $ (0, \frac{3}{2}) $ & $ (1, 1, 1, 0) $ & $ (0, \frac{1}{2}) $ \\ \hline
		$ (1, 1, 1, 1) $ & $ 3(0, \frac{1}{2}) \oplus (\frac{1}{2}, 0) $ & $ (1, 1, 1, 2) $ & $ 2(0,\frac{1}{2}) \oplus 2(0,\frac{3}{2}) \oplus (\frac{1}{2},1) $ \\ \hline
		$ (1, 1, 2, 0) $ & $ (0, \frac{3}{2}) $ & $ (1, 1, 2, 1) $ & $ 2(0,\frac{1}{2}) \oplus 2(0,\frac{3}{2}) \oplus (\frac{1}{2},1) $ \\ \hline
		$ (1, 1, 2, 2) $ & $ \!\! 4(0,\frac{1}{2}) \oplus 3(0,\frac{3}{2}) \oplus (\frac{1}{2},0) \oplus (\frac{1}{2},1) \!\! $ & $ (1, 2, 0, 0) $ & $ (0, \frac{3}{2}) $ \\ \hline
		$ (1, 2, 0, 1) $ & $ (0,\frac{1}{2}) \oplus (0,\frac{3}{2}) $ & $ (1, 2, 0, 2) $ & $ (0,\frac{1}{2}) \oplus (0,\frac{3}{2}) $ \\ \hline
		$ (1, 2, 1, 0) $ & $ (0,\frac{1}{2}) \oplus (0,\frac{3}{2}) $ & $ (1, 2, 1, 1) $ & $ \!\! 4(0,\frac{1}{2}) \oplus 2(0,\frac{3}{2}) \oplus (\frac{1}{2},0) \oplus (\frac{1}{2},1) \!\! $ \\ \hline
		$ (1, 2, 1, 2) $ & $ 7(0,\frac{1}{2}) \oplus 4(0,\frac{3}{2}) \oplus 2(\frac{1}{2},0) \oplus 2(\frac{1}{2},1) $ & $ (1, 2, 2, 0) $ & $ (0,\frac{1}{2}) \oplus (0,\frac{3}{2}) $ \\ \hline
		$ (1, 2, 2, 1) $ & $ 7(0,\frac{1}{2}) \oplus 4(0,\frac{3}{2}) \oplus 2(\frac{1}{2},0) \oplus 2(\frac{1}{2},1) $ & $ (1, 2, 2, 2) $ & $ 17(0,\frac{1}{2}) \oplus 8(0,\frac{3}{2}) \oplus 6(\frac{1}{2},0) \oplus 6(\frac{1}{2},1) \oplus (1,\frac{1}{2}) $ \\ \hline
		$ (2, 1, 0, 2) $ & $ (0, 2) $ & $ (2, 1, 1, 1) $ & $ (0, 0) \oplus (0, 1) $ \\ \hline
		$ (2, 1, 1, 2) $ & $ \!\! (0,0) \oplus 2(0,1) \oplus 3(0,2) \oplus (\frac{1}{2},\frac{3}{2}) \!\! $ & $ (2, 1, 2, 0) $ & $ (0, 2) $ \\ \hline
		$ (2, 1, 2, 1) $ & $ \!\!(0,0) \oplus 2(0,1) \oplus 3(0,2) \oplus (\frac{1}{2},\frac{3}{2}) \!\! $ & $ (2, 1, 2, 2) $ & $ 4(0,0) \oplus 8(0,1) \oplus 6(0,2) \oplus 2(0,3) \oplus 2(\frac{1}{2},\frac{1}{2}) \oplus 2(\frac{1}{2},\frac{3}{2}) \oplus (\frac{1}{2},\frac{5}{2}) $ \\ \hline
		$ (2, 2, 0, 0) $ & $ (0, 2) $ & $ (2, 2, 0, 1) $ & $ (0, 1) \oplus (0, 2) $ \\ \hline
		$ (2, 2, 0, 2) $ & $ \!\! (0,0) \oplus 2(0,1) \oplus 2(0,2) \oplus (0,3) \!\! $ & $ (2, 2, 1, 0) $ & $ (0,1) \oplus (0,2) $ \\ \hline
		$ (2, 2, 1, 1) $ & $ 3(0,0) \oplus 6(0,1) \oplus 3(0,2) \oplus (\frac{1}{2},\frac{1}{2}) \oplus (\frac{1}{2},\frac{3}{2}) $ & $ (2, 2, 1, 2) $ & $ 8(0,0) \oplus 16(0,1) \oplus 10(0,2) \oplus 2(0,3) \oplus 4(\frac{1}{2},\frac{1}{2}) \oplus 4(\frac{1}{2},\frac{3}{2}) \oplus (\frac{1}{2},\frac{5}{2}) $ \\ \hline
		$ (2, 2, 2, 0) $ & $ (0,0) \oplus 2(0,1) \oplus 2(0,2) \oplus (0,3) $ & $ (2, 2, 2, 1) $ & $ 8(0,0) \oplus 16(0,1) \oplus 10(0,2) \oplus 2(0,3) \oplus 4(\frac{1}{2},\frac{1}{2}) \oplus 4(\frac{1}{2},\frac{3}{2}) \oplus (\frac{1}{2},\frac{5}{2}) $ \\ \hline
		$ (2, 2, 2, 2) $ & \multicolumn{3}{C{68.5ex}|}{$ 29(0,0) \oplus 52(0,1) \oplus 30(0,2) \oplus 7(0,3) \oplus 22(\frac{1}{2},\frac{1}{2}) \oplus 18(\frac{1}{2},\frac{3}{2}) \oplus 4(\frac{1}{2},\frac{5}{2}) \oplus 2(1,0) \oplus 3(1,1) \oplus (1,2) $} \\ \hline
	\end{tabular}
	\caption{Spectrum of the Wilson loop operator in the representation $ \mathbf{r} = [-1, 0, 0] $ in the 6d minimal $ SU(3) $ theory for $ d_i \leq 2 $. Here, $ \mathbf{d} = (d_1, d_2, d_3, d_4) $ labels the BPS states with fugacity $ Q_1^{-1/3} Q_3^{1/3} Q_4^{1/3} \prod_i Q_i^{d_i} $.}\label{table:6d_SU3}
\end{table}

\paragraph{Codimension-4 defect: 2d fermion} We shall now consider codimension-4 defect introduced by coupling a 2d free fermion to the 6d bulk $SU(3)$ gauge field. The partition function of this 6d/2d coupled system can be written as
\begin{align}\label{eq:6d-su3}
Z^{6d/2d} &= Z^{\rm vec}_{\rm pert} \times e^{\phi_0} \sum_{k=0}^\infty e^{-3k\phi_0} Z_k^{6d/2d} \, , \nonumber \\
Z^{\rm vec}_{\rm pert} &= \PE\bigg[-\frac{1+p_1 p_2}{(1-p_1)(1-p_2)} \frac{1}{1-q} \sum_{i<j}^3(e^{a_i-a_j} + q e^{-a_i+a_j}) \bigg]  \, , \nonumber \\
Z^{2d}_{\rm pert} &= e^{\phi_0} Z^{6d/2d}_{k=0} = e^{\phi_0} \prod_{j=1}^3 \frac{\theta_1(z-a_j)}{i\eta} \, ,
\end{align}
where $a_i$ are the $SU(3)$ gauge holonomies that can be expressed as $(a_1,a_2,a_3)=(\phi'_1,\phi'_2-\phi_1',-\phi_2')$ in terms of the parameters $\phi_i'$ in the $SU(3)$ Dynkin basis and these $SU(3)$ parameters are related to the geometric K\"ahler parameters as $(\phi_1',\phi_2')=(\phi_1-\phi_0,\phi_2-\phi_0)$. So the primed parameters here will denote the parameters in the 6d field theory basis respecting the 6d $SU(3)$ gauge algebra. $Z^{\rm vec}_{\rm pert}$ is the $SU(3)$ vector multiplet contribution and $Z^{2d}_{\rm pert}$ is the perturbative contribution from the 2d fermion with mass $z$ in the $SU(3)$ fundamental representation. Again, the $e^{\phi_0}$ factor is multiplied so that the partition function of the 6d/2d coupled system satisfies the blowup equations given below.

One can formulate two blowup equations for the 6d/2d partition function with magnetic fluxes
\begin{align}
n'_0 \in \mathbb{Z} \pm 1/6 \, , \quad
n'_1 \in \mathbb{Z} \, , \quad
n'_2 \in \mathbb{Z} \, , \quad
B_{\tau} = 0 \, , \quad
B_z = 0 \ ,
\end{align}
where $n'_i$ are the magnetic fluxes for the tensor and the vector fields of $(\phi_0,\phi_1',\phi_2')$. We insert the 6d/2d partition function of the form \eqref{eq:6d-su3} into the blowup equations and compute order by order the self-dual string contribution $Z_k^{6d/2d}$. The blowup equations at $0$-string can be solved to fix the $ \Lambda_0 $ factor as
\begin{align}
\Lambda_0 = \underset{n_1'=n_0' \mp 1/6 \in \mathbb{Z}}{\sum_{\vec{n}'=(n_0', n_1', n_1')}} (-1)^{\abs{\vec{n}}'} e^{-V(\vec{n}')} p_1^{-n_0'} = \left\{
\begin{array}{ll}
\vartheta \genfrac[]{0pt}{1}{1/6}{1/2} (3\tau, -2\epsilon_1 - \epsilon_2) &  \ \text{for}  \ n'_0 \in \mathbb{Z} + 1/6 \\
\vartheta \genfrac[]{0pt}{1}{-1/6}{1/2} (3\tau, -2\epsilon_1 - \epsilon_2) & \ \text{for} \ n'_0 \in \mathbb{Z} - 1/6 \, ,
\end{array}\right.
\end{align}
where $ \vartheta\genfrac[]{0pt}{1}{\alpha}{\beta}(\tau, x) $ is the theta function with characteristics defined in \eqref{eq:theta_char}. We then compute the 1-string contribution as
\begin{align}
Z_{1}^{6d/2d} = \frac{\vartheta\genfrac[]{0pt}{1}{-1/6}{1/2} (3\tau, 2\epsilon_-) F_{1/6} - \vartheta\genfrac[]{0pt}{1}{1/6}{1/2} (3\tau, 2\epsilon_-) F_{-1/6}}{
\vartheta\genfrac[]{0pt}{1}{1/6}{1/2} (3\tau, -2\epsilon_1 \!-\! \epsilon_2) \vartheta\genfrac[]{0pt}{1}{-1/6}{1/2} (3\tau, 2\epsilon_-\!) \!- \vartheta\genfrac[]{0pt}{1}{-1/6}{1/2} (3\tau, -2\epsilon_1\! -\! \epsilon_2) \vartheta\genfrac[]{0pt}{1}{1/6}{1/2} (3\tau, 2\epsilon_-\!)}\ ,
\end{align}
where
\begin{align}
F_{\alpha} &= \vartheta\genfrac[]{0pt}{1}{\alpha}{1/2} (3\tau, -4\epsilon_-) Z_0^{6d/2d}(\phi_1', \phi_2', \tau) Z_1(\phi'_1, \phi'_2, \tau; \epsilon_1 - \epsilon_2, \epsilon_2) \nonumber \\
& \quad + \sum_{i \neq j, k\neq i, j}^3  \frac{q^{1/2} \eta(\tau)^6  \vartheta\genfrac[]{0pt}{1}{\alpha}{1/2}(3\tau, \epsilon_1 + 2\epsilon_2 + 3(a_i - a_j))}{\theta_1(a_i - a_j)\theta_1(a_j-a_k)\theta_1(a_k-a_i) \theta_1(\epsilon_{1,2}+a_i-a_j) \theta_1(2\epsilon_++a_i-a_j)} \nonumber \\
& \qquad \quad \times \qty[\prod_{l=1}^3 \frac{\theta_1(z-a_l-(\delta_{li} - \delta_{lj})\epsilon_1)}{i\eta(\tau)}] \, .
\end{align}
The contributions at higher strings can also be computed iteratively.

We shall relate the 6d/2d partition function to the Wilson loop expectation value we computed above. One can recast the perturbative contribution from the 2d fermion as
\begin{align}\label{eq:6d_SU3_cls_factorize}
Z^{2d}_{\rm pert}
&= -i \frac{\theta_1(3\tau, 3z)}{\eta(\tau)^3} \mathcal{W}_0(\phi_i', \tau) + i q^{1/2} z \frac{\theta_1(3\tau, 3z + \tau)}{\eta(\tau)^3} \mathcal{W}_1(\phi_i', \tau) \nonumber \\
& \quad  + i q^{1/2} z^{-1} \frac{\theta_1(3\tau, 3z - \tau)}{\eta(\tau)^3} \mathcal{W}_2(\phi_i', \tau) \, .
\end{align}
In this expression, we separate the terms depending on the dynamical K\"ahler parameters $\phi_i'$ from the $z$-dependent terms. We have checked this separation up to 10-th order in $e^{-\tau}$ expansion. Recall now that the blowup equations are linear equations in the 6d/2d partition function and that the magnetic fluxes for $ \tau $ and $ z $ are not activated. One can then deduce from this fact that each $ \mathcal{W}_{0,1,2} $ together with its string correction, which we define as an expectation value
\begin{align}
\langle \mathcal{W}_{0,1,2}\rangle = \mathcal{W}_{0,1,2}+\sum_{k=1}^\infty e^{-3k\phi_0} \mathcal{W}^{(k)}_{0,1,2} \ ,
\end{align}
independently solves the blowup equations. It turns out that the expectation value $\langle \mathcal{W}_0\rangle$ is precisely the fundamental Wilson loop VEV we considered above in \eqref{eq:6d_su3_Wcls}:
\begin{align}
\langle \mathcal{W}_{0}\rangle (\phi_i') = \langle W_{[-1,0,0]}\rangle(\phi) \ .
\end{align}
Furthermore, the other two expectation values are also related to this fundamental Wilson loop VEV by exchanging K\"ahler parameters such as
\begin{align}
\langle\mathcal{W}_1\rangle(\phi_0, \phi_1, \phi_2, \tau) &= q^{-1/3} \langle \mathcal{W}_{0}\rangle (\phi_1 - \tau/3, \phi_0 + \tau/3, \phi_2, \tau) \, , \nonumber \\
\langle\mathcal{W}_2\rangle(\phi_0, \phi_1, \phi_2, \tau) &= q^{-1/3} \langle \mathcal{W}_{0}\rangle (\phi_2 - \tau/3, \phi_1, \phi_0 + \tau/3, \tau) \, .
\end{align}
Here, we used the geometric K\"ahler parameters $\phi_{0,1,2}$. This implies $ \langle\mathcal{W}_1\rangle $ and $ \langle\mathcal{W}_2\rangle $ are VEVs of the Wilson loop operators in the representations $ \mathbf{r} = [0, -1, 0] $ and $ \mathbf{r} = [0, 0, -1] $, respectively. We thus conclude that the expectation value of the codimension-4 defect is a linear combination of three minimal Wilson loop operators each corresponding to a primitive non-compact curve intersecting an $ \mathbb{F}_1 $ surface at one point.

\subsection{\texorpdfstring{$SU(3)_9$}{SU(3)9} theory}

The 5d $ SU(3) $ gauge theory at the CS-level 9 is a KK theory arising from a $\mathbb{Z}_2$ twisted compactification of the 6d minimal $SU(3)$ gauge theory \cite{Jefferson:2017ahm,Jefferson:2018irk}. This theory can be engineered in M-theory compactified on a local 3-fold embedding $ \mathbb{F}_{10} $ and $ \mathbb{F}_0 $ surfaces glued as follows:
\begin{align}
\begin{tikzpicture}
\draw (0, 0) node {$ \mathbb{F}_{10} $}
(4, 0) node {$ \mathbb{F}_0 $};
\draw[thick] (0.5, 0) -- (3.5, 0);
\draw (0.7, 0.2) node {$ _e $}
(3.1, 0.2) node {$ _{h+4f} $};
\end{tikzpicture} \, .
\end{align}
We introduce K\"ahler parameters $ \{Q_1, Q_2, Q_3 \} $ for the $ e $ curve in $ \mathbb{F}_0 $ and fiber curves in two surfaces. These parameters can be written in terms of the K\"ahler parameters $\phi_1,\phi_2$ of two surfaces and the mass parameter $m$ as
\begin{align}
Q_1 = e^{-(m - 4\phi_1 + 2\phi_2)} \, , \qquad
Q_2 = e^{-(2\phi_1 - \phi_2)} \, , \qquad
Q_3 = e^{-(-\phi_1 + 2\phi_2)} \, .
\end{align}
The parameters $\phi_1,\phi_2$ are identified with the Coulomb branch parameters of the 5d $SU(3)$ gauge theory.

The effective prepotential on the Coulomb branch of the 5d theory is given by
\begin{align}\label{eq:su3_9_E}
\mathcal{E} &= \frac{1}{\epsilon_1 \epsilon_2} \qty(\mathcal{F} - \frac{\epsilon_1^2 + \epsilon_2^2}{12}(\phi_1 + \phi_2) + \epsilon_+^2(\phi_1 + \phi_2)) \, , \\
6\mathcal{F} &= 8\phi_1^3 + 24\phi_1^2 \phi_2 -30\phi_1 \phi_2^2 + 8\phi_2^2 + 6m(\phi_1^2 - \phi_1 \phi_2 + \phi_2^2)\ ,
\end{align}
where $ m $ is the inverse gauge coupling in the 5d $SU(3)$ gauge theory. Using this effective prepotential and the consistent magnetic fluxes
\begin{align}\label{eq:su3_9_flux1}
n_1, n_2 \in \mathbb{Z} \, , \quad
B_m = 0 \, ,
\end{align}
in the 5d $SU(3)$ frame, or
\begin{align}\label{eq:su3_9_flux2}
n_1 \in \mathbb{Z} + \frac{h}{3} \, , \quad
n_2 \in \mathbb{Z} + \frac{2h}{3} \, , \quad
B_m = 0 \, ,
\end{align}
with $h=0,1,2$ in the 6d frame, one can formulate the blowup equations for the partition function without loop operators. These blowup equations are solved to calculate the BPS spectrum of this KK theory in \cite{Kim:2020hhh}.

\paragraph{Representation $ \mathbf{r} = [0, -1] $}

\begin{table}
	\centering
	\begin{tabular}{|c|C{28ex}||c|C{28ex}|} \hline
		$ \mathbf{d} $ & $ \oplus \tilde{N}_{j_l, j_r}^{\mathbf{d}} (j_l, j_r) $ & $ \mathbf{d} $ & $ \oplus \tilde{N}_{j_l, j_r}^{\mathbf{d}} (j_l, j_r) $ \\ \hline
		$ (1, 0, 0) $ & $ (0, 0) $ & $ (1, 0, 1) $ & $ (0, 1) $ \\ \hline
		$ (1, 0, 2) $ & $ (0, 2) $ & $ (1, 1, 0) $ & $ x(0, 0) $ \\ \hline
		$ (1, 1, 1) $ & $ (0,0) \oplus (0,1) $ & $ (1, 1, 2) $ & $ (0,1) \oplus (0,2) $ \\ \hline
		$ (1, 2, 1) $ & $ (0, 1) $ & $ (1, 2, 2) $ & $ (0,0) \oplus (0,1) \oplus (0,2) $ \\ \hline
		$ (1, 3, 0) $ & $ x(0, 0) $ & $ (1, 3, 1) $ & $ (0, 1) $ \\ \hline
		$ (1, 3, 2) $ & $ x(0,0) \oplus (0,1) \oplus (0,2) $ & $ (1, 4, 0) $ & $ (0, 0) $ \\ \hline
		$ (1, 4, 1) $ & $ (0, 0) \oplus (0, 1) $ & $ (1, 4, 2) $ & $ (0, 0) \oplus (0, 1) \oplus (0, 2) $ \\ \hline
		$ (2, 0, 1) $ & $ (0, 2) $ & $ (2, 0, 2) $ & $ (0,2) \oplus 2(0,3) \oplus (\frac{1}{2},\frac{7}{2}) $ \\ \hline
		$ (2, 1, 1) $ & $ (0,1) \oplus (0,2) $ & $ (2, 1, 2) $ & $ (0,1) \oplus 4(0,2) \oplus 3(0,3) \oplus (\frac{1}{2},\frac{5}{2}) \oplus (\frac{1}{2},\frac{7}{2}) $ \\ \hline
		$ (2, 2, 1) $ & $ (0,0) \oplus (0,1) \oplus (0,2) $ & $ (2, 2, 2) $ & $ (0,0) \oplus 4(0,1) \oplus 5(0,2) \oplus 3(0,3) \oplus (\frac{1}{2},\frac{3}{2})\oplus (\frac{1}{2},\frac{5}{2})\oplus (\frac{1}{2},\frac{7}{2}) $ \\ \hline
		$ (2, 3, 0) $ & $ y(0, 0) $ & $ (2, 3, 1) $ & $ \! (1+x)(0,0) \oplus {(2+x)(0,1)} \oplus (0,2)\oplus x(\frac{1}{2},\frac{1}{2}) \! $ \\ \hline
		$ (2, 3, 2) $ & $ 3(0,0)\oplus (6+x)(0,1) \oplus {(6+x)(0,2)} \oplus 3(0,3) \oplus (\frac{1}{2},\frac{1}{2}) \oplus (1+x)(\frac{1}{2},\frac{3}{2}) \oplus (\frac{1}{2},\frac{5}{2}) \oplus (\frac{1}{2},\frac{7}{2}) $ & $ (2, 4, 1) $ & $ (2+x)(0,0) \oplus (4+x)(0,1) \oplus (0,2) \oplus (2+x)(\frac{1}{2},\frac{1}{2}) \oplus (1,0) $ \\ \hline
		$ (2, 4, 2) $ & \multicolumn{3}{C{68.5ex}|}{$ (5+x)(0,0) \oplus (10+2x)(0,1) \oplus (9+x)(0,2) \oplus 3(0,3) \oplus {(3+x)(\frac{1}{2},\frac{1}{2})} \oplus (3+x)(\frac{1}{2},\frac{3}{2}) \oplus (\frac{1}{2},\frac{5}{2}) \oplus (\frac{1}{2},\frac{7}{2}) \oplus 1(1,1) $} \\ \hline
	\end{tabular}
	\caption{Spectrum of the fundamental Wilson loop operator in the $ SU(3)_9 $ theory for $ (d_1, d_2, d_3) \leq (2, 4, 2) $. Here, $ \mathbf{d} = (d_1, d_2, d_3) $ labels the BPS states with fugacity $ Q_2^{-1/3} Q_3^{-2/3} \prod Q_i^{d_i}$. Here, the coefficients $x$ and $y$ are conjectured to be 0.}\label{table:SU3_9_fund}
\end{table}

Consider the fundamental Wilson loop in the 5d $ SU(3)_9 $ theory. Its classical expectation value is
\begin{align}
\langle W_{[0, -1]}^{\mathrm{cls}} \rangle = e^{\phi_2} + e^{\phi_1 - \phi_2} + e^{-\phi_1} \, .
\end{align}

The instanton corrections to the VEV of this Wilson loop operator can be computed by solving the blowup equations with $ \mathbf{r}_1 = [0, -1] $, $ \mathbf{r}_2 = \emptyset $ and background magnetic fluxes given above. The spectrum of the Wilson loop operator is summarized in Table~\ref{table:SU3_9_fund}. In the table, there are two unknown BPS degeneracies which we denote by $ x = \tilde{N}_{0,0}^{(1, 1, 0)} $ and $ y = \tilde{N}_{0,0}^{(2,3,0)} $. We could not fix them until $ (d_1, d_2, d_3) \leq (3, 10, 5) $ order in the K\"ahler parameter expansion. We note that their BPS charge is $ m - 2\phi_1 $ and $ 2m - 2\phi_1 $, respectively. Our assertion is however that a Wilson loop VEV in a minimal representation cannot have in its spectrum BPS states with electric charge $e_i \le 0$ for all $i$'s other than the primitive state. Therefore, we claim that $ x = y = 0 $. We will see below that the codimension-4 defect partition function in the 6d theory also suggests this.

\paragraph{Representation $ \mathbf{r} = [-1, 0] $}
We next consider the Wilson loop operator in the anti-fundamental representation of $ SU(3) $ whose classical expectation value is given by
\begin{align}
\langle W_{[-1, 0]}^{\mathrm{cls}}\rangle = e^{\phi_1} + e^{-\phi_1 + \phi_2} + e^{-\phi_2} \, .
\end{align}

\begin{table}
	\centering
	\begin{tabular}{|c|C{28ex}||c|C{28ex}|} \hline
		$ \mathbf{d} $ & $ \oplus \tilde{N}_{j_l, j_r}^{\mathbf{d}} (j_l, j_r) $ & $ \mathbf{d} $ & $ \oplus \tilde{N}_{j_l, j_r}^{\mathbf{d}} (j_l, j_r) $ \\ \hline
		$ (1, 1, 0) $ & $ (0, 0) $ & $ (1, 1, 1) $ & $ (0, 1) $ \\ \hline
		$ (1, 1, 2) $ & $ (0, 2) $ & $ (1, 2, 1) $ & $ (0, 0) \oplus (0, 1) $ \\ \hline
		$ (1, 2, 2) $ & $ (0, 1) \oplus (0, 2) $ & $ (1, 3, 1) $ & $ (0, 1) $ \\ \hline
		$ (1, 3, 2) $ & $ (0, 0) \oplus (0, 1) \oplus (0, 2) $ & $ (1, 4, 0) $ & $ (0, 0) $ \\ \hline
		$ (1, 4, 1) $ & $ (0, 0) \oplus (0, 1) $ &$ (1, 4, 2) $ & $ (0, 0) \oplus (0, 1) \oplus (0, 2) $ \\ \hline
		$ (1, 5, 1) $ & $ (0, 1) $ & $ (1, 5, 2) $ & $ (0, 1) \oplus (0, 2) $ \\ \hline
		$ (2, 1, 1) $ & $ (0, 2) $ & $ (2, 1, 2) $ & $ (0,2) \oplus 2(0,3) \oplus (\frac{1}{2},\frac{7}{2}) $ \\ \hline
		$ (2, 2, 1) $ & $ (0, 1) \oplus (0, 2) $ & $ (2, 2, 2) $ & $ (0,1) \oplus 4(0,2) \oplus 3(0,3) \oplus (\frac{1}{2},\frac{5}{2}) \oplus (\frac{1}{2},\frac{7}{2}) $ \\ \hline
		$ (2, 3, 1) $ & $ (0,0) \oplus (0,1) \oplus (0,2) $ & $ (2, 3, 2) $ & $ (0,0) \oplus 4(0,1) \oplus 5(0,2) \oplus 3(0,3) \oplus (\frac{1}{2},\frac{3}{2}) \oplus (\frac{1}{2},\frac{5}{2}) \oplus (\frac{1}{2},\frac{7}{2}) $ \\ \hline
		$ (2, 4, 1) $ & $ 2(0,0) \oplus 3(0,1) \oplus (0,2) \oplus (\frac{1}{2},\frac{1}{2}) $ & $ (2, 4, 2) $ & $ 3(0,0) \oplus 7(0,1) \oplus 7(0,2) \oplus 3(0,3) \oplus (\frac{1}{2},\frac{1}{2}) \oplus 2(\frac{1}{2},\frac{3}{2}) \oplus (\frac{1}{2},\frac{5}{2}) \oplus (\frac{1}{2},\frac{7}{2}) $ \\ \hline
		$ (2, 5, 1) $ & $ 3(0,0) \oplus 4(0,1) \oplus (0,2) \oplus 2(\frac{1}{2},\frac{1}{2}) $ & $ (2, 5, 2) $ & $ 5(0,0) \oplus 11(0,1) \oplus 9(0,2) \oplus 3(0,3) \oplus 3(\frac{1}{2},\frac{1}{2}) \oplus 3(\frac{1}{2},\frac{3}{2}) \oplus (\frac{1}{2},\frac{5}{2}) \oplus (\frac{1}{2},\frac{7}{2}) $ \\ \hline
	\end{tabular}
	\caption{Spectrum of the anti-fundamental Wilson loop operator in the $ SU(3)_9 $ theory for $ (d_1, d_2, d_3) \leq (2, 5, 2) $. Here, $ \mathbf{d} = (d_1, d_2, d_3) $ labels the BPS states with fugacity $ Q_2^{-2/3} Q_3^{-1/3} \prod Q_i^{d_i} $.}\label{table:SU3_9_antifund}
\end{table}

The instanton corrections to the VEV of this Wilson loop operator can be computed by solving the blowup equations with $ \mathbf{r}_1 = [ -1,0] $, $ \mathbf{r}_2 = \emptyset $ and background magnetic fluxes given above. The spectrum of the Wilson loop operator is summarized in Table~\ref{table:SU3_9_antifund}. We again remark that the fundamental and the anti-fundamental Wilson loop operators have completely distinct spectra although their classical expectation values are complex conjugate to each other, as expected from the absence of the charge conjugation symmetry in the presence of the Chern-Simons term.

\paragraph{Codimension-4 defect: 2d fermion}
Consider the 6d/2d coupled system by coupling 2d degrees of freedom to the 6d $SU(3)$ theory on a circle with $ \mathbb{Z}_2 $ twist. The 6d partition function with $\mathbb{Z}_2$ twist before the insertion of the 2d degrees of freedom can be written as
\begin{align}
Z &= Z_{\mathrm{pert}}^{\mathrm{vec}} \times Z_{\mathrm{str}} \, , \nonumber \\
Z_{\mathrm{pert}}^{\mathrm{vec}} &= \PE\qty[-\frac{1+p_1p_2}{(1-p_1)(1-p_2)} \frac{1}{1-q} \qty(e^{-2\phi_1'} + (e^{\phi_1'} + e^{-\phi_1'}) (q^{1/4} + q^{3/4}) + q e^{2\phi_1'})] \, , \nonumber \\
Z_{\mathrm{str}} &= 1 + \sum_{k=0}^\infty e^{-3k\phi_0'} Z_k \, ,
\end{align}
where $ Z_{\mathrm{pert}}^{\mathrm{vec}} $ is the perturbative contribution from the vector multiplet and $ Z_k $ is the $ k $-string elliptic genus. Here we used the K\"ahler parameters in the 6d basis that are related to the 5d parameters as
\begin{align}\label{eq:k=9_su3z2_basis}
\phi_1 = \phi_0' + \frac{\tau}{6} \, , \qquad
\phi_2 = 2\phi_0' + \phi_1' + \frac{\tau}{12} \, , \qquad
m = \frac{\tau}{2} \ ,
\end{align}
where $ \phi_0' $ and $ \phi_1' $ are VEVs of scalars in the tensor and the vector multiplets in the 6d theory, respectively. 

There are three independent blowup equations constructed with three sets of consistent magnetic fluxes
\begin{align}\label{eq:su3z2_flux}
n_0' \in \mathbb{Z} + \frac{h}{3} \, , \qquad
n_1' \in \mathbb{Z} \, , \qquad
B_\tau = 0 \, ,
\end{align}
with $h=0,1,2$. One can solve these blowup equations order by order in the string number $ k $-expansion and compute the exact form of $ Z_k $ at each $k$. For example, the solution of the blowup equations at $ k=1 $ order is
\begin{align}
Z_1 = \frac{\sum_{i, j, k=1}^3 \epsilon_{ijk} F_{i/3} \vartheta\genfrac[]{0pt}{1}{j/3}{1/2} (3\tau, 2\epsilon_1 - \epsilon_2) \vartheta\genfrac[]{0pt}{1}{k/3}{1/2} (3\tau, -\epsilon_1 + 2\epsilon_2)}{\sum_{i, j, k=1}^3 \epsilon_{ijk} \vartheta\genfrac[]{0pt}{1}{i/3}{1/2} (3\tau, -\epsilon_1 - \epsilon_2) \vartheta\genfrac[]{0pt}{1}{j/3}{1/2} (3\tau, 2\epsilon_1 - \epsilon_2) \vartheta\genfrac[]{0pt}{1}{k/3}{1/2} (3\tau, -\epsilon_1 + 2\epsilon_2) } \ ,
\end{align}
where $ \epsilon_{ijk} $ is the Levi-Civita symbol, $ \vartheta\genfrac[]{0pt}{1}{\alpha}{\beta}(\tau, x) $ is defined in \eqref{eq:theta_char} and
\begin{align}
F_\alpha = \frac{-q^{1/4} \eta^6 \vartheta\genfrac[]{0pt}{1}{\alpha}{1/2} (3\tau, 6\phi_1' + 4\epsilon_+)}{\theta_1(2\phi_1') \theta_1(2\phi_1' + \epsilon_{1,2}) \theta_1(2\phi_1' + 2\epsilon_+) \theta_1(\phi_1' \pm \tau/4) } + (\phi_1' \to -\phi_1') \, .
\end{align}

We now introduce a particular codimension-4 defect by coupling two 2d fermions in the fundamental representation of the invariant subalgebra $\mathfrak{su}(2)$ after $\mathbb{Z}_2$ twist. The partition function of the 6d/2d couple system is given by 
\begin{align}
Z^{6d/2d} &= Z_{\mathrm{pert}}^{\mathrm{vec}} \times e^{2\phi_0'} \sum_{k=0}^\infty e^{-3k\phi_0'} Z_k^{6d/2d} \ , \nonumber \\
Z_{\mathrm{pert}}^{2d} &= e^{2\phi_0} Z_{k=0}^{6d/2d} = e^{2\phi_0} \frac{\theta_1\big(z \pm (\pm \phi_1' - \tau/4)\big)}{(i \eta)^4} \ ,
\end{align}
where $Z^{\rm 2d}_{\rm pert}$ is the perturbative contribution from the 2d fermions and $z$ is the chemical potential for the diagonal $U(1)$ flavor symmetry of the two fermions. The factor $e^{2\phi_0'}$ is introduced so that the 6d/2d partition function solves the blowup equations given below. As one sees, the fermions carry fractional KK charges. This 2d degrees of freedom can arise from a $\mathbb{Z}_2$ twist of the 6d $SU(3)$ theory coupled to two 2d fermions: one in the fundamental representation and the other in the anti-fundamental representation of the $SU(3)$ gauge group. The $\mathbb{Z}_2$ action exchanges the fundamental and the anti-fundamentals of $SU(3)$, and therefore exchanges these two fermions with each other. The twisted compactification then leaves two 2d $\mathfrak{su}(2)$ fundamental fermions with fractional KK-charges, and two singlets which we ignore in the 6d/2d partition function computation.

There is another way to understand this defect through the Higgsing of the 6d $ G_2 $ gauge theory with 1 fundamental to the $SU(3)$ theory with $\mathbb{Z}_2$ twist discussed in \cite{Kim:2019dqn}. Consider the $ G_2 $ gauge theory coupled to a 2d fundamental fermion. The perturbative partition function of this 2d fermion is given by
\begin{align}
Z_{\mathrm{pert}, G_2}^{2d} = \prod_{w\in {\bf 7}} \frac{\theta_1(z+w\cdot a)}{i\eta}= \frac{\theta_1(z)}{i\eta} \prod_{i=1}^3 \frac{\theta_1(z \pm a_i)}{(i\eta)^2} \, ,
\end{align}
where $a$ denotes the $G_2$ chemical potential and $ z $ is the chemical potential for the 2d fermion number. The Higgsing to the twisted $SU(3)$ theory is realized in the 6d partition function by tuning the parameters as \cite{Kim:2021cua}
\begin{align}
a_1 \to \frac{\tau}{2} \, , \qquad
a_2 \to -\phi_1' - \frac{\tau}{4} \, , \qquad
a_3 \to \phi_1' - \frac{\tau}{4} \, .
\end{align}
This Higgsing when applied to the 2d perturbative leads to the perturbative part of the 2d fermions $Z^{2d}_{\rm pert}$ in the twisted $SU(3)$ theory up to the $\mathfrak{su}(2)$ singlet terms.

The blowup equations for the 6d/2d partition function can be formulated by using the magnetic fluxes in \eqref{eq:su3z2_flux} together with $ B_z = 0 $. We solve the blowup equations with $ h=0, 1 $ to calculate the self-dual string contributions $ Z_k^{6d/2d} $. At $ k=0 $ order, we solve
\begin{align}
\Lambda_0 = \sum_{\vec{n} = (n_0, 0)} (-1)^{\abs{\vec{n}}} e^{-V(\vec{n})} p_1^{-2n_0} = \left\{
\begin{array}{ll}
\vartheta\genfrac[]{0pt}{1}{0}{1/2} (3\tau, -3\epsilon_1 - \epsilon_2) & \ \text{for} \ n_0 \in \mathbb{Z} \, , \\
\vartheta\genfrac[]{0pt}{1}{1/3}{1/2} (3\tau, -3\epsilon_1 - \epsilon_2) & \ \text{for} \ n_0 \in \mathbb{Z} + 1/3 \, .
\end{array}\right.
\end{align}
At $ k=1 $ order, the solution is given by
\begin{align}
Z_1^{6d/2d} = \frac{\vartheta\genfrac[]{0pt}{1}{0}{1/2} (3\tau, - \epsilon_2) G_{1/3} - \vartheta\genfrac[]{0pt}{1}{1/3}{1/2} (3\tau, - \epsilon_2) G_0}{\vartheta\genfrac[]{0pt}{1}{1/3}{1/2} (3\tau, -3\epsilon_1 - \epsilon_2) \vartheta\genfrac[]{0pt}{1}{0}{1/2} (3\tau, - \epsilon_2) - \vartheta\genfrac[]{0pt}{1}{0}{1/2} (3\tau, -3\epsilon_1 - \epsilon_2) \vartheta\genfrac[]{0pt}{1}{1/3}{1/2} (3\tau, - \epsilon_2) }
\end{align}
where
\begin{align}
G_\alpha &= \vartheta\genfrac[]{0pt}{1}{\alpha}{1/2} (3\tau, -3\epsilon_1 + 2\epsilon_2) Z_0^{6d/2d}(\phi_1', \tau) Z_1(\phi_1', \tau; \epsilon_1 - \epsilon_2, \epsilon_2) \nonumber \\
& \quad - \Bigg( \frac{q^{1/4} \eta^6 \vartheta\genfrac[]{0pt}{1}{\alpha}{1/2} (3\tau, 6\phi_1' + 2\epsilon_2) Z_0^{6d/2d}(\phi_1' + \epsilon_1, \tau)}{\theta_1(2\phi_1') \theta_1(2\phi_1' + \epsilon_{1,2}) \theta_1(2\phi_1' + 2\epsilon_+) \theta_1(\phi_1' \pm \tau/4)} \nonumber \\
& \qquad \qquad + \frac{q^{1/4} \eta^6 \vartheta\genfrac[]{0pt}{1}{\alpha}{1/2} (3\tau, -6\phi_1' + 2\epsilon_2) Z_0^{6d/2d}(\phi_1' - \epsilon_1, \tau)}{\theta_1(2\phi_1') \theta_1(2\phi_1' - \epsilon_{1,2}) \theta_1(2\phi_1' - 2\epsilon_+) \theta_1(\phi_1' \pm \tau/4)} \Bigg) \, .
\end{align}
One can compute the higher string contributions iteratively.

We now relate this 6d/2d partition function to the Wilson loop expectation values in the 5d $SU(3)_9$ theory. Using the identities \eqref{eq:theta_three_identity} and \eqref{eq:theta_shift-tau}, we can recast the 2d perturbative part as
\begin{align}
Z_{\mathrm{pert}}^{2d} = \frac{q^{1/3} \theta_3(0) \theta_3(\tau/2) \theta_3(2z)}{2\eta^4} \mathcal{W}_1 - \frac{q^{5/24} \theta_2(0) \theta_2(\tau/2) \theta_2(2z)}{2\eta^4} \mathcal{W}_2\ ,
\end{align}
where
\begin{align}
\mathcal{W}_1 = q^{-1/3} e^{2\phi_0'} \theta_3(2\phi_1') \, , \quad
\mathcal{W}_2 = q^{-5/24} e^{2\phi_0'} \theta_2(2\phi_1') \, .
\end{align}
In this expression, the terms depending on the dynamical K\"ahler parameters $\phi'_i$ are separated from the $z$-dependent terms. This implies that the expectation value of each $\mathcal{W}_{1,2}$ defined as
\begin{align}
\langle \mathcal{W}_{1,2}\rangle = \mathcal{W}_{1,2} + \sum_{k=1}^\infty e^{-3k\phi_0'} \mathcal{W}_{1,2}^{(k)} \, 
\end{align}
independently solves the same blowup equations for the 6d/2d partition function.

It turns out that $ \langle \mathcal{W}_2 \rangle $ is the VEV of the fundamental Wilson loop in the 5d $ SU(3)_9 $ theory. We find using the parameter map (\ref{eq:k=9_su3z2_basis}) that $ \langle \mathcal{W}_2 \rangle $ is expanded as
\begin{align}
\langle \mathcal{W}_2 \rangle = e^{\phi_2} + e^{-\phi_2 + \phi_1} + e^{-\phi_1} + \cdots \, ,
\end{align}
where $ \cdots $ involves the 5d instanton corrections that precisely match the spectrum of the fundamental Wilson loop given in Table~\ref{table:SU3_9_fund} with $ x=y=0 $. This may justify our conjecture above fixing $ x=y=0 $.
 On the other hand, $ \langle \mathcal{W}_1\rangle $ is expanded as
\begin{align}
\langle \mathcal{W}_1\rangle = e^{2\phi_1} + e^{-2\phi_1 + 2\phi_2} + e^{-2\phi_2} + \cdots \, .
\end{align}
This expansion suggests that $ \langle \mathcal{W}_1\rangle $ is the VEV of the Wilson loop operator in $ \mathbf{r} = [-2, 0] $ representation, i.e., the rank-2 anti-symmetric representation of $ SU(3) $. We summarize the instanton correction of the Wilson loop operator in Table~\ref{table:SU3_9_symm}. This illustrates the relationship between the codimension-4 defect in the twisted $SU(3)$ theory and the Wilson loop operators in the dual 5d theory.

\begin{table}
	\centering
	\begin{tabular}{|c|C{28ex}||c|C{28ex}|} \hline
		$ \mathbf{d} $ & $ \oplus \tilde{N}_{j_l, j_r}^{\mathbf{d}} (j_l, j_r) $ & $ \mathbf{d} $ & $ \oplus \tilde{N}_{j_l, j_r}^{\mathbf{d}} (j_l, j_r) $ \\ \hline
		$ (1, 2, 1) $ & $ (0, 0) \oplus (0, 1) \oplus (\frac{1}{2}, \frac{1}{2}) $ & $ (1, 2, 2) $ & $ (0, 1) \oplus (0, 2) \oplus (\frac{1}{2}, \frac{3}{2}) $ \\ \hline
		$ (1, 2, 3) $ & $ (0, 2) \oplus (0, 3) \oplus (\frac{1}{2}, \frac{5}{2}) $ & $ (1, 3, 1) $ & $ (0,0) \oplus (0,1) \oplus (\frac{1}{2},\frac{1}{2}) $ \\ \hline
		$ (1, 3, 2) $ & $ (0,0) \oplus 2(0,1) \oplus (0,2) \oplus (\frac{1}{2},\frac{1}{2}) \oplus (\frac{1}{2},\frac{3}{2}) $ & $ (1, 3, 3) $ & $ (0,1) \oplus 2(0,2) \oplus (0,3) \oplus (\frac{1}{2},\frac{3}{2}) \oplus (\frac{1}{2},\frac{5}{2}) $ \\ \hline
		$ (2, 2, 0) $ & $ (0, 0) $ & $ (2, 2, 1) $ & $ (0,1) \oplus (0,2) \oplus (\frac{1}{2},\frac{3}{2}) $ \\ \hline
		$ (2, 2, 2) $ & $ (0,0) \oplus 3(0,2) \oplus 2(0,3) \oplus (\frac{1}{2},\frac{3}{2}) \oplus 2(\frac{1}{2},\frac{5}{2}) \oplus (\frac{1}{2},\frac{7}{2}) \oplus (1,3) $ & $ (2, 2, 3) $ & $ (0,1) \oplus 2(0,2) \oplus 5(0,3) \oplus 4(0,4) \oplus (\frac{1}{2},\frac{3}{2}) \oplus 2(\frac{1}{2},\frac{5}{2}) \oplus 5(\frac{1}{2},\frac{7}{2}) \oplus 2(\frac{1}{2},\frac{9}{2}) \oplus (1,3) \oplus 2(1,4) \oplus (1,5) \oplus (\frac{3}{2},\frac{9}{2}) $ \\ \hline
		$ (2, 3, 1) $ & $ (0,0) \oplus 2(0,1) \oplus (0,2) \oplus (\frac{1}{2},\frac{1}{2}) \oplus (\frac{1}{2},\frac{3}{2}) $ & $ (2, 3, 2) $ & $ (0,0) \oplus 4(0,1) \oplus 6(0,2) \oplus 3(0,3) \oplus (\frac{1}{2},\frac{1}{2}) \oplus 4(\frac{1}{2},\frac{3}{2}) \oplus 4(\frac{1}{2},\frac{5}{2}) \oplus (\frac{1}{2},\frac{7}{2}) \oplus (1,2) \oplus (1,3) $ \\ \hline
		$ (2, 3, 3) $ & \multicolumn{3}{C{68.5ex}|}{$ \! (0,0) \oplus 4(0,1) \oplus 10(0,2) \oplus 13(0,3) \oplus 6(0,4) \oplus (\frac{1}{2},\frac{1}{2}) \oplus 4(\frac{1}{2},\frac{3}{2}) \oplus 10(\frac{1}{2},\frac{5}{2}) \oplus 10(\frac{1}{2},\frac{7}{2}) \oplus 3(\frac{1}{2},\frac{9}{2}) \oplus (1,2) \oplus 4(1,3) \oplus 4(1,4)\oplus (1,5) \oplus (\frac{3}{2},\frac{7}{2}) \oplus (\frac{3}{2},\frac{9}{2}) \! $} \\ \hline
	\end{tabular}
	\caption{Spectrum of the Wilson loop operator in the rank-2 anti-symmetric representation in the $ SU(3)_9 $ theory for $ (d_1, d_2, d_3) \leq (2, 3, 3) $. Here, $ \mathbf{d} = (d_1, d_2, d_3) $ labels the BPS states with fugacity $ Q_2^{-4/3} Q_3^{-2/3} \prod Q_i^{d_i} $.}\label{table:SU3_9_symm}
\end{table}

\section{Conclusion}\label{sec:conclusion}

In this paper, we proposed blowup equations for the partition functions with Wilson loops in generic 5d/6d theories including non-Lagrangian theories by extending the ordinary blowup equations for the partition functions on the $\Omega$-background. We presented how to formulate and solve the blowup equations which enables one to compute VEVs of Wilson loop operators in various representations. When it takes the form of a 1d index, we conjecture that the solution of the blowup equations correctly counts the spectrum of 1d BPS bound states to the loop operator. We tested this proposal with many interesting examples, which involves Wilson loops in 5d gauge theories with $G_2$ and $F_4$ gauge symmetries and codimension-4 defects in 6d SCFTs as well as loop operators in the simplest non-Lagrangian theory from the CY 3-fold of a local $\mathbb{P}^2$, by showing that the results from the blowup method match the known partition functions of loop operators and also agree with those expected from dualities. Our work may provide a universal framework to compute Wilson loop expectation values in arbitrary representations in 5d and also in 6d field theories.

Some open questions deserve further investigation. First, there is a problem in solving the blowup equations for Wilson loops in the representations of large electric charges. We found that the solutions for Wilson loops in certain big representations, for example, the rank-5 or higher rank symmetric representation in the $SU(2)$ gauge group, contain unphysical states in the first few terms in the K\"ahler parameter expansion. These unphysical states, which depend on dynamical K\"ahler parameters, however, either do not form a representation of the $SO(4)$ Lorentz rotation or have negative degeneracies. Interestingly, all the 1d bound states after a few terms in the expansion properly form the representations of the Lorentz rotation with positive degeneracies. In addition, we observed that the blowup equations with different background fluxes are solved independently to produce the same result after so few terms. We think these observations signal that the solution to the blowup equations correctly captures the spectrum of the Wilson loop states except a few leading states in the expansion. To consolidate our blowup approach for generic loop operators, we do need to understand the origin of these unphysical states and how to exclude them from the solution while solving the blowup equations.

A natural generalization of the blowup approach is now to formulate the blowup equations for other types of defects. The blowup equations for the surface defects in the 4d $\mathcal{N}=2$ $SU(2)$ gauge theory with four fundamental hypermultiplets have been proposed in \cite{Nekrasov:2020qcq,Jeong:2020uxz} and they are used to establish new relations for the Painlev\'e VI tau-function. Similarly, we can consider the blowup equations for co-dimension-2 defect operators in 5d and also in 6d. This may provide us a new way of defining supersymmetric codimension-2 defects in higher dimensional theories and systematically calculate their spectra.

Finally, we can consider a 4d reduction of the blowup formula for Wilson loop operators. Under the reduction along a circle, the Wilson loop operators in a 5d gauge theory wrapping a circle will be reduced to chiral operators in the resulting 4d $\mathcal{N}=2$ gauge theory, which are among the most important observables that can be studied. The expectation values of such chiral operators in the $SU(N)$ (or $U(N)$) gauge theories were calculated using so-called qq-characters in \cite{Nekrasov:2015wsu}, while a systematic computational tool for those in other gauge groups is still lacking. The 4d reductions of our blowup equations may offer a complementary approach to computing the expectation values of 4d chiral operators. This would be an interesting future research to pursue.

\acknowledgments
We would like to thank Kimyeong Lee, Jaewon Song, Marcus Sperling and Futoshi Yagi for useful discussions. The research of HK and MK is supported by the National Research Foundation of Korea (NRF) Grant 2018R1D1A1B07042934. The research of SK is partially supported by the Fundamental Research Funds for the Central Universities 2682021ZTPY043.

\appendix

\section{Elliptic functions}\label{appendix:elliptic}

We summarize the definitions and some properties of the elliptic functions used in this paper.

Let $ \tau $ be the complex structure of a torus, $ q=e^{-\tau} $. The Dedekind eta function is defined to be
\begin{align}
\eta(\tau) = q^{1/24} \prod_{n=1}^\infty (1-q^n) \, .
\end{align}
The theta function with characteristics is
\begin{align}\label{eq:theta_char}
\vartheta \genfrac[]{0pt}{1}{\alpha}{\beta} (\tau, x) = \sum_{n \in \mathbb{Z}} q^{\frac{1}{2}(n + \alpha)^2} y^{n+\alpha} e^{2\pi i \beta (n+\alpha)} \, ,
\end{align}
where $ y = e^{-x} $. The Jacobi theta functions are defined to be
\begin{alignat}{2}
&\theta_1(\tau, x) = -\vartheta \genfrac[]{0pt}{1}{1/2}{1/2} (\tau, x) \, , \qquad
&&\theta_2(\tau, x) = \vartheta \genfrac[]{0pt}{1}{1/2}{0} (\tau, x) \nonumber \\
&\theta_3(\tau, x) = \vartheta \genfrac[]{0pt}{1}{0}{0} (\tau, x) \, , \qquad
&&\theta_4(\tau, x) = \vartheta \genfrac[]{0pt}{1}{0}{1/2} (\tau, x) \, .
\end{alignat}
It follows from \eqref{eq:theta_char} that they are given by
\begin{alignat}{2}
&\theta_1(\tau, x) = -i \sum_{n \in \mathbb{Z}} (-1)^{n} q^{\frac{1}{2}(n+1/2)^2} y^{n+1/2} \, , \quad
&&\theta_2(\tau, x) = \sum_{n \in \mathbb{Z}} q^{\frac{1}{2}(n+1/2)^2} y^{n+1/2} \, , \nonumber \\
&\theta_3(\tau, x) = \sum_{n \in \mathbb{Z}} q^{\frac{n^2}{2}} y^n \, , \quad
&&\theta_4(\tau, x) = \sum_{n \in \mathbb{Z}} (-1)^n q^{\frac{n^2}{2}} y^n \, .
\end{alignat}
The infinite product representations of the Jacobi theta functions are
\begin{align}
\theta_1(\tau, x) &= -i q^{1/8} y^{1/2} \prod_{n=1}^\infty (1-q^n) (1-y q^n) (1-y^{-1} q^{n-1}) \, , \nonumber \\
\theta_2(\tau, x) &= q^{1/8} y^{1/2} \prod_{n=1}^\infty (1-q^n) (1+y q^n) (1+y^{-1} q^{n-1}) \, , \nonumber \\
\theta_3(\tau, x) &= \prod_{n=1}^\infty (1-q^n) (1+y q^{n-1/2}) (1+y^{-1} q^{n-1/2}) \, , \nonumber \\
\theta_4(\tau, x) &= \prod_{n=1}^\infty (1-q^n) (1-y q^{n-1/2}) (1-y^{-1} q^{n-1/2}) \, .
\end{align}

There are various identities which are useful for simplifying equations. The 6d perturbative partition function can be expressed in terms of elliptic functions using
\begin{align}\label{eq:PE_vec_theta}
\PE\qty[\qty(y+\frac{q}{y}) \frac{1}{1-q}]
= \prod_{n=0}^{\infty} \frac{1}{(1-y q^n)(1-y^{-1} q^{n+1})} = \frac{i q^{1/12} \eta(\tau)}{y^{1/2} \theta_1(\tau, x)}
\end{align}
for vector multiplet and
\begin{align}\label{eq:PE_mat_theta}
\PE\qty[-\qty(y+\frac{q}{y}) \frac{1}{1-q}]
= \prod_{n=0}^\infty (1-y q^n)(1-y^{-1} q^{n+1}) = -\frac{i y^{1/2} \theta_1(\tau, x)}{q^{1/12} \eta(\tau)}
\end{align}
for hypermultiplet. However, the hypermultiplet part is affected by constructing hatted partition function, $ (-1)^F \to (-1)^{2J_R} $. Such replacement effectively affects $ x \to x - \pi i $, or $ y \to y e^{\pi i} $ in \eqref{eq:PE_mat_theta}. The following half-period shift identities are useful to convert unhatted partition function to hatted partition function, or vice versa:
\begin{align}\label{eq:theta_shift}
&\theta_1(\tau, x - \pi i) = \theta_2(\tau, x) \, , \qquad
\theta_2(\tau, x - \pi i) = -\theta_1(\tau, x) \, , \nonumber \\
&\theta_3(\tau, x - \pi i) = \theta_4(\tau, x) \, , \qquad
\theta_4(\tau, x - \pi i) = \theta_3(\tau, x) \, .
\end{align}
The other half-period shift relations are
\begin{alignat}{2}\label{eq:theta_shift-tau}
&\theta_1(\tau, x + \tau/2) = i q^{-1/8} y^{-1/2} \theta_4(\tau, x) \, , \quad
&&\theta_2(\tau, x + \tau/2) = q^{-1/8} y^{-1/2} \theta_3(\tau, x) \, , \nonumber \\
&\theta_3(\tau, x + \tau/2) = q^{-1/8} y^{-1/2} \theta_2(\tau, x) \, , \quad
&&\theta_4(\tau, x + \tau/2) = i q^{-1/8} y^{-1/2} \theta_1(\tau, x) \, .
\end{alignat}
Some miscellaneous identities that used in this paper are as follows. First, the relation between theta functions with modular parameter $ \tau $ and $ 2\tau $ is
\begin{align}\label{eq:theta_identity}
&\theta_1(\tau, x) \theta_1(\tau, y) = \theta_3(2\tau, x+y) \theta_2(2\tau, x-y) - \theta_2(2\tau, x+y) \theta_3(2\tau, x-y) \, .
\end{align}
Second, the three-term Weierstrass addition identities that we used are
\begin{align}\label{eq:theta_three_identity}
&(-1)^{i-1} \theta_1(\tau, x_1+x_2) \theta_1(\tau, x_1-x_2) \theta_1(\tau, x_3+x_4) \theta_1(\tau, x_3-x_4)  \nonumber \\
& \qquad \qquad = \theta_i(\tau, x_1 + x_3) \theta_i(\tau, x_1 - x_3) \theta_i(\tau, x_2 + x_4) \theta_i(\tau, x_2 - x_4) \nonumber \\
&\qquad \qquad \quad - \theta_i(\tau, x_1 + x_4) \theta_i(\tau, x_1 - x_4) \theta_i(\tau, x_2 + x_3) \theta_i(\tau, x_2 - x_3) \quad (i = 1, 2, 3, 4) \, , \nonumber \\
&\theta_2(\tau, x_1+x_2) \theta_2(\tau, x_1-x_2) \theta_2(\tau, x_3+x_4) \theta_2(\tau, x_3-x_4)  \nonumber \\
&\qquad \qquad = \theta_3(\tau, x_1 + x_3) \theta_3(\tau, x_1 - x_3) \theta_3(\tau, x_2 + x_4) \theta_3(\tau, x_2 - x_4) \nonumber \\
&\qquad \qquad \quad - \theta_4(\tau, x_1 + x_4) \theta_4(\tau, x_1 - x_4) \theta_4(\tau, x_2 + x_3) \theta_4(\tau, x_2 - x_3) \, .
\end{align}
Other useful identities as well as these identities can be found in \cite{Kharchev_2015}. We frequently use $ \eta = \eta(\tau) $, $ \theta_i(x) = \theta_i(\tau, x) $, $ \theta_i(x_1 \pm x_2) = \theta_i(x_1 + x_2) \theta_i(x_1 - x_2) $, $ \theta_i(x_1 + x_{2,3}) = \theta_i(x_1 + x_2) \theta_i(x_1 + x_3) $, etc., for simplicity.

\bibliographystyle{JHEP}
\bibliography{refs}
\end{document}